\documentclass[aps,pra,twocolumn,groupedaddress]{revtex4}
\usepackage{comment,psfrag,appendix,dcolumn,epsf,float,amssymb}
\usepackage{verbatim}
\usepackage{epsfig}
\usepackage{graphicx}
\usepackage{amsmath}
\usepackage{bm}

\begin{document}
\title{Vortex Molecules in Spinor Condensates}

\author{Ari M. Turner$^{*\dagger}$ and Eugene Demler$^*$}
\affiliation{$^*$Department of Physics, Harvard University, Cambridge MA 02138}
\affiliation{$^\dagger$%
Department of Physics, University of California, Berkeley CA 94720}
\date{\today}
\begin{abstract}Condensates of atoms with spins 
can have vortices of several types; these
are related to the symmetry group of the atoms' ground state.  
We discuss how, when a condensate is placed in a small magnetic field that
breaks the spin symmetry, these vortices may form bound states.  Using
symmetry classification of vortex-charge and rough estimates for
vortex interactions,
one can show that some configurations that are stable
at zero temperature can
decay  at finite temperatures by crossing over energy barriers.  Our
focus is cyclic spin 2 condensates, which have tetrahedral symmetry.
\end{abstract}
\maketitle
In an image of a nematic liquid crystal by polarized light,
one can identify topological defects of various topological
charges (see Ref. \cite{chandrasekharbook}).
Bose condensates (see the
books\cite{pethickbook, stringaribook}) are starting to provide
another context for studying topological defects: in the texture
formed by the phase \emph{and}
spin of a spinor condensate\cite{spindomains, ho98,ohmi98}.
(See also \cite{lewenstein06,bloch} for reviews of the theory and
experimental techniques being applied to spinor condensates.)

A topological defect in the \emph{phase} 
\emph{of a superfluid} is a quantized vortex.
The discontinuity in the phase as the defect is encircled must be $2\pi n$
for an integer $n$, and the circulation of the vortex is then
$n\frac{h}{m}$.
In a single-component superfluid, multiply quantized vortices ($|n|>1$)
are usually
not stable. The widely known explanation is that the energy of a vortex
is proportional to $n^2$. 
Thus a doubly quantized vortex ($n^2=4$) can lower
its energy by splitting into two singly quantized vortices. Similar
arguments can be formulated for multicomponent condensates, but we will find
that some vortices in these condensates can be very long-lived
in spite of having large energies.
Our predictions are about condensates of atoms with spin in which the rotational
symmetry has been
weakly disrupted by a small parameter $q$, 
such as the interaction
between the spins and the magnetic field.  (We focus
on the \emph{cyclic} condensates of spin 2 atoms, see Ref. \cite{ciobanu}.)

%Long-lived multiply quantized
%vortices are a special case of another phenomenon:
%vortices may be bound together to produce vortices with
%composite cores, and some of the \emph{composite vortices}
%are multiply quantized.
%Any cluster of vortices can be regarded
%as a composite vortex with a net topological charge that is the
%(noncommutative) sum of all the charges of the vortices that make it up.
%The behavior of the order parameter at infinity is determined by this 
%net charge, hence from far away, a composite
%vortex cannot be distinguished from a point vortex, and it is surprising
%to find a vortex with two quanta of circulation that is stable, the
%component with smaller charges are hidden.
%Symmetry violating fields provide the force that
%binds such a group of vortices together.

Long-lived multiply quantized
vortices are particular examples
of composite vortices, which
are made up of several vortices bound together.  Ground states with
complicated symmetries have many types of vortices\cite{mermin79}.
Symmetry violating fields provide a force that
can bind some groups types vortices together so that they form
a  ``composite
core"  for a larger vortex.  
%Now any cluster of vortices can be regarded,
%from far away,
%as a composite vortex with a net topological charge that is the
%(noncommutative) sum of all the charges of the vortices that make it up.
The composite vortex might have two quanta of circulation (or have
some other extra-large topological charge).
Such a vortex
would be surprising if observed under a low resolution, because
the smaller vortices that make it up would be hidden in its core;
the behavior of the order parameter at infinity is determined entirely
by the \emph{net} topological charge.
%that the composite vortex has separate components.

Some earlier theories about
vortices in multicomponent condensates also describe
vortices with
asymmetrical cores which can be regarded as collections of closely spaced
vortices.  The parameter $q$ makes
this interpretation even more meaningful: 
when $q$ is very small, we find that the component
vortices move far apart, so that their cores do not overlap, while still
remaining bound.
%For the vortices studied here, with spin 2 atoms in a magnetic field these
% cores consist of tightly bound $q=0$ vortices while the
%vortex as a whole corresponds to a nonzero value of $q$.
Ref. \cite{kasamatsu}
describes vortices that can occur in a condensate of two atomic states
when there is an RF-field producing coherent transitions between the
states.  The binding of these vortices
also comes from an asymmetry, but the asymmetry comes
from the dependence of the
interaction strength
on the internal states of the atoms, rather than from an external magnetic field.
Ref. \cite{isoshima} studies vortices of spinor atoms in
a magnetic field, like us, but focuses on spin 1,
and describes composite vortices as well.  
Since the scenario involves rotation
as well as a magnetic field, the vortices would be held close to the
\emph{axis} of the condensate by rotational confinement 
without the magnetic field.  With just a magnetic field,
we show that the vortices are attracted to \emph{one another}.

Vortices and bound states are not hard to picture, by taking advantage of 
the fact that the state of a spinor atom can be represented by a geometrical 
figure.  The appropriate 
shape depends on the type of condensate.  For a ferromagnetic
condensate, a stake pointing in the direction of the magnetization
could represent the local state of the condensate.  Other condensates can
be represented by more complicated shapes.  Now imagine a plane filled with
identical shapes (tetrahedra, for the cyclic phase), 
with orientations varying continuously as a function
of position.  This shape field (or ``spin texture") together with a phase
field would represent a nonuniform state of a condensate.  If the shapes
rotate around a fixed symmetry axis as some point is encircled
then the spin texture has a topological defect at this point.  Such
configurations generalize vortices,
because some of them are accompanied by persistent spin or charge
currents. For each symmetry of the
tetrahedron, there will be a vortex when the Hamiltonian is $SU_2$ symmetric.
(Each discrete subgroup of $SU_2$ describes the vortices of some
phase for atoms of some spin\cite{inert}; to find vortices for more
even more complicated groups like $SO_5$ or $SO_7$, one might want to
study gases of spin $\frac{3}{2}$ atoms\cite{threehalf,tsvelik}.) The
vortex spectrum is decimated when a
magnetic field is applied, since the field favors a particular orientation
of the order parameter
--e.g.,
tetrahedra may want to have an order-three axis aligned with the magnetic field.
The spectrum of vortices is then reduced from the full spectrum
of ``tetrahedral" vortices (based on arbitrary symmetries of the tetrahedron)
to ``field-aligned" vortices, where the tetrahedra must rotate
around this order three axis, so as not to lose their alignment with
the magnetic field. 
We may introduce the ground state 
space $\mathcal{M}_q$ (of tetrahedra aligned with
the magnetic field), where the interaction energy is $V=V_{min}$,
and the space of ground states of
the $SU_2$ invariant part of the Hamiltonian,
$\mathcal{M}$ (of tetrahedra  with arbitrary orientations).  
A set of vortices can be assigned a topological charge
based on the loop
traced out in one of these spaces by the values of the order
parameter on a circle around the set.  Specifically,
the topological charge describes the symmetry transformation
that brings the order parameter back to its initial value
as the set of
vortices is circumambulated. (The relation between topology and
symmetry is presented in \cite{mermin79}.)
From the general point of view, the reduction of the
charge types from the tetrahedral to the field-aligned ones
when $q$ is turned on results because
the smaller ground state space at nonzero $q$, $\mathcal{M}_q$,
has fewer closed loops\footnote{This is a
small oversimplification.  Decreasing the size of the space
$\mathcal{M}$ also stabilizes vortices that would be unstable at $q=0$--removing
a portion of $\mathcal{M}$ can leave a new hole for a vortex-circuit to wrap around.}.

%a system on a circle surrounding the vortex core.  Although there are many
%ways to add wiggles to a given circuit,
% only the topological structure is important,
%leading to a discrete set of possible vortex charges. The ground state
%space of the $SU_2$ invariant part of the Hamiltonian is $\mathcal{M}$,
%with all orientations of tetrahedra; closed
%loops in this space correspond to arbitrary symmetries of a tetrahedron.
%As one travels around a loop containing a vortex core, the order parameter 
%has to rotate continuous around a symmetry axis so
%that the corresponding sequence of spinors in $\mathcal{M}$ returns
%to the original spinor.  There are fewer vortices when $q\neq 0$ because
%the ground state space then, $\mathcal{M}_q$, is smaller: it only consists
%of orientations of the tetrahedra that are aligned with the magnetic field.
%Some closed loops in $\mathcal{M}$ cannot be pushed into $\mathcal{M}_q$ 
%(which has lower magnetic energies) without going into states of much higher
%energies outside of $\mathcal{M}$. A vortex whose field passes
%through $\mathcal{M}$ has a very large energy because the cost of
%leaving $\mathcal{M}_q$ is incurred on each circle out to infinity.

\begin{figure}
\psfrag{Gamma}{$\Gamma$}
\psfrag{Q}{$Q$}
\includegraphics[width=.45\textwidth]{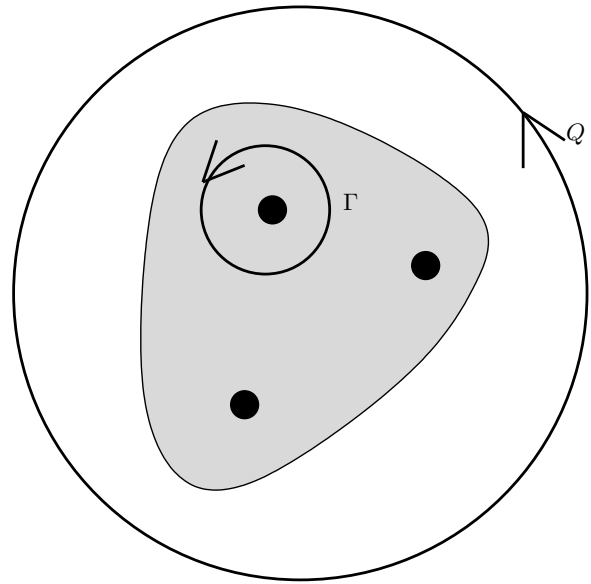}
\caption{\label{fig:cherimoya}
 A composite vortex reflects 
the hierarchy of the order parameter space.  In the white, grey, and black 
regions the order parameter moves from $\mathcal{M}_q$ to $\mathcal{M}$ 
to $\mathcal{H}$ which has the highest energy of all.  The charges of the subvortices are tetrahedral charges, represented by $\Gamma$ and the charges 
of the composite vortex is a ``field-aligned" charge.}
\end{figure}

%Vortices can be understood and classified in terms of order parameter
%spaces--this more general perspective notes that along rays
%from a vortex, the order parameter approaches ground states of the system.
%A vortex can therefore be classified by the topology of the circuit 
%traced out in the manifold of degenerate ground states of the system.
If $q$ is very small, the excess energy of an order
parameter in $\mathcal{M}$ rather than $\mathcal{M}_q$ is small.  
There is then a hierarchy of spaces $\mathcal{M}_q$, $\mathcal{M}$
and $\mathcal{H}$ (the whole Hilbert space, corresponding
to arbitrarily \emph{distorted} tetrahedra) with increasing energy scales. %based on the topology of the
%circuit traced out in this abstract space  on a circle containing
%the vortex.  Although there are many ways to add wiggles to a given
%circuit, only the topological structure is important, leading to a discrete
%set of possible vortex charges.  
%For a small $q$,
%the order parameter space can be stratified.   In increasing order,
%it is built of the $\mathcal{M}_q$, the space of absolute ground
%states when the magnetic field is turned on (for
%the cyclic state, the tetrahedra are aligned with the magnetic
%field), $\mathcal{M}$, the space of ground
%states for the $SU_2$ symmetric part of the Hamiltonian (tetrahedra
%with arbitrary orientations) and finally, 
%the entire
%Hilbert space, $\mathcal{H}$ (order parameters represented by arbitrary
%shapes).  Field-aligned and tetrahedral charges correspond
%to closed loops in $\mathcal{M}$ and $\mathcal{M}_q$ respectively.
When $q\neq 0$, a vortex has to have a field-aligned
charge \emph{at infinity} because the tetrahedra must
eventually move into $\mathcal{M}_q$ 
to avoid too big of an energy cost.
But if $q$ is small, the hierarchy of the order parameter space
is reflected in the fact that
a vortex with a field-aligned charge
can have a \emph{composite} core that may contain tetrahedral
vortices, as
in Fig. \ref{fig:cherimoya}; the core is sort of like the pulp
and
seeds of a fruit.  Inside the fruit is a texture of arbitrarily
oriented tetrahedra, almost as if $q$ were equal to zero.
The seeds are
then the cores of tetrahedral vortices and the pulp is qualitatively
the same as the texture that would
surround these vortices in the absence
of the magnetic field.  The net charge of
the vortices has to be field-aligned so that the loop of order parameters
that surrounds the whole core
can move into $\mathcal{M}_q$ as $r\rightarrow\infty$. This picture
becomes more accurate as $q$ becomes small, since the vortex
cores within the composite core are far apart in that case--
the size of a region where
the order parameter is in $\mathcal{M}-\mathcal{M}_q$ or $\mathcal{H}-\mathcal{M}$
is inversely
proportional to the energy scale for each space.
As $q\rightarrow 0$,
the binding of the tetrahedral charges  becomes weaker and weaker,
until they become free from each other; each vortex therefore is described by
independent degrees of freedom when $q=0$.

The optimal size $L_q$ of a composite vortex results
from competition between two forces--a confinement
force from the anisotropy term and the familiar
logarithmic interaction of vortices.
The symmetry breaking energy favors minimizing the area
over which the order parameter
leaves $\mathcal{M}_q$, pushing the component vortices
toward one another. On the other hand, it cannot compress them
to a point since the Coulomb-like repulsion one expects of
vortices keeps them apart.

Some of these vortex molecules will turn out to be
metastable.  Ref. \cite{isoshima} mentions an interesting clue to
such a  phenomenon; namely 
there are multiple
steady state wave functions describing a condensate with a given
magnetization and rotational frequency;
these local minima of the energy function
can maybe be analyzed using the group theoretic
metastability conditions we discuss in Section
\ref{sec:metastable}. For a spinor condensate,
wave functions for states besides the ground
state are experimentally important, since
the experiments of Ref. \cite{tiedye}, as well as the liquid
crystal experiments of Ref. \cite{ike}, 
reveal complicated textures produced by chance;
an initial fluctuation around a uniform excited state becomes
unstable and evolves into an intricate \emph{nonequilibrium} texture.
So it is useful to analyze 
spin textures which are only 
\emph{local} minima of the Gross-Pitaevskii energy functional (like the metastable
vortex
molecules considered here)
as well as unstable equilibria (which take a long
time to fluctuate out of their initial configuration).
Examples of unstable equilibrium 
are described for single-component condensates by \cite{pietila}.
The process by which textures form out of uniform initial states
has been discussed in theoretical
articles, including \cite{quantumquenchedvortices,quantumquenches} 
(on the statistics of the spin
fluctuations and vortices that are
produced from this random evolution), 
\cite{cherng1} (on the spectrum of instabilities)
 and \cite{ferrodynamics,lewentropy}
on the dynamics of spinor condensates.  The experiments
described in \cite{vengdipole} show that the patterns that evolve
in rubidium condensates are probably affected by dipole-dipole interactions,
though we are not considering these.  Dipole-dipole interactions lead
to antiferromagnetic phases\cite{meystredipole}, which maybe can
be described as \emph{ground state} configurations of vortices.

Besides just hoping for unusual types of
vortices to form,
one can make a vortex lattice by rotating a condensate.  The effects of
rotating on a spinor condensate have been investigated theoretically
in \cite{ryan} as well as the review \cite{multicompvortexreview}.
Experiments can also make a single vortex of a prescribed 
type\cite{twistvortex,corelessspinor,shinvortexsplit}.  Excited by these
possibilities, physicists have come up with several types of vortices
and topological defects they would be interested in seeing: skyrmions\cite{skyrmions,moreskyrmions}, monopoles\cite{monopoles}, textures whose order
parameter-field lines make linked loops\cite{knots}, 
as well as the noncommutative vortices of the cyclic phase that
we will be expanding on here\cite{makela}.  %Vortices in a spinor
%condensate can also have interesting effects on the Kosterlitz-Thouless
%transition in such 
%condensates\cite{subrotovortices,dannyvortices,aleinervortices}.

\begin{figure}
\psfrag{a}{a)}
\psfrag{b}{b)}
\psfrag{x}{$x$}
\psfrag{y}{$y$}
\psfrag{z}{$z$}
\includegraphics[width=.45\textwidth]{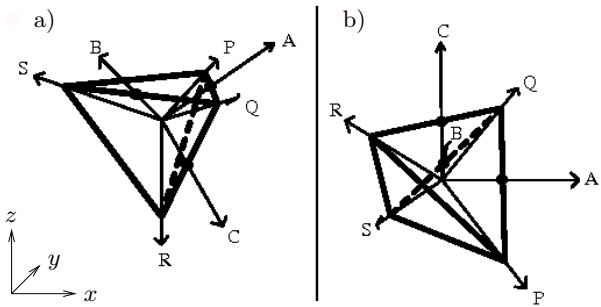}
\caption{\label{fig:bottler} Geometrical representation of
the tetrahedral superfluid phases of spin 2 atoms.
Two orientations of a tetrahedron, corresponding to the spin roots
of the ground state spinor, are illustrated
and the symmetry axes are labelled. (The orientation of the coordinate
axes, shown on the side, is the same in both cases.)
a) An orientation with
the vertical $z$-axis along an order three axis, corresponding
to the ground state $\sqrt{n_0}\chi_3$ (see Eq. (\ref{eq:order3})).  
b) An orientation
with the $z$-axis along an order two axis,
corresponding to the ground state $\sqrt{n_0}\chi_2$ (see Eq. (\ref{eq:chi2})).
For a magnetic field along the $z$ axis,
orientation a) is preferred when $c>4$ and orientation b) is
preferred when $c<4$. Note that the three order $2$ axes, labelled
$A,B,C$ can be used as a set of three orthogonal body axes.}
\end{figure}
The key to our discussion of vortices will be a geometrical
representation of the order parameter, allowing us to
visualize a texture of the cyclic phase as a field of tetrahedra
with different orientations. (See Fig. \ref{fig:bottler}.)  Without
this representation, a spin texture would be given by a spinor field 
$(\psi_2(x,y),\psi_1(x,y),\psi_0(x,y),\psi_{-1}(x,y),\psi_{-2}(x,y))^T$; the
fact that this spinor lies in the ground state manifold $\mathcal{M}$
would have
to be described by a set of
polynomial relations between the five components.
A more revealing way to represent a spinor is to draw
a geometrical figure consisting of ``spin-roots" (as in
\cite{icosa}),
and in the cyclic phase, these spin roots form the vertices of
a tetrahedron. (A similar construction can be used to classify vortices 
in condensates of spin 3 atoms, see Ref. \cite{spin3}.)
Even without using the spin-root interpretation,
one can justify using tetrahedra to
represent order parameters in the cyclic phase because they are a concrete
way of representing the symmetry of this phase.
Ref. \cite{makela} worked out the symmetry group
of a state in the cyclic phase by finding all the pairs $\bm{\hat{n}},\alpha$
such that
\begin{equation}
e^{-i\alpha\bm{F}\cdot\bm{\hat{n}}}\chi_3\propto\chi_3
\label{eq:syms}
\end{equation}
where 
\begin{equation}
\chi_3=\left(\begin{array}{c}\sqrt{\frac{1}{3}}\\0\\0\\\sqrt{\frac{2}{3}}\\0\end{array}\right)
.\label{eq:order3}
\end{equation}
The spinor $\sqrt{n_0}\chi_3$ is
a representative cyclic state, if $n_0$ is the density of the condensate.
These symmetries are the same as the symmetries 
of a tetrahedron oriented as in Fig. \ref{fig:bottler}a; hence
we may
represent
the state $\sqrt{n_0}\chi_3$ by this tetrahedron.  Any other ground
state should be represented by a tetrahedron oriented so that
its symmetry axes and symmetry axes of the spinor coincide.  The
appropriate orientation of the tetrahedron for a given ground state
can be determined in an automatic way by calculating the spin roots.

The Hamiltonian for spin 2 atoms in a magnetic field, 
$\mathcal{H}=\iint 
d^2\mathbf{u} [\frac{\hbar^2}{2m}\nabla\psi^{\dagger}\nabla\psi +V_{tot}(\psi)]$
has a simple expression\cite{ciobanu}
in terms of the density $n=\psi^{\dagger}\psi$, 
magnetization $\mathbf{m}=\psi^{\dagger}\mathbf{F}\psi$, and singlet-pair
amplitude $\theta=\psi_{t}^{\dagger}\psi.$ 
($\psi_t$ stands for the time reversal of $\psi$.)
\begin{equation}
%\mathcal{H}=\iint \frac{1}{2}(\alpha n^2+\beta \mathbf{m}^2+c\beta
%|\theta|^2)
%-q\psi^{\dagger}F_z^2\psi d^2\mathbf{x}
V_{tot}(\psi)=\frac{1}{2}(\alpha n^2+\beta \mathbf{m}^2+c\beta|\theta|^2)
-q\psi^{\dagger}F_z^2\psi-\mu\psi^{\dagger}\psi
\label{eq:hamiltonian}
\end{equation}
The first three terms describe the rotationally symmetric interactions
of pairs of atoms.
The first one describes repulsion between a pair of atoms
and the next two terms describe additional, smaller interactions,
that depend on the spin states of the two colliding atoms.  These
terms determine the spinor ground state in the absence of a magnetic
field\cite{ciobanu}.
(The spin-dependent interaction  strengths $\beta,c\beta$ 
can be expressed in terms of the scattering lengths.)  
The properties of spin 2 atoms, which are described
by this Hamiltonian, and of spin 1 atoms 
have been investigated experimentally in Refs.
\cite{resonance,chang1and2,widera,schmaljohann,tiedye}; 
\cite{tunnelingreview} reviews
more experimental phenomena. Ref. \cite{widera,chang1and2} found values for
$\alpha,\beta$ and $c$ for $^{87}$Rb
that are consistent with theoretical predictions,
although even the sign of $c$ is not known for sure, because $c$ is small.
The final term contains the chemical potential $\mu$.

If $\beta$ and $c$ are positive, the ground state of a condensate
of spin 2 atoms is 
cyclic.
The deformation of a cyclic state due to
a magnetic field is the simplest if we assume that $\alpha\gg |\beta|$
and that $c$ is on the order of $1$.  We therefore
assume $c\sim 1$, though 
$c\ll 1$ for rubidium.
The magnetic field influences the atoms
through the fourth term of the energy, breaking the rotational symmetry of
the Hamiltonian; this term
describes the quadratic Zeeman shift due to a magnetic
field $B$ along the $z$-axis and $q\propto B^2$.
(See \cite{pethickbook} for the explanation of why the quadratic Zeeman 
term $q$ but not the linear 
term is relevant if the condensate's
initial magnetization is zero. A nonzero magnetization is described
by a Lagrange multiplier term
$-p\iint d^2\mathbf{x} \psi^{\dagger}F_z\psi$, which
looks like a linear Zeeman coupling. We assume $p=0$; a nonzero
$p$ should have similar consequences as a nonzero $q$, since
both break the rotational symmetry.) 
%If $q=0$, then
%as long as $c$
%and $\beta$ are positive, the phase is cyclic and an element $M$
%can  
%be regarded as an orientation of a tetrahedron (together with a phase\cite{makela,spin2sym}). 
%Each symmetry of the tetrahedron leads to a number of vortices.

Introducing a small $q$ 
reduces the ground state space from $\mathcal{M}$ to $\mathcal{M}_q$.
This can be explained (see Sec. \ref{sec:cubeshape}) by 
finding the modulation of the energy of an arbitrary
tetrahedral state as a function
of the orientation of the corresponding tetrahedron:
\begin{equation}
V_{eff}=(c-4)\frac{3q^2}{4c\beta}
(\cos^4\alpha_1+\cos^4\alpha_2+\cos^4\alpha_3),
\label{eq:effective}
\end{equation}
where $\alpha_1,\alpha_2$ and $\alpha_3$ are the angles between
the $z$-axis and three body-axes $A,B,C$ fixed to the tetrahedron (see Fig. 
\ref{fig:bottler}a).
The orientations that minimize this energy are the true ground
states at nonzero $q$.
If $c>4$ (assumed until Sec. \ref{sec:sampler}), 
then the tetrahedron in its ground state
is oriented with the $z$-axis
perpendicular to a face as in Fig. \ref{fig:bottler}a, with
a ground state energy 
%(corresponding to $\cos\alpha_i=\frac{1}{\sqrt{3}$)
of $V_{min}=\frac{(c-4)q^2}{4c\beta}$.
Thus the absolute ground state space $\mathcal{M}_q=\mathcal{M}_{q3}$ 
contains all
the wave functions that are arbitrary
rotations about the $z$-axis (combined with rephasings)
of $\sqrt{n_0}\chi_3$.
When $c<4$, the ground states $\mathcal{M}_{q2}$ are rotations about $z$
of the tetrahedron illustrated in Fig. 
\ref{fig:bottler}b.  In particular, when $q\neq 0$, there is a phase transition
at $c=4$, though there is nothing special about $c=4$ in \emph{zero} magnetic
field.  In this paper, we will mostly assume that $c>4$ because
our initial example occurs under this condition.

Consider first a single 
vortex associated with the rotation through $180^{\circ}$ of
the ground state tetrahedron (see Fig. \ref{fig:bottler}a)
about the $A$ axis, 
$(\sqrt{\frac{2}{3}},0,\sqrt{\frac{1}{3}})$.
Such a vortex is described at large distances by
\begin{equation}
\psi(r,\phi)=e^{-i\frac{\phi}{2}\frac{1}{\sqrt{3}}(\sqrt{2}F_x+F_z)}\sqrt{n_0}
\chi_3
\label{eq:bisect}
\end{equation}
where $r,\phi$ are polar coordinates centered on the core of the vortex.
This vortex has an excess energy (relative to the ground state energy
of the condensate) which diverges with the condensate size.  In fact,
its Zeeman
energy density (with $V_{min}$ subtracted) is given
according to Eq. (\ref{eq:effective}) by\footnote{The $\alpha$'s for this
vortex may be calculated by noting that,
relative to the tetrahedron, the $z$-axis of space rotates $180^{\circ}$
about $A$ so that $(\cos\alpha_1,\cos\alpha_2,\cos\alpha_3)$, its
coordinates relative to the body axes of the tetrahedron are easy
to work out as functions of $\phi$.}
%\begin{equation}
%\frac{(c-4)(q^2)}{12\beta}(3+\sin^2\phi)
%\label{eq:galacticarms}
%\end{equation}
\begin{equation}
\frac{(c-4)(q^2)}{6\beta c}(\sin^2\phi).
\label{eq:galaxyarms}
\end{equation}
The tetrahedron is pointing in the wrong direction except along
the positive and negative $x$-axis; 
and the integrated energy is proportional
to the area:
\begin{equation}
E_{misalign}\sim\frac{q^2R^2}{\beta c}
\label{eq:misalign}
\end{equation}
where $R$ is the condensate's radius.
Such a vortex cannot be the only vortex in an infinite condensate!

\begin{figure}
\includegraphics[width=.45\textwidth]{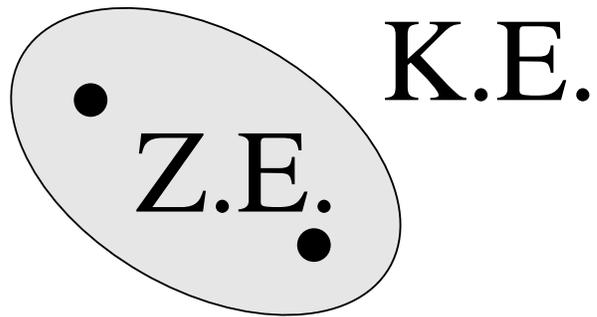}
\caption{\label{fig:murky}Illustration of the energy costs in
a vortex molecule.  The energy in the
region occupied by the two vortices is dominated by the quadratic
Zeeman cost and the energy outside it is dominated by kinetic energy.
Increasing the molecule size decreases the kinetic energy and increases
the Zeeman energy.  The equilibrium size $L$ is determined by
minimizing the sum.}
\end{figure}

The vortex described by Eq. (\ref{eq:bisect}) can
form a partnership with another vortex of the same type, producing
a molecule that can exist without costing too much energy. 
This is because the net charge of the two vortices is compatible with
the magnetic field.  
The combination of two $180^{\circ}$
rotations of the tetrahedral order parameter about the $A$ axis is a
$360^{\circ}$ rotation about the $A$-axis, but
since any rotation axis is a $360^{\circ}$ symmetry of the tetrahedron,
the rotation axis on circles of radius $r$ enclosing the vortices can be tilted continuously relative to
the tetrahedron as $r$ increases until
it becomes the $R$ axis instead.
  Then
the tetrahedra align with the magnetic field far from the
vortices and stay in $\mathcal{M}_{q3}$.
Each vortex screens the part of the other vortex's charge that produces
the large Zeeman cost, as illustrated in Fig. \ref{fig:murky}.
The Zeeman energy in the region around the vortices,
where the tetrahedra are still tilted,
may be estimated by replacing the total condensate size
$R$  in Eq. (\ref{eq:misalign}) by the diameter $L$
of this region.
The Zeeman energy tries to pull the vortices toward one another
but the elastic energy cost of
rapid changes in the order parameter (the gradient term
in the Hamiltonian) opposes this tendency: 
The tetrahedra rotate
twice as fast around circles beyond $L$, 
where the two vortices act in concert.  
Therefore the elastic energy
increases as $L$ becomes smaller; this leads to the
Coulomb repulsion between the vortices,
$2\pi\frac{n_0\hbar^2}{m}\ln\frac{R}{a}-\pi \frac{n_0\hbar^2}{m}\ln\frac{L}{a_c}$,
where $m$ is the mass of the atoms in the condensate.
The energy of the vortex molecule is therefore
\begin{equation}
E=k\frac{q^2L^2}{\beta}-\pi \frac{n_0\hbar^2}{m}\ln\frac{L}{a_c}+cnst.
\label{eq:barrier}
\end{equation}
where $k$ is a numerical constant.  The
equilibrium size can be determined by minimizing over $L$.  The
quadratic Zeeman force binds this molecule together while
the repulsion keeps the vortices from merging.

If the vortices \emph{were} to coalesce, then they could
react to form a set of vortices that are not bound by the Zeeman
energy; one possible set of
decay products is 
three vortices each involving a rotation through $120^{\circ}$
about the $R$ axis (which have the same net $360^{\circ}$ rotation as
the original pair of vortices).  
The vortex molecule of the two $A$ rotations is
\emph{metastable}
because thermal fluctuations may overcome their Coulomb
repulsion and push them together, leading to such a fission process.
%by randomly driving the two component vortices
%together.

The rest of this paper elaborates: it gives a qualitatively correct
expression for
the spin texture surrounding the molecule just described and determines
how this molecule \emph{actually} decays.  In order to do this,
we will give some more general results: first Section \ref{sec:dali},
summarizes the non-commutative group theory of combining
vortex charges and a classification of the tetrahedral vortices (the
vortices that occur by themselves when $q=0$ and in clusters when $q\neq 0$);
then Section \ref{sec:carnot}
estimates the elastic energies and Zeeman energies
of such clusters
as functions of these vortex charges; finally Section \ref{sec:lavoisier}
gives
criteria determining which types of tetrahedral
vortices form bound states or metastable states 
(see Section \ref{sec:metastable}). 
The last two sections
illustrate the criteria with a few additional surprising
examples (see Section
\ref{sec:sampler}) and give
some basic ideas about how to observe metastable vortices 
(Section \ref{sec:fieldwork}).

\section{\label{sec:dali}Topological Charges}
%The energy for a single component superfluid is given
%by $V(\psi)=\frac{1}{2}\alpha|\psi|^4-\mu|\psi|^2$.  This
%potential has the Mexican hat form, with minima along the circle in
%$\psi$-space,
%defined by $|\psi|=\sqrt{n_0}$ (with $n_0=\frac{\mu}{\alpha}$). energy,
%charge as winding in a complicated space, charge combinations as group multiplication.

\begin{figure}
\psfrag{M}{$\mathcal{N}$}
\psfrag{Gamma}{$\Gamma$}
\includegraphics[width=.45\textwidth]{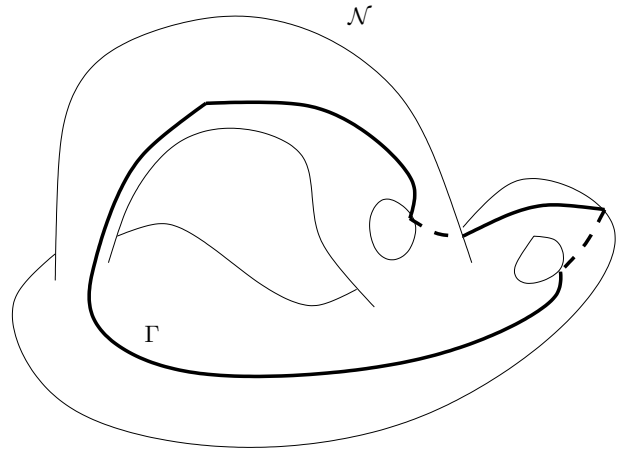}
\caption{\label{fig:iron}  The order parameter at infinity of a vortex winds
around a loop $\Gamma$ in the order parameter space $\mathcal{M}$.  Realistic
order parameter spaces are usually more symmetrical.}
\end{figure}

Vortices are simplest to understand
for an interaction energy that has a single manifold, $\mathcal{N}$, 
of ground states and no hierarchy.
The order parameter must move into $\mathcal{N}$ far away from any spin texture.
Vortices are classified by the topology
of the circuit traced out in $\mathcal{N}$ by the order
parameter on a large circle containing a vortex or set of vortices.
Although there are many ways to add wiggles to a given circuit, only
the topological structure is important, leading to a discrete
set of possible vortex charges.  Fig. \ref{fig:iron} shows how
a circuit may be tangled with holes in the space $\mathcal{N}$.
As a vortex evolves, such a circuit can evolve only into other circuits
that are tangled in the same way; thus the vortex charge is conserved.

This generalizes circulation-conservation in a single-component
condensate:
The circulation quantum
number $n$ for a set of vortices
is also the number
of times that the wave function at infinity winds around the circle
that minimizes the Mexican hat potential\cite{nelsonbook}. The tangling
in
a multicomponent condensate can be described by a group of generalized
winding numbers around the ground state manifold $\mathcal{N}$,
$\pi_1(\mathcal{N})$, the ``fundamental group"
(see Fig. \ref{fig:iron}).

Besides the conservation of topological charge, two further properties
of vortices follow from the geometry of the internal ground-state space.
First of all, the \emph{net winding} or topological ``charge"
of a set of vortices can be found by multiplying their charges together,
using the definition of multiplication for the fundamental group.
(In a general space, the fundamental group has a multiplication operation
 defined by splicing circuits together.)
This rule
is the generalization of adding the $n$'s of the individual vortices
in the scalar order-parameter case.  (The fundamental group 
is noncommutative
so one has to multiply the
charges together in the right order, see Appendix \ref{app:catstring}.)  

Secondly, the \emph{energy} of a vortex
can be estimated as a function of the winding behavior at infinity.  For
a vortex
that \emph{minimizes} this energy, the order parameter travels along
a geodesic in $\mathcal{N}$.
Since the interaction energy
$V$ is constant (and equal to its minimum $V_{min}$) at infinity,
the energy of a vortex is determined by the elastic energy, that is,
the cost of variations in $\psi$ as a function of the azimuthal angle $\phi$.
The closed loop traced out by $\psi(R\cos\phi,R\sin\phi)$ will relax to
make this energy-cost small.  When it
shrinks as much as is possible without leaving $\mathcal{N}$, it
becomes
a \emph{geodesic} in $\mathcal{N}$.
The energy of the vortex is related to the length $l$ of this
geodesic by 
\begin{equation}
E=(\frac{\hbar^2 n_0}{m}\ln\frac{R}{a_c})\frac{l^2}{4\pi }\label{eq:chocolatepudding}
\end{equation}
where $R$ and $a_c$ are the
radii of the condensate and the vortex core.
(Note that replacing $l$ by $2\pi n$ 
gives the standard expression
for a vortex in a scalar condensate.) 

A small magnetic field introduces hierarchies into the order parameter space, leading to spin textures that
wind around one manifold at an intermediate length scale 
and around a smaller manifold
far away.

%Last, a conservation law can be formulated:  the net charge of
%a set of vortices is the product of the charges 
%corresponding to circles enclosing each vortex separately.
%(This statement relies on the fact that a multiplication rule can 
%be defined for the fundamental group.)  This net charge corresponds
%to a circle containing all the vortices, and hence it cannot
%change as the vortices move around.
\subsection{\label{sec:doublestar}Spin Textures and their Vortices}
\begin{figure}
\psfrag{0}{$\phi=0$}
\psfrag{2pi}{$\phi=2\pi$}
\psfrag{a}{a)}
\psfrag{b}{b)}
\psfrag{g1}{$\Gamma_1$}
\psfrag{g2}{$\Gamma_2$}
\psfrag{g3}{$\Gamma_3$}
\psfrag{Q}{$Q$}
\psfrag{lm}{$\lambda$}
\includegraphics[width=.45\textwidth]{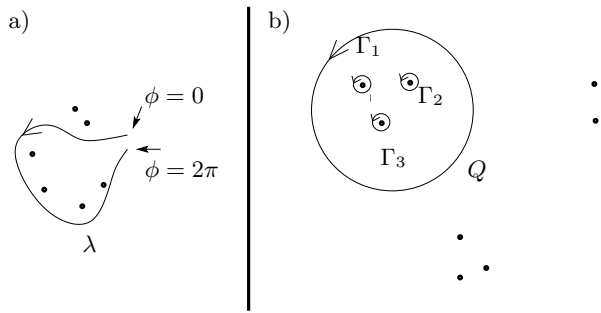}
\caption{\label{fig:plumpudding} Classifying vortex topologies. a)
The combined topology of a set of vortices can be found be seeing
how the order parameter changes around a closed loop parameterized
by $\phi$.  The loop is
drawn with a small gap between $\phi=0$ and $\phi=2\pi$ to indicate
that $g(\phi)$ and $\theta(\phi)$ vary continuously only between
$0$ and $2\pi$.  The topological charge is defined as the jump in
the values of $g$ and $\theta$ across the gap. b) 
Vortices that cannot exist by
themselves in the presence of a magnetic field can still sometimes appear
in clusters.
The charges $\Gamma_1,\Gamma_2$,$\Gamma_3$ in one of the clusters
can involve rotations around arbitrary
axes, but the combined charge $Q$ has to use the direction of the magnetic
field for its rotational axis to avoid a large energy-cost.}
\end{figure}

Let us consider how nested vortices hold themselves together in
a continuous ``spin texture."  
We  will try to give a general
argument showing how topology and the energy-hierarchy of the
order parameter spaces implies that any texture is made up
of a set of composite vortices which in turn are made up
of clusters of nearly point-like vortices.  
%A symmetry breaking term reduces the dimension
%of the space of ground states.  When it
%is small, the
%order parameter space can be viewed hierarchically. The order parameter
%$\psi$
%will not always have to be in an absolute ground state, with the minimal
%energy $V(\psi)=V_{min}$.  It can leave the
%ground state, if forced by topology, over regions whose area is inversely
%proportional (roughly) to the scale of the energy cost of
%the departure from the ground state.  (This is like saving
%eggnog--with all that fat!--for one week out of the year.)
%The first level of the hierarchy
%is the entire Hilbert space containing the spinor (of $2F+1$ dimensions,
%when $F$ is the hyperfine spin of the atoms),
%in which the energy $V(\psi)-V_{min}$ varies from $0$ up to
%a typical value of $\beta n_0^2$. (As long as the density of the
%condensate does not change, $\beta$ rather than $\alpha$ sets the
%energy scale, see Eq. (\ref{eq:hamiltonian}).)  In the Hilbert space is contained the four dimensional
%space of ground states $\mathcal{M}$
%of the $SU_2$ invariant Hamiltonian; here
%$V(\psi)-V_{min}$ varies from $0$ to on the order of $\epsilon(q)$,
%where $\epsilon(q)$ is small if $q$ is.   Smallest of all is 
%the two dimensional
%space $\mathcal{M}_q$ of ground states of the Hamiltonian
%with $q$ turned on, where $V-V_{min}=0$.  For a particular subspace 
%$\mathcal{M}_i$
%in this
%sequence, where $V(\psi)\sim\epsilon_i$, the size of the regions
%where $\psi$ is in $\mathcal{M}_i$ is typically 
%$L_i\sim\sqrt{\frac{K}{\epsilon_i}}$ where $K=\frac{\hbar^2 n_0}{m}$ is
%the ``superfluid stiffness".  

The order parameter subspaces in order 
of increasing energies are the two-dimensional
$\mathcal{M}_q$, the
four-dimensional $\mathcal{M}$
and the $4F+2$ dimensional $\mathcal{H}$ (for spin $F$), 
with corresponding energy scales
$V-V_{min}=0,\ \epsilon(q)=\frac{q^2}{\beta}$ (see Eq. (\ref{eq:effective})) and
$\beta n_0^2$ (see Eq. (\ref{eq:hamiltonian}); as long as the density
of the condensate does not vary, $\beta$ rather than $\alpha$
sets the energy scale).  The condensate can move out of
the ground state into one of the higher-energy subspaces if
forced to by topology, spending less space in the manifolds 
with the greater energies.  (This is like saving
eggnog--with all that fat!--for one week out of the year.)%, eating lowfat swiss
%once a week, and living off celery the rest of the time.)
For a particular subspace
$\mathcal{M}_i$
in this
sequence, in which $V(\psi)-V_{min}\sim\epsilon_i$, the size $L_i$
of the regions
where $\psi$ is in $\mathcal{M}_i$ is typically
$L_i\sim\sqrt{\frac{K}{\epsilon_i}}$, where $K=\frac{\hbar^2 n_0}{m}$ is
the elasticity of the condensate.  
This relation can be guessed at without
understanding anything about the field configuration in such a region; just
assume equipartition between kinetic and interaction energies,
so that $\epsilon_i$ is equal to
$\frac{K}{L_i^2}$, a typical scale for the kinetic energy density,
$\frac{\hbar^2}{2m}|\nabla\psi|^2$. Consequently, the composite vortex
cores in Fig. \ref{fig:cherimoya}, where $\psi$ is in $\mathcal{M}-\mathcal{M}_q$, have diameters 
\begin{equation}
L_q\sim\sqrt{\frac{K\beta}{q^2}}\sim\frac{\hbar}{q}\sqrt{\frac{n_0\beta}{m}}
\label{eq:betterlate}
\end{equation}
where the energy scale is taken from Eq. (\ref{eq:effective}) 
and the component cores, where $\psi$ varies
through $\mathcal{H}-\mathcal{M}$, have typical diameters 
$a_c\sim\sqrt{\frac{K}{n_0^2\beta}}$.  

The more complete version of the equipartition argument, given
in Sec. \ref{sec:molecules}, estimates
the total energy of the vortex as follows:  the kinetic
energy outside the composite core is about $K\frac{l^2}{4\pi}\ln\frac{R}{L}$,
according to Eq. (\ref{eq:chocolatepudding}), the magnetic field energy inside
the core is about $\epsilon(q) L^2$.  Minimizing the sum, which has 
the form $-c_1K\ln{L}+\epsilon(q)L^2+c_2$, gives
$L\sim\sqrt{\frac{K}{\epsilon(q)}}$.  
The analogous argument also
applies to 
vortices in a single-component condensate, giving the core size 
$\sqrt{\frac{K}{n_0^2\alpha}}$.
Because the composite cores in a cyclic
condensate have some substructure (the component vortices), the
actual coefficient
$c_1$ of the logarithm for the cyclic condensate will turn out
to be less than $\frac{l^2}{4\pi}$
once the kinetic energy of the vortices inside
the core is included (see Section \ref{sec:enindex}).

If $q$ is very small, the component cores are much smaller than
the composite vortices, and may be regarded as points.
The wave function can therefore be approximated by a field that always
stays in $\mathcal{M}$ except at ``singularities" corresponding
to the component cores:
\begin{eqnarray}
&&\psi(x,y)\approx e^{-i\alpha(x,y)\mathbf{\hat{n}}(x,y)\cdot\mathbf{F}}e^{i\theta(x,y)}
\sqrt{n_0}\chi_3\nonumber\\
&&\ \ \ \ \ \ \ \ \ \ \ \ \ \ \mbox{ except at ``points"},\label{eq:Manifold}
\end{eqnarray}
where $\chi_3$ is the spinor
defined in Eq. (\ref{eq:order3}).
Over distances much greater than $L_q$ even the composite cores
seem to dwindle to points, suggesting the following approximation:
\begin{eqnarray}
&&\psi(x,y)\approx e^{-i\alpha(x,y) F_z}e^{i\theta(x,y)}\sqrt{n_0}\chi_3\nonumber\\
&&\ \ \ \ \ \ \ \ \ \ \ \ \ \ \mbox{ far from clusters}.
\label{eq:alphax}
\end{eqnarray}
These expressions generalize\cite{ryanconv} 
the ``phase-only approximation" for
ordinary superfluids\cite{nelsonbook}; for example, Eq. (\ref{eq:Manifold})
parameterizes elements of the space $\mathcal{M}$ in terms of the symmetries
of the $q=0$ Hamiltonian,
rotations $\psi\rightarrow e^{-i\alpha\mathbf{F}\cdot\bm{\hat{n}}}\psi$, and
rephasings, $\psi\rightarrow e^{i\theta}\psi$, which are applied
to a representative state.  
Vortices are points which look like singularities at the level of resolution
exposed by one of these
approximations, although what one identifies as vortex singularity
depends on which level of resolution one uses!
% if the expressions in Eqs. \ref{eq:Manifold}, \ref{eq:alphax}   were assumed to apply all
%the way to zero radius of a vortex,
%the parameters of the symmetries
%would
%vary rapidly.  (For a single-component condensate, a vortex appears
%to be a singularity until one looks close enough at the core and
%realizes that the spinor vanishes there.) Of course,
%what one identifies as a vortex singularity
%depends on which level of resolution one uses!

Composite vortices result from the order parameter being forced
out of the minimum-energy space $\mathcal{M}_q$ by topology:  
If the order parameter goes around
a hole in $\mathcal{M}_q$ on some circle in the condensate, 
then inside the circle, the order parameter has to leave $\mathcal{M}_q$.
If Eq. (\ref{eq:alphax}) continued to hold all the way to the center
of the circle, then there would be a singularity since the angular variables
$\alpha$ and $\theta$ would run rapidly through multiples of $2\pi$.  So we can
 think of the circle as the core of an extended vortex in
the field in Eq. (\ref{eq:alphax}).  The singularity is actually filled
in by a field of tetrahedra with higher-energy orientations, described by
Eq. (\ref{eq:Manifold}), just as the wave
function of a vortex in a single-component condensate avoids singularities
by vanishing in the core.
In the filled-in region of freely-oriented tetrahedra, there are also
some elementary ``point"-vortices, actually spread out
over the distance $a_c$.  These
result when the order parameter gets tangled around holes
in $\mathcal{M}$, and then has to move outside of \emph{this} space.
These cores are smaller than the composite core because of the large
energy scale associated with $\mathcal{M}$.
The fields of these vortices can be described qualitatively by
setting $q=0$ since kinetic energy surpasses
the anisotropy energy very close to the cores.

%The cores of the component vortices are filled to
%almost the same density $n_0$ as the rest of the condensate (the 
%spin-independent repulsion $\alpha$ favors uniformity since $\alpha\gg\beta$). Vortices
%with this property are often
%referred to as ``coreless."   But an alternative
%definition implies that these vortices \emph{do} have cores:
%a core is a region where the wave function departs from its ground state form.
%(In the core of a ``coreless" vortex in a ferromagnetic condensate,
%the order parameter becomes polar rather than ferromagnetic.)
%The ``true" vortex cores are therefore the 
%\emph{composite} core regions (grey in Fig. \ref{fig:cherimoya}),
%where $\psi$ leaves $\mathcal{M}_q$.  If we instead define departures
%from $\mathcal{M}$ to be cores, then the black
%spots surrounded by tetrahedral windings are entitled ``cores." 
%The sphere $\mathcal{S}$ (defined by $|\psi_0|=\sqrt{n_0}$) can also
%be inserted
%in the hierarchy of subspaces; departures from $\mathcal{S}$
%correspond to ordinary cores--regions
%where the order parameter has to vanish to avoid discontinuity; in spinor
%condensates with $\beta<<\alpha$, however, 
%vortices of this type are not stable, since 
%the sphere in a multidimensional Hilbert space
%is simply-connected.

The cores of the component vortices are filled to
almost the same density $n_0$ as the rest of the condensate (the 
spin-independent repulsion $\alpha$ favors uniformity since $\alpha\gg\beta$). 
Vortices
with this property are often
referred to as ``coreless," but we use the word ``core" in a more general way,
so that every vortex can have one.
Instead of defining core as a region where the \emph{density} vanishes, 
our definition identifies a
core as a region where the wave function departs from any particular form.
(In the core of a ``coreless" vortex in a ferromagnetic condensate,
the order parameter becomes polar rather than ferromagnetic.)
The core of the component tetrahedral vortices is the region where the
order parameter leaves $\mathcal{M}$.
Using the absolute ground state space $\mathcal{M}_q$ 
leads to a different
definition of vortex cores of field-aligned vortices: 
the core is where Eq. (\ref{eq:alphax}) breaks
down.  This is why we call the region surrounding the tetrahedral
vortices a ``composite core." The ordinary definition of a core
arises when we focus on another space, $\mathcal{S}$, the sphere defined 
by $|\psi|=\sqrt{n}_0$. (This space minimizes the largest of
the interaction terms,
$\frac{1}{2}\alpha n^2-\mu n$.)  From the perspective of $\mathcal{S}$, a core 
would be a region where the
wave function vanishes to avoid being discontinuous; however, in a spinor
condensate with $\alpha\gg\beta$, the wave function never has to vanish
because a closed loop cannot get snagged in the surface of
the simply-connected sphere $\mathcal{S}$.

Now we have to classify both the topologies of the
point vortices (or ``tetrahedral charges") and the composite
vortices (or ``aligned charges"), using
the method described
e.g. in \cite{mermin79}.  For the point vortices we must parameterize
the symmetry group that generates the composite core spin textures
by a \emph{simply connected} group $G^*$. We will take
$G^*=\{(g,\theta)|g\in SU_2\ \mathrm{and\ } \theta\in\mathbb{R}\}$
where multiplication is
defined by $(g_1,\theta_1)(g_2,\theta_2)=(g_1g_2,\theta_1+\theta_2)$.
We parameterize $G^*$ as follows,
\begin{equation}
D(e^{-i\alpha\frac{\bm{\hat{n}}\cdot\bm{\sigma}}{2}},\theta)=
e^{i\theta-i\alpha\mathbf{\hat{n}}\cdot\mathbf{F}},
\label{eq:D}
\end{equation}
allowing us to regard $G^*$ as a (redundant) description of the
symmetries of the Hamiltonian.
Using this representation, one can define the net vortex charge
inside any closed loop $\lambda$(see Fig. \ref{fig:plumpudding}a);
 one simply expresses the wave function
along the loop in the form
\begin{equation}
\psi(\phi)=D(g(\phi),\theta(\phi))\sqrt{n_0}\chi_3
\label{eq:rays}
\end{equation}
where $\phi$ parameterizes the loop ($0<\phi<2\pi$). 
In order for Eq. (\ref{eq:rays}) to be continuous when
the circuit is closed, $\psi(\phi=0)=\psi(\phi=2\pi)$, or
\begin{equation}
D(g(0),\theta(0))\chi_3=D(g(2\pi),\theta(2\pi))\chi_3.
\label{eq:netcloses}
\end{equation}
It follows that
 the ``classifying group element" or net ``topological charge" inside the
curve
\begin{equation}
\Gamma(\lambda)=
(e^{-i\alpha_{\lambda}\frac{\bm{\hat{n}_{\lambda}}\cdot\bm{\sigma}}{2}},\theta_{\lambda})
\equiv(g(0)^{-1}g(2\pi),\theta(2\pi)-\theta(0))
\label{eq:wrongdivision}
\end{equation}
is a symmetry
describing the net rotation and rephasing around the closed loop.  (Let us
also use the briefer notation
$g_\lambda=
e^{-i\alpha_{\lambda}\frac{\bm{\hat{n}_{\lambda}}\cdot\bm{\sigma}}{2}}$ for
the net rotation.)
Since
the tetrahedron has only $24$ symmetries (see Sec. \ref{sec:namingofcats}), this makes for a tractable
classification of the topologies.  %Because we
%have to use $G^*$ rather than $SO_3\times U(1)$ to classify charges,
%$(id,0)$, $(-id,0)$ and $(id,2\pi)$ (where $id$ is the identity matrix
%in $SU_2$ describe three different vortex types, although these
%group elements all act by the identity on the Hilbert space.  The continuous
%process of building up to one of these elements as $\phi$ goes from
%$0$ to $2\pi$ are different in each case.
These topologies form the group of tetrahedral charges, multiplied together
according to $(g_1,\theta_1)(g_2,\theta_2)=(g_1g_2,\theta_1+\theta_2)$.

When $q$ is very large then Eq. (\ref{eq:alphax}) has to apply basically
everywhere and when $q$ is small
then Eq. (\ref{eq:alphax}) has to apply outside of bound
clusters of vortices.
In order to classify the topology of vortices in the former case or
of vortex clusters in the latter,
we need a simply connected group that parameterizes the $q\neq 0$
symmetry group $e^{i\theta}e^{-i\alpha F_z}$; we take $G_q^*
=\{(\alpha,\theta)|\alpha,\theta\in \mathbb{R}\}$ and use the mapping
$D(\alpha,\theta)=e^{i\theta-i\alpha F_z}$.  Then we write 
$\psi(\phi)=D_q(\alpha(\phi),\theta(\phi))\sqrt{n_0}\chi_3$
and define 
\begin{equation}
Q(\lambda)=(\alpha_{\lambda},\theta_{\lambda})_3=
(\alpha(2\pi)-\alpha(0),\theta(2\pi)-\theta(0))_3
\label{eq:tiedup}
\end{equation}
where the subscript $3$ indicates that the magnetic field favors the 3-fold
symmetric
orientation illustrated in Fig. \ref{fig:bottler}a 
(since we are assuming $c>4$), and the rotation
through $\alpha_{\lambda}$ is understood to be around the field axis.  
The possible values for $Q$ can be
referred to as ``field-aligned charges" since they describe the
topologies of vortex-fields in which the tetrahedra keep
the orientation favored by the magnetic field.

Now let us consider the form of the fields near 
a tetrahedral vortex core.
Near vortex $i$, the spin texture will be rotationally
symmetric, and given by
\begin{equation}
\psi=e^{i\frac{\theta_i}{2\pi}\phi}e^{-i\frac{\alpha_i}{2\pi}\phi
\bm{\hat{n}_i'}\cdot\mathbf{F}}\sqrt{n_0}\chi_0
\label{eq:lathes}
\end{equation}
where $\phi$ is now the azimuthal angle $\phi$ centered at this
vortex.  Because the tetrahedra near this vortex may be tilted,
we use $\sqrt{n_0}\chi_0$, a \emph{generic}
member of the cyclic order parameter space. The vortex is azimuthally
symmetric since
the parameters
$\theta_i,\alpha_i, \bm{\hat{n}_i'}$
are constants.  The rotation axis $\bm{\hat{n}_i'}$
is a local symmetry axis for the 
possibly \emph{tilted} tetrahedra.
To reduce the possible symmetries to a finite
set, we should relate
the spinor $\chi_0$ to
the spinor $\chi_3$ corresponding to the tetrahedron
as oriented in Fig. \ref{fig:bottler}a.
If we write $\chi_0=D(R,\xi)\chi_3$ for an appropriate rotation $R$ and
phase $\xi$,
the vortex in Eq. (\ref{eq:lathes})
can be written
\begin{equation}
\psi(\phi)=D(R,\xi)e^{i\frac{\theta_i}{2\pi}\phi}e^{-i\frac{\alpha_i}{2\pi}\phi
\bm{\hat{n}_i}\cdot\mathbf{F}}\chi_3,
\label{eq:lathes2}
\end{equation}
where
\begin{equation}
\bm{\hat{n}_i}=R^{-1}(\bm{\hat{n}_i'})\label{eq:anotherpictureonanotherwall}
\end{equation}
and we have used the transformation rule for angular momentum:
\begin{equation}
 D(R,\xi)^{\dagger}F^iD(R,\xi)=\sum_{j=1}^3R_{ij}F_j.
\label{eq:transfnrule}
\end{equation}
(The right-hand side uses the $SO_3$ matrix $R_{ij}$ associated with
the rotation
$R$.) 
Continuity implies that
the rotation axis $\bm{\hat{n}_i}$ is one of the finitely many symmetry
axes illustrated in Fig. \ref{fig:bottler}a; furthermore, according
to the above scheme, 
the group element that classifies this vortex is
$\Gamma_i=(e^{-i\alpha_{i}\frac{\bm{\sigma}\cdot\bm{\hat{n}_{i}}}{2}}
,e^{i\theta_{i}})$. There are only a discrete
set of possible charges when we use $\bm{\hat{n}_i}$, the axis relative
to the body axes of the tetrahedron rather than $\bm{\hat{n}_i'}$, the
axis relative to the lab coordinates.

Eq. (\ref{eq:lathes2}) expresses the vortex as a product
of a constant matrix $D(R,\xi)$ (the phase $\xi$ is unimportant)
and a standardized vortex configuration.
The transformation $D(R,\xi)$ rotates the standardized configuration in spin
space, changing both the rotation axis (from $\bm{\hat{n}_i}$ to
the local axis $\bm{\hat{n}_i'}$)
and the orientation of the tetrahedra.

In the distant surroundings of a cluster of vortices,
the spin texture will again have
a rotationally symmetric form.
The group element describing
the change in the order parameter as one tours the loop $\lambda$
enclosing the entire cluster is
given by
\begin{equation}
\Gamma(\lambda)=\prod_i\Gamma_i.
\label{eq:berlin}
\end{equation}
This algebraic law has a few consequences.  First,
it leads to a conservation law that constrains vortex alchemy: the net charge
$\Gamma(\lambda)$ has to be conserved as the vortices inside the loop combine
and metamorphose, since the topology on the loop cannot change
suddenly.  (There is only a discrete set of
vortex charges because charges are defined using the body coordinates.) 
Second, Eq. (\ref{eq:berlin}) restricts the types
of vortices which can form a cluster when $q\neq 0$.
 (See Fig. \ref{fig:plumpudding}b.)  Outside the cluster,
the tetrahedra must be aligned with the field (see Eq. (\ref{eq:alphax}),
so the charge is described by a field-aligned order parameter
$Q(\lambda)=(\alpha,\theta)_3$.  Eq. (\ref{eq:berlin})
requires that
\begin{equation}
\prod_i g_i=e^{-i\alpha \frac{\sigma_z}{2}}
\label{eq:compatibility}
\end{equation}
and $\sum_i \theta_i\equiv\theta (\mathrm{mod\ }2\pi)$.  
The group elements must multiply to
form a rotation about the $z$-axis
to avoid the large Zeeman cost
outside the clusters, or ``composite cores" of the vortex molecules.

Let us now review the example in the previous section:  the vortex molecule
was made out of two vortices of type
$\Gamma_1=(e^{-i\pi\frac{1}{2\sqrt{3}}(\sqrt{2}\sigma_x+\sigma_z)},0)$;
since $\Gamma_1^2=(e^{-i\pi\frac{\sqrt{2}\sigma_x+\sigma_z}{\sqrt{3}}},0)
=(-id,0)=(e^{-i\pi\sigma_z},0)$, the spin texture
can align with the magnetic field outside the pair of vortices and
$Q=(2\pi,0)_3$.  This uses the fact that, in $SU_2$, 
all $2\pi$ rotations are equal to $-id$, where
$id$ is the $2\times2$ identity matrix.

%Vortices
%with this property are often
%referred to as ``coreless."   The definition of 
%cores we use implies that these vortices \emph{do} have cores.
%Instead of focusing on the density, the definition used here identifies a
%core as a region where the wave function departs from its ground state form.
%The core of the component tetrahedral vortices is the region where the
%order parameter leaves $\mathcal{M}$; since Eq. (\ref{eq:Manifold}) breaks down
%near every vortex, all tetrahedral vortices \emph{will} have cores.  Applying
%the same idea to the absolute ground state space $\mathcal{M}_q$ 
%leads to a different
%definition of vortex cores of field-aligned vortices: 
%the core is where Eq. (\ref{eq:alphax}) breaks
%down.  Thus the region surrounding the tetrahedral
%vortices can be called a ``composite core."

\subsection{\label{sec:namingofcats}Notation for Vortices with and without a Magnetic Field}

Let us first assign names to the \emph{tetrahedral}
charges corresponding to closed loops
in $\mathcal{M}$.
At $q=0$ or within vortex clusters where $q$ can be neglected, 
the topological charges are
described by
a pair $\Gamma=(g,\theta)$ (defined in Eq. (\ref{eq:wrongdivision})) where $g=
e^{-i\alpha\frac{\mathbf{\hat{n}}\cdot\mathbf{\sigma}}{2}}\in SU_2$ and 
$\theta\in\mathbb{R}$.  Note that for the phase of
the order parameter to be continuous, as
Eq. (\ref{eq:netcloses}) requires, 
the allowed values of $\theta$ and $g$ must be
correlated\cite{makela,spin3}, so only a discrete
sequence of phases may accompany a given rotational symmetry.  Also
note that
in $SU_2$, rotation angles are defined
modulo $4\pi$ rather than $2\pi$. %Any rotation can be described
%as a rotation through an angle $\alpha$ around some axis,
%where $-2\pi<\alpha<2\pi$, and a negative angle refers to a clockwise
%rotation.  
The twelve symmetries of
the tetrahedron according to the ordinary method of counting become
$24$ because, e.g., 
a clockwise
$240^{\circ}$ rotation around an axis is distinguished from the counterclockwise
$120^{\circ}$ 
rotation.  This is more than a technical point: the vortex  
where the order parameter 
rotates through $-240^{\circ}$ about an order three axis
cannot deform
continuously into one where the order parameter rotates through
 $120^{\circ}$, and it has more
energy as well.  On the other hand, the $\alpha=4\pi$ ``vortex" can
relax continuously to a state free of vortices. The necessity of using $SU_2$ 
instead of $SO_3$ is the biggest surprise to come out of the topological
theory.

Since $g$ must be a symmetry of the tetrahedron corresponding to 
$\sqrt{n_0}\chi_3$, illustrated in Fig. \ref{fig:bottler}a,
we can describe $g$ by indicating its rotation axis
using the labels from the figure.
 We refer to the minimal rotation around a given
axis using just the label of the axis; hence
$S,P,Q,R$  refer to the
rotations through $120^{\circ}$ counterclockwise as viewed from the
tips of the corresponding arrows, and $A,B,C$ 
refer to counterclockwise rotations
through $180^{\circ}$, about $A,B,C$. Rotations
through larger angles can be written as powers of these rotations.
Therefore $P^2$ is a $240^{\circ}$ rotation; also $P^3=A^2=-id$
since $360^{\circ}$ rotations around any axis correspond to $-id$
in $SU_2$.  We find it convenient to describe
each rotation as a 
rotation through an angle $\alpha$ around some axis,
where $-2\pi\leq\alpha\leq 2\pi$. (Positive and
negative $\alpha$'s
correspond to counterclockwise and 
clockwise rotations respectively.)  An arbitrary rotation angle
can be replaced by an angle in this interval using the
fact that $720^{\circ}$ rotations in $SU(2)$ are equivalent to the identity.
For example,  the $480^{\circ}$ counterclockwise rotation
$P^4$ is the same as the $-240^{\circ}$ clockwise 
rotation $P^{-2}$ because $P^6$, a rotation through two
full turns, corresponds to the identity of $SU(2)$.
On the other hand, the $240^{\circ}$ counterclockwise
rotation
$P^2$ is not equivalent to the clockwise $120^{\circ}$ rotation
around the same axis, since the corresponding
$SU(2)$ matrices differ by
a minus sign.

Now we can list all the pairs of rotations and phases which
are allowed by the continuity condition, Eq. (\ref{eq:netcloses}).
The possible vortices according to Ref. \cite{makela,spin3}
are $(R^m,\frac{2\pi m}{3}+2\pi n)$ and $(A,2\pi m)$ 
where $n$ and $m$ are integers, as well as the corresponding
vortices with $R$ replaced by $P$, $Q$, or $S$ and $A$ replaced
by $B$ or $C$.
%
%We can parameterize the $q=0$ ground states in the form 
%$\psi=U(R)\sqrt{n_0}\chi_3$
%where $R$ is a rotation of three dimensional space and $U$ is the corresponding
%spin rotation. To visualize this spinor, we take the tetrahedron illustrated
%on the left of Fig. \ref{fig:bottler} and rotate it with $R$. Any shape
%can be used to keep track of the order parameter, but the tetrahedron
%works best.  For example the $\frac{2\pi}{3}$ rotation about the
%$z$-axis brings  the spinor back to itself.  Using an arbitrary shape
%in place of the tetrahedron would imply that more than one orientation
%of the shape corresponds to the same state of the atom.  But the two orientations of the tetrahedron are indistinguishable; all the other symmetries of the
%spinor also correspond to indistinguishable orientations of the tetrahedron.

One can work out explicit expressions for the $SU(2)$
elements corresponding to given rotations.  For example,
let us find the $SU(2)$ element corresponding to $A$;
since the rotation angle is $180^{\circ}$,
\begin{equation}
A=e^{-i\pi\frac{\bm{\hat{a}}\cdot\bm\sigma}{2}}=-i\bm{\hat{a}}\cdot\bm{\sigma},
\label{eq:Aapple}
\end{equation}
where $\bm{\hat{a}}$ is the $A$-axis.
Note that
 the tetrahedron has its vertices 
at $(0,0,-1)$,$(-\frac{2\sqrt{2}}{3},0,\frac{1}{3})$,
$(\frac{\sqrt{2}}{3},\sqrt{\frac{2}{3}},\frac{1}{3})$,
$(\frac{\sqrt{2}}{3},-\sqrt{\frac{2}{3}},\frac{1}{3})$. The
 $A$ axis bisects the segment connecting the last pair of points;
the midpoint of these two points is
\begin{equation}
\frac{1}{2}(\bm{\hat{p}}+\bm{\hat{q}})=(\frac{\sqrt{2}}{3},0,\frac{1}{3}).
\end{equation}
The unit vector $\bm{\hat{a}}$ is obtained by normalizing this vector,
so
\begin{equation}
\bm{\hat{a}}=(\sqrt\frac{2}{3},0,\sqrt\frac{1}{3}).
\label{eq:ahat}
\end{equation}
Hence 
\begin{equation}
A=-i\sqrt{\frac{2}{3}}\sigma_x-i\sqrt\frac{1}{3}\sigma_z.
\label{eq:Ahimself}
\end{equation}
%These symmetries can be found using the formulae
%for spin $2$ rotation matrices\cite{makela}.
%The correspondence between the tetrahedron
%and the spinor also has a more concrete interpretation \cite{spin2sym}.

The net charge, Eq. (\ref{eq:berlin}), of
a set of vortices results from multiplying the matrices 
$g$ for the vortices of the set.  
The result can be identified as one of the
rotations $A^n$,$B^n$,$P^n$, etc.
This procedure completely determines
the $SU(2)$ product element, whereas the geometric method
of applying the appropriate sequence of rotations
to a tetrahedron does not determine the \emph{sign} of the $SU(2)$ matrix.

%The notation $(g,\theta)$ can refer either to the net charge of a set of
%vortices inside a curve $\lambda$  (see Fig. \ref{fig:plumpudding}a)
%or can be used more narrowly to describe a single point vortex.
We will describe a vortex reaction with the following notation,
\begin{equation}
(-id,0)\rightarrow (R,\frac{2\pi}{3})*(R,\frac{2\pi}{3})*(R,-\frac{4\pi}{3}).
\end{equation}
Each factor describes the charge of an individual vortex-\emph{atom}, rather
than a cluster of atoms, although $(g,\theta)$ can be used to describe
the net charge of a set of vortices as well, as in Fig. \ref{fig:plumpudding}a.
This reaction describes onne point-vortex breaks up into three
vortices.  The $*$ is just a separator between the different reaction
products, reminding one to
check conservation
of charge by multiplying both
sides of the reaction out.

Another useful cyclic spinor is
\begin{equation}
\chi_2=\left(\begin{array}{c}\frac{1}{2}\\0\\-\frac{i}{\sqrt{2}}\\0\\\frac{1}{2}\end{array}\right).
\label{eq:chi2}
\end{equation}
\emph{This} spinor corresponds to the
tetrahedron in Fig. \ref{fig:bottler}b.  Its vertices are
at the points of the form 
$(\pm\frac{1}{\sqrt{3}},\pm\frac{1}{\sqrt{3}},\pm\frac{1}{\sqrt{3}})$
if we restrict the choices of signs so that there are always
$0$ or $2$ minus signs.  The fact that the
$A$, $B$, and $C$ axes of this tetrahedron correspond to the $\bm{\hat{x}},
\bm{\hat{y}}$ and $\bm{\hat{z}}$ coordinate vectors makes the spinor 
$\sqrt{n_0}\chi_2$ especially convenient for determining the consequences
of the quadratic Zeeman term in the next section.
The orientation of the tetrahedron in Fig. \ref{fig:bottler}b is also very
convenient for working out the group of charges, since the expressions
for the symmetry axes are so simple. 
(E.g., $\bm{\hat{p}}=(\frac{1}{\sqrt{3}},-\frac{1}{\sqrt{3}},
-\frac{1}{\sqrt{3}})$, since vertex $P$ is in the $x>0,y<0,z<0$
octant.  Thus $P=e^{-i\frac{\pi}{3}\bm{\hat{p}}\cdot\bm{\sigma}}=\frac{1-i\sigma_x+i\sigma_y+i\sigma_z}{2}$.)

At nonzero magnetic field, only
rotations 
around the $z$-axis are symmetries.
When $c>4$, the ground state space is $\mathcal{M}_{q3}$, consisting
of rotations and rephasings of $\sqrt{n_0}\chi_3$, as in Eq. (\ref{eq:alphax}).
Vortices are described as in Eq. (\ref{eq:tiedup}) by 
an ordered pair $(\alpha,\theta)_3$ describing the 
rotation and rephasing
angles of the vortex.  The subscript $3$ is used to indicate that
the $z$-axis is an order three symmetry of the $c>4$ 
ground state tetrahedron. 
The continuity of the phase limits the vortex types to the form
\begin{equation}
Q=(\alpha,\theta)_3=(\frac{2\pi m}{3},2\pi (\frac{m}{3}+n))_3.
\label{eq:type3}
\end{equation}

When $c<4$, minimizing Eq. (\ref{eq:effective}) implies that
the magnetic field axis is an order two
symmetry, and the ground state space is $\mathcal{M}_{q2}$, the rotations
and rephasings of $\sqrt{n_0}\chi_2$.  Now
 $(\alpha,\theta)_2$ specifies the vortex types.  The
possibilities are
\begin{equation}
Q=(\alpha,\theta)_2=(\pi m,2\pi n)_2.
\label{eq:type2}
\end{equation}

\section{\label{sec:carnot} Energies and Symmetries}
This section considers the effect of the magnetic field, 
which binds vortices,
and the kinetic energy, which keeps the bound vortices from
merging altogether.  These are both included in the full energy
function
\begin{equation}
\mathcal{H}=\iint d^2\mathbf{r}\frac{\hbar^2}{2m}\nabla\psi^{\dagger}\nabla\psi
+V_{tot}(\psi),
\label{eq:howtousebatteries}
\end{equation}
where $V_{tot}$ is given by Eq. (\ref{eq:hamiltonian}).
When $B\neq 0$,
the Zeeman effect introduces an extra phase boundary dividing the
cyclic phase into two phases, with a 
phase transition at $c=4$ where the tetrahedron changes its
orientation relative to the magnetic field. (This result applies
for moderate magnetic fields; very low magnetic fields cause
different transitions\cite{ryan}.)
\subsection{\label{sec:cubeshape} The Anisotropy potential 
from the Quadratic Zeeman effect}

Let us determine which orientations
of the tetrahedron will be preferred by a magnetic field along the $z$ axis.
The preferred orientation can be calculated for small $q$ from an effective
potential which is
a function of the orientation of the tetrahedron. As long as 
\begin{equation}
q\ll n_0\beta,
\label{eq:smallfield}
\end{equation}
the tetrahedron will be only slightly deformed.  It will move into
a space
$\mathcal{M}'$ displaced by a distance on
the order of $\frac{q}{n_0\beta}$ from the space $\mathcal{M}$ of
arbitrarily oriented
perfect tetrahedra.  
The spinors in the distorted
space are given by
\begin{equation}
\psi= \sqrt{n_0}D(R,\xi)\chi_2+\delta\psi,
\label{eq:squish}
\end{equation}
where $D(R,\xi)$ is the spin two rotation matrix corresponding
to the rotation $R$ of space, multiplied by a phase.  The distortion
$\delta\psi$ depends on the orientation $R$.
Eq. (\ref{eq:Manifold}), which omits the deformation,
is a harmless shorthand description, emphasizing the orientation
of the tetrahedron. 
%where $U$ is the spin 2 matrix corresponding to the rotation $R$.
In this section, we use $\sqrt{n_0}\chi_2$ as the standard
spinor orientation instead of $\sqrt{n_0}\chi_3$ to simplify calculating
the energy; conveniently, the body axes $A,B,C$ of
the corresponding tetrahedron for the \emph{former} state
are aligned with 
the coordinate axes $\mathbf{\hat{x}},
\mathbf{\hat{y}},\mathbf{\hat{z}}$.  The body axes of the rotated state
$D(R)\sqrt{n_0}\chi_2$ 
(which make the angles $\alpha_1,\alpha_2,\alpha_3$ with the
$z$-axis) are thus $R(\mathbf{\hat{x}})$,$R(\mathbf{\hat{y}})$,
$R(\mathbf{\hat{z}})$. Therefore the $z$ component of the spin, 
in terms of the components
of the spin along the body axes, is 
\begin{equation}
D(R)^{\dagger}F_zD(R)=\cos\alpha_1 F_x+\cos\alpha_2 F_y+\cos\alpha_3 F_z.
\label{eq:qmechspecial}
\end{equation}

At first order 
the quadratic Zeeman effect does not have any dependence 
on the orientation of the tetrahedron because a tetrahedral
spinor is ``pseudoisotropic," i.e., $\chi_2^{\dagger}F_iF_j\chi_2=2\delta_{ij}$.
The first order energy is thus given
by
\begin{eqnarray}
<qF_z^2>&\approx& n_0q\chi_2^{\dagger}D(R)^{\dagger}F_z^2D(R)\chi_2\nonumber\\
&\approx &n_0q\sum_{i,j=1}^3 \cos\alpha_i\cos\alpha_j
\chi_2^{\dagger}F_iF_j\chi_2
\nonumber\\
&\approx &2n_0q,
\label{eq:tievote}
\end{eqnarray}
which does not prefer any orientation of the tetrahedron. In the last step
we used
\begin{equation}
 \sum_{i=1}^3\cos^2\alpha_i=1
\label{eq:pythagoras}
\end{equation}
which follows from the fact that $\bm{\hat{z}}$, a unit vector,
has body-coordinates
$(\cos\alpha_1,\cos\alpha_2,\cos\alpha_3)$.

To find the second order energy due to the quadratic Zeeman effect, which
will break the tie,
we have to find the deformed state and its energy.  The deformation
$\delta\psi$ is determined by minimizing the total interaction and Zeeman
energy in Eq. (\ref{eq:hamiltonian}) for each given orientation $R$.  
Now if the deformation is not restricted
somehow, the ``deformation" which minimizes the energy will be very large,
involving the tetrahedron rotating all the way to the absolute ground
state.%\footnote{As an analogy, imagine someone is fixing his house up,
%trying to save money.  One can decide to save money by not using fancy paint
%or by not getting new furniture, but if one tries to save the maximum
%money on everything, the best option is to do nothing.  One has to
%set some partly arbitrary limits to one's thriftiness.}.
We therefore allow deformations only of the form
\begin{alignat}{1}
\delta&\psi=d D(R,\xi)\chi_2\nonumber\\
&+\frac{1}{\sqrt{2}}(aD(R,\xi)F_x\chi_2+bD(R,\xi)F_y\chi_2+cD(R,\xi)
F_z\chi_2)\nonumber\\
&+
(e+if)D(R,\xi)\chi_{2t},
\label{eq:modes}
\end{alignat}
where $a,b,c,d,e,f$ are real numbers. These terms
correspond to the excitation modes found by \cite{uedamodes}.
This correction only perturbs $\psi$ in $6$ of
the $10$  directions in the Hilbert space.
The other $4$ directions are accounted for by
the rotation $R$ and the phase $\xi$ which would be
Goldstone modes when $q=0$. (The energy remains $\xi$-independent
even when $q\neq0$.) The particular 
six stiff deformations in Eq. (\ref{eq:modes})
are chosen because they are orthogonal to infinitesimal
rotations and rephasings of the tetrahedral state.
We have to find the deformations that minimize $V_{tot}$, Eq.
(\ref{eq:hamiltonian}),
for each rotation $R$.
%find and substitute into the Hamiltonian
%the values of $a,b,c,d,e,f$ which minimize the Hamiltonian
%for a given $R$. 

To evaluate $V_{tot}$, note that $D(R,\xi)$ cancels
from all the terms in the energy except for the Zeeman term, 
where one
can use Eq. (\ref{eq:qmechspecial}). 
The resulting expression for the energy density reads
\begin{alignat}{1}
V_{tot}(\psi)=\frac{1}{2}&\alpha(\tilde{\psi}^{\dagger}\tilde{\psi})^2+
\frac{1}{2}\beta(\tilde{\psi}^{\dagger}\mathbf{F}\tilde{\psi})^2
+\frac{1}{2}\gamma|\tilde{\psi}_t^{\dagger}\tilde{\psi}|^2-
\mu\tilde{\psi}^{\dagger}\tilde{\psi}\nonumber\\
&-q\sum_{i,j=1}^3\cos\alpha_i\cos\alpha_j\tilde{\psi}^{\dagger}F_iF_j\tilde{\psi}
\label{eq:rotatedfield}
\end{alignat}
where $\tilde{\psi}=\sqrt{n_0}\chi_2
+d\chi_2+\frac{1}{\sqrt{2}}(aF_x\chi_2+bF_y\chi_2+c
F_z\chi_2)]+
(e+if)\chi_{2t}$ is the perturbed wave function without the rotation.
The effective potential Eq. (\ref{eq:effective}) is obtained by minimizing
$V_{tot}$ over $a,b,\dots$ while keeping $R$ fixed.  Further details
are in Appendix \ref{sec:feelingofdealing}. (Working with $\tilde{\psi}$
is equivalent to fixing the orientation of the tetrahedron and rotating
the magnetic field.)

The effective potential
suggests an analogue of the magnetic and charge
healing length in condensates without magnetic fields (the ``tetrahedron
tipping length")--when the tetrahedra
are rotated out of the appropriate ground state orientation
at the edge of a condensate, the energy in Eq. (\ref{eq:effective}) returns
the order parameter to $\mathcal{M}_q$ within the distance 
$L_q\sim \frac{\hbar}{q}\sqrt{\frac{n_0\beta}{m}}$.

Now the ground states can be found as a function of $c$.
When $c>4$, a short calculation
shows that Eq. (\ref{eq:effective}) has its minimum at
$\cos\alpha_i=\pm \frac{1}{\sqrt{3}},\ i=1,2,3$; i.e., when
the magnetic field is along the line connecting a vertex of the tetrahedron
to the opposite face or vertex.  Hence the order parameter space $\mathcal{M}_q$
is as given in Eq. (\ref{eq:alphax}).
When $c<4$, the effective potential is
minimized by an orientation in which the field points parallel to
the line joining a pair of opposite edges (see Fig. 
\ref{fig:bottler})\footnote{Hence, when 
$c>4$, the fluid is ferromagnetic--i.e., it becomes spontaneously
magnetized either parallel or antiparallel to the external magnetic field
with an $\mathbf{m}$ that can be calculated from the expressions in Appendix
\ref{sec:feelingofdealing}.
This magnetization spontaneously breaks a symmetry since the quadratic
Zeeman effect is invariant under $F_z\rightarrow -F_z.$
For zero net magnetization,
domains will form that have magnetization in both directions.
Even in the
presence of a small cubic Zeeman term, one can show that there is
 a discontinuous
phase transition (as the thermodynamical variable conjugate
to magnetization is varied) although the symmetry
between $\pm\hat{z}$ is not exact.}. 

The dependence of the ground-state orientation on $c$ can
be understood intuitively using the geometrical representation of Ref. 
\cite{icosa}. In a tetrahedral state, 
$\mathbf{m}=\psi^{\dagger}\mathbf{F}\psi=0$ and
$\theta=\psi_t^{\dagger}\psi=0$, minimizing the
interaction energies in Eq. (\ref{eq:hamiltonian}).
When a magnetic field is applied, the base of
the tetrahedron
illustrated in Fig. \ref{fig:bottler}a gets pushed toward the vertex
at $-\bm{\hat{z}}$.  The tetrahedron
in Fig. \ref{fig:bottler}b, on the other hand, has its upper and lower edges
pushed together, toward the $xy$-plane. Both these deformations
move the spin roots (see \cite{icosa}) away from
the north and south poles, which increases
the probability that $F_z=\pm 2$ in the corresponding spinors, as favored by
the quadratic Zeeman effect.   Now these two types of deformed
tetrahedra have different spin-dependent energies.  In the first
case, the magnetization $\mathbf{m}$ of the spinor becomes nonzero,
increasing the $\beta$ term of Eq. (\ref{eq:hamiltonian}).
In the second case, the magnetization of the
spinor is still zero, by symmetry, but one can check that the spinor
is no longer orthogonal to its time-reversal, so $\theta$
is nonzero\footnote{Using the equations,
the quadratic Zeeman term distorts
the ground state $\sqrt{n_0}\chi_3$ with increasing magnetic
field $B$ into $\psi_3(B)=\sqrt{n_0}
(X(B),0,0,Y(B),0)^T$
with $X>\sqrt{\frac{1}{3}}, Y<\sqrt{\frac{2}{3}}$
because the quadratic
Zeeman field favors large values of $|F_z|$.  The ground state
$\sqrt{n_0}\chi_2$ distorts 
into $\psi_2(B)=\sqrt{n_0}(X'(B),0,-iY'(B),0,X'(B))$ 
with
$X'>\frac{1}{2}$ and $Y'<\frac{1}{\sqrt{2}}$.  The deformed state
$\psi_3(B)$ has a nonzero $\mathbf{m}$ and a vanishing $\theta$ and
$\psi_2(B)$ is the other way around.  Incidentally,
when $B$ becomes large enough,
the ground state will become more symmetrical, with $X$ and $X'=1$ (compare
with Ref. \cite{ryan}).}. Therefore
the orientation of the ground state
tetrahedron is determined by whether the $\bm{m}^2$ term or
the $|\theta|^2$ has a larger coefficient in the Hamiltonian.  If the 
coefficient of the magnetization term is very large, then the state without
any magnetization is preferred.  The
detailed calculation shows that the relevant comparison is between $c\beta$
and $4\beta$.

The main source of anisotropy is different
at sufficiently low magnetic fields\cite{ryan};
the cubic Zeeman effect, proportional to $B^3$, then dominates over 
the effective potential in Eq. (\ref{eq:effective}), proportional to $B^4$.
Nevertheless the
the order $B^4$ effect we have calculated can dominate 
over the $B^3$ effect, even when the magnetic field
is small enough for the perturbation theory just described to apply.
This is possible because the denominator $n_0\beta$ in Eq. 
(\ref{eq:effective})
is small compared to the hyperfine energy splitting $A_{HF}$.  
For spin 2 atoms
the effect of the magnetic field is given by
\begin{equation}
V_{nz}=\psi^{\dagger}\sqrt{(\mu_B B)^2+A_{HF}^2+A_{HF}\mu_BBF_z}\psi
\end{equation}
where $A_{HF}$ is the hyperfine coupling. It follows that
the quadratic and cubic Zeeman effects are $qF_z^2=\frac{\mu_B^2B^2}{8A}F_z^2$
and
$\frac{\mu_B^3B^3}{16A^2}F_z^3$. The analysis we have given 
applies when the magnetic field is weak enough that the wave function is
not drastically distorted, Eq. (\ref{eq:smallfield}), but
strong enough for the second order effect of the quadratic Zeeman
term
to dominate over the cubic Zeeman term. These
conditions combine into
\begin{equation}
n_0\beta<<\mu_BB<<\sqrt{A_{HF}n_0\beta}
\end{equation}
For $^{87}$Rb at density $5\times10^{20}/m^3$,  
$n_0\beta=3$ nK and $A_{HF}=160$ mK while $\mu_B=67\mu$K/Gauss;
hence the anisotropy potential used here is actually valid
for a wide range of magnetic fields, between $.04$ mG and $.3$ G.
%The real ground-state space is the space of tetrahedron orientations $R\psi_0$
%which minimize $F$. From Eq. () we see that if $4\beta<\gamma$, then
%one of the tetrahedron's order three axes lines up with the magnetic field. 
%If $4\beta>\gamma$, then in the ground state, one of the tetrahedron's
%order two axes lines up with the magnetic field. That is, $R$ must be chosen so
%that $R\psi_0$ corresponds to a rotation about the z-axis of the tetrahedron shown in either the left or right frame of Fig. \ref{fig:bottler}. Equivalently, we can take representatives $R_3\psi_0=\psi_3$
%and $R_2\psi_0=\psi_2$ of these two ground state spaces, and then
%the ground state spaces are given by
%\begin{equation}
%\psi=e^{-i\alpha F_z}e^{ix}\psi_{2,3}.
%\label{eq:reallylight}
%\end{equation}
%Rotations involving other axes 
%besides the $z$ axis have a non-zero but small ($O(q^2)$) energy costs,
%hence vortices involving these axes can exist provided they are screened by other vortices at not
%too great a distance. 
%For the orientations of the tetrahedron shown in Fig. \ref{fig:bottler} (with
%the plane of the image parallel to the $xz$-plane), the wave functions are
%\begin{eqnarray}
%\psi_3=\left(\begin{array}{c}\sqrt{\frac{1}{3}}\\0\\0\\\sqrt{\frac{2}{3}}\\0\end{array}\right)\\
%\psi_2=\left(\begin{array}{c}-\frac{i}{\sqrt{2}}\\0\\\frac{1}{2}\\0\\\frac{i}{\sqrt{2}}\end{array}
%\right)\\
%\label{eq:2ndandthirdpose}
%\end{eqnarray}

\subsection{\label{sec:enindex}
Kinetic Energy}
Now that we have estimated the energy due to misalignments with
the quadratic Zeeman field, let us determine the kinetic energy of
vortices in terms of the rotations and rephasings, Eq. (\ref{eq:wrongdivision}).
This energy will be the source of the repulsion that keeps the vortices apart
within the vortex molecules. The kinetic energy is defined by
the gradient terms in the Hamiltonian $\mathcal{H}$; the analogue for
liquid crystals is the elastic energy favoring alignment of the order parameter.

In order for the kinetic energy of a vortex to be minimal, 
the field on a circle around it should trace 
out a geodesic, as mentioned above. For
the cyclic phase, geodesics take the form given
by Eq. (\ref{eq:lathes}).
To see that vortices are geodesics as functions
of $\phi$, suppose 
the field of a vortex is given far away by the radius independent
expression 
\begin{equation}
\psi(r,\phi)=\sqrt{n_0}F(\phi),
\label{eq:rayfield}
\end{equation}
for an appropriate spinor function $F(\phi)$.  The main
contribution to the energy of this vortex is from the kinetic energy,
which can be estimated by integrating from the core radius
$a_c\sim\hbar\sqrt{\frac{4\pi}{n_0\beta m}}$. This
is the same as the spin healing length (see \cite{tiedye} for the
definition).
The kinetic energy is
\begin{eqnarray}
E&\approx&\iint d^2\mathbf{r} \frac{\hbar^2}{2m}\nabla\psi^{\dagger}\cdot\nabla\psi\nonumber\\
&\approx&\int_0^{2\pi}n_0|F'(\phi)|^2d\phi\int_{a_c}^R\frac{\hbar^2}{2mr^2}rdr
\nonumber\\
&\approx&\frac{\hbar^2n_0}{2m}\ln\frac{R}{a_c}\int_0^{2\pi}|F'(\phi)|^2d\phi
\label{eq:kespinor1}
\end{eqnarray}
Now the curve parameterized by $F(\phi)$ adjusts itself
so as to minimize the last integral, while
maintaining the topology of the circuit traced out by $F(\phi)$
in the order parameter space. One can show
that an integral of this form
is minimized when $F(\phi)$ traces out a closed \emph{geodesic} in the
ground state space.  The length of a closed curve is defined
by $\int_0^{2\pi}|F'(\phi)|d\phi$, so
it is not surprising that
the geodesic of charge $\Gamma$, which
minimizes this expression, also minimizes Eq. 
(\ref{eq:kespinor1}). Furthermore, if the geodesic has length $l_{\Gamma}$,
then
$\int_0^{2\pi}d\phi|F'(\phi)|^2=\frac{l_{\Gamma}^2}{2\pi}$.

A geodesic in the cyclic order parameter space (which
has the local geometry of a perfect sphere 
in four dimensions)
can be described by a rotation
at a fixed rate around a single axis as in
Eq. (\ref{eq:lathes}).  
(For a shape less isotropic than a tetrahedron, 
the order parameter would rotate around a wobbling axis according
to the rigid-rotation equations.)
Substituting the symmetrical $F(\phi)$ into Eq. (\ref{eq:kespinor1})
gives\cite{ryanconv}
\begin{alignat}{1}
E\approx\frac{\hbar^2 n_0}{2m}\ln\frac{R}{a_c}\int &d\phi
\psi^{\dagger}(-\alpha\frac{\mathbf{\hat{n}\cdot F}}{2\pi}+
\frac{\theta}{2\pi})^2\psi\nonumber\\
\approx \frac{\hbar^2n_0}{4\pi m}\ln\frac{R}{a_c}
[\alpha^2 &(n_i n_j Q_{ij}+\frac{s(s+1)}{3})\nonumber\\
&-2\alpha \theta n_i M_i
+\theta^2]
\label{eq:kespinor}
\end{alignat}
where the general expression for any spin and phase has been given in
terms of the quantum fluctuation matrix
$Q_{ij}=\frac{1}{n_0}(\psi^{\dagger}\frac{F_iF_j+F_jF_i}{2}\psi)-\frac{s(s+1)}{3}\delta_{ij}$ 
and the magnetization per particle $M_i=\frac{m_i}{n_0}$.
The spin 2 tetrahedron state appears to be isotropic as long as one
does not go beyond second order correlators, as seen from the following
calculations:
\begin{eqnarray}
<\chi_3|F_i|\chi_3>=0\label{eq:sphere?1}\\
<\chi_3|F_iF_j|\chi_3>=2\delta_{ij}
\label{eq:sphere?}
\end{eqnarray}
and hence $M=Q=0$. Eq. \ref{eq:kespinor} implies that the
energy is proportional to $l_{\Gamma}^2=(\theta^2+2\alpha^2)$, a generalization
of the Pythagorean theorem showing how to combine the amount of rephasing
and rotation to get the total geodesic length.  A spin
rotation costs twice as much energy as a rephasing by the same angle.

In order to study vortex stability and Coulomb forces,
let us define the ``energy index," 
\begin{equation}
I_E=(\frac{l_{\Gamma}}{2\pi})^2=(\frac{\theta}{2\pi})^2+2(\frac{\alpha}{2\pi})^2,
\label{eq:index}
\end{equation}
which is a fraction for each tetrahedral charge from Sec. 
\ref{sec:namingofcats}.
The energy of a vortex is given as $\frac{\pi\hbar^2n_0}{m}\ln\frac{R}{a_c}
I_E(\Gamma)$, a multiple of the energy of an ordinary phase vortex.
The force between a pair of vortices can be expressed very
simply in terms of $I_E$. 

\begin{figure}
\psfrag{X}{$X$}
\includegraphics[width=.45\textwidth]{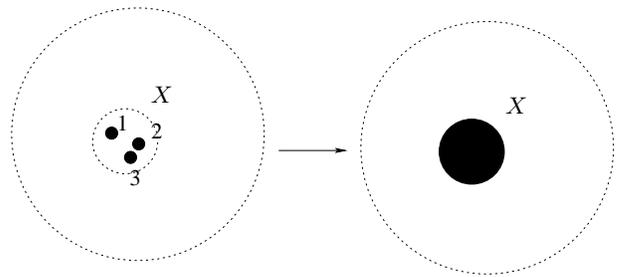}
\caption{\label{fig:maizy} A group of vortices is combined
into the core of a composite vortex $X$. 
The energy outside the blacked-out ``core"
is calculated from the resultant winding number of the vortices inside, and
the energy inside the ``core" is calculated by looking inside the core
to see which vortices are actually there.}
\end{figure}

The force follows from an
estimate of the energy of a cluster of vortices. Using ideas
from Ref. \cite{chandrasekhar},
we think of the cluster
as forming the ``core" of a bigger vortex, as illustrated in 
Fig. \ref{fig:maizy}. (We are not necessarily assuming that the vortices are
bound together.) Draw a circle of radius $X$ just around the group
of vortices.
%The three nearby vortices can be thought
%of as the core of a vortex whose winding numbers are the resultant winding
%numbers of the three vortices as illustrated in the second image.
%Away from the three clustered vortices, the field is determined by
%the combined winding number, so the energy of the white region in the
%second figure is determined by the composite winding numbers.  Within 
%the "core" of the imaginary vortex, the three vortices can be resolved.
The kinetic energy can then be found as the sum of the energies outside
and inside of $X$; for the case illustrated in the figure this
energy is approximately
\begin{equation}
\pi \frac{n_0\hbar^2}{m}
([I_E(X)\ln\frac{R}{L_X}]+
[(I_E(1)+I_E(2)+I_E(3))\ln\frac{L_X}{a_c}]),
\label{eq:witches}
\end{equation}
%\begin{equation}
%\pi \frac{n_0\hbar^2}{m}
%([I_E(B)\ln\frac{R}{L_B}]+[(I_E(4)+I_E(5))\ln\frac{L_B}{a}+
%I_E(A)\ln\frac{L_B}{L_A}]
%\\+[(I_E(1)+I_E(2)+I_E(3))\ln\frac{L_A}{a}])
%\end{multline}
where $L_X$ is the diameter of the group being combined together and $R$
is the radius of whole system. The first term describes the energy
outside of $X$.
%This result is obtained by adding the energy inside the circle labelled $A$,
% and the energy outside $A$. 
Sufficiently far outside of $X$,
the field should have the form of a rotationally symmetric vortex.
The rotation and rephasing of this vortex, measured along a circle outside
$X$,
is obtained by multiplying the group elements which
describe all the individual vortices according to the rule
for vortex unification Eq. (\ref{eq:berlin}). Hence 
the energy outside $X$
is given by an expression like in
Eq. (\ref{eq:kespinor}), except that the integral
must start at the radius $L_X$ of the circle, so $a_c$ must
be replaced by $L_X$.
%The values of $x$ and $\alpha$ to be used here
%are obtained by multiplying the group elements 
The energy inside $X$ is approximated by adding the
energies of the three vortices in it, which are calculated
like in Eq. (\ref{eq:kespinor}) but now
with $R$ replaced by $L_X$.

This approximation
makes a small error (compared to $\ln\frac{L_X}{a_c}$)
by ignoring the region where the vortices' fields overlap.  This
error has a special scaling form
if $q=0$ and if there are no other vortices in the condensate.  In this
case, the kinetic energy gives a complete description of the vortices
outside their cores, and has a symmetry
under rescaling.  Therefore (as in \cite{chandrasekhar}), one can
show that
the difference between Eq. (\ref{eq:witches})
and the actual energy  has the form 
\begin{equation}
\Delta E=\frac{\pi n_0\hbar^2}{m}f(\frac{L_{12}}{L_{13}},\frac{L_{23}}{L_{13}})+O(\frac{a_c}{L_X})\label{eq:chandrasekhar}
\end{equation}
where $L_{ij}$ refer to the sidelengths
of the triangle formed by the vortices.  This correction
can be ignored relative to the logarithmically divergent
terms kept in Eq. (\ref{eq:witches}), \emph{as long as the side-lengths all
have the same order of magnitude}, since
in this case $f$ has no singularity.

The general expression for the energy of a set of vortices
with charges $\Gamma_i$, after rearranging the formula to emphasize
the dependence on the diameter of the set $L$, is:
\begin{alignat}{1}
E_{K}\approx\frac{\hbar^2\pi n_0}{m}[\sum_i &I_E(\Gamma_i)-I_E(\prod_i\Gamma_i)]
\ln\frac{L}{a_c}\nonumber\\
&+\frac{\hbar^2\pi n_0}{m}I_E(\prod_i\Gamma_i)\ln\frac{R}{a_c}
\label{eq:nerffusion}
\end{alignat} 
As long as the vortex cores are well-separated, the error in
this estimate depends only on the ratios of the distances
between the vortices, as in the three-vortex case, Eq. (\ref{eq:chandrasekhar}).
The energy function has a similar form to the two-dimensional
Coulomb interaction between vortices in an ordinary single-component
condensate, and shows that the vortices repel or attract each other according
as the index of the combined vortex is greater or less than the sum
of the separate vortices' indices.  An interesting difference
is that the interaction
is \emph{not} a sum of interactions of all pairs (consider three
$(A,0)$ vortices, for example, or almost any other set of three vortices).
%\footnote{  A set of three $(A,0)$ vortices
%are attracted to each other collectively, since the combination of
%three clockwise $180^{\circ}$ rotations is an anticlockwise rotation:
%$I_E((A,0)^3)=I_E((A^{-1},0)=\frac{1}{2}<3I_E(A,0)$. But any two of
%the vortices repel each other, since two $180^{\circ}$ rotations around
%the same axis make a $360^{\circ}$ rotation which is
% classified as $-id$ in $SU_2$ and has energy index 2.} 

When there are just two vortices, the force between them can
be calculated by differentiating Eq. (\ref{eq:nerffusion}),
giving $-\frac{\hbar^2\pi n_0}{mL}[I_E(\Gamma_1)+I_E(\Gamma_2)-
I_E(\Gamma_1\Gamma_2)]$.
This is the \emph{exact} (if $L\gg a_c$ and $q=0$)
expression for the force between
the two vortices because the error (see Eq. (\ref{eq:chandrasekhar})) 
reduces for two vortices to a constant.
Improving on Eq. (\ref{eq:nerffusion}) depends on
finding the spin texture in the overlap regions, and these
equations are nonlinear when  
not all the
vortices use the same symmetry axes.

In the next section, we will apply Eq. (\ref{eq:nerffusion}) more broadly.
When there is a small cluster of vortices of size $L$, 
but there are some other vortices
besides, the energy of just the vortices \emph{in the cluster} 
can be estimated by
replacing $R$ in Eq. (\ref{eq:nerffusion}) by the distance $D$ to the nearest
vortex not in the cluster.
Likewise, for $q\neq 0$ the energy of a cluster with a net charge that is
field-aligned can be estimated 
by replacing
$R$ by $L_q$.  As long as $L\ll L_q$, the detailed expression
for the energy has the form given by Eq. (\ref{eq:chandrasekhar}),
but
when $L$ approaches $L_q$, the anisotropy energy starts to compete with
the kinetic energy and to distort the spin texture around the vortices.

\section{\label{sec:lavoisier} Chemistry of Vortices}

In this section we will first discuss stability of
isolated tetrahedral vortices and then determine
when these vortices can combine to form molecules (and what the spin
texture around a molecule looks like).  Some of these molecules
are only metastable and can each break up in several ways.

Bound states of vortices will be formed out of stable tetrahedral
vortices.  These are the vortices based on $120^{\circ}$ symmetries of
a tetrahedron, 
accompanied by a phase shift of $\frac{2\pi}{3}$ or $-\frac{4\pi}{3}$,
and vortices based on $180^{\circ}$ symmetries without a phase shift. Vortices
with larger rotations or phase shifts will not occur as components of 
molecules.
The net charge of the bound state must be field-aligned, with
a schematic relation
\begin{equation}
Q=\prod_i \Gamma_i\label{eq:hoax}
\end{equation}
between the field-aligned charge and
the component charges.
A composite vortex is stable if Coulomb
repulsions can prevent the collapse of
any subset of its components; since the Coulomb interaction
is not a sum of pairwise interactions, 
it is not enough that every pair of vortices repel each other.
Finally any bound state whose charge $Q=(\alpha,\theta)_3$ has big enough
rotation or rephasing angles can in principle decay into 
molecules with less energy, but this can only occur if thermal energy
overcomes the Coulomb repulsion between the component vortices.
The following expands on this general picture and points out some 
interesting sidelights.

\subsection{\label{sec:list}Stable Tetrahedral Vortices}
Only stable vortices will be found in the core-region of composite vortices.
Since a weak anisotropy term can be neglected near component vortices,
we can enumerate the stable vortex types assuming that $q=0$.

%These are the
%vortices that cannot break up (in a way consistent with the topology
%at infinity)
%into a set of vortices with less kinetic energy. In a non-zero
%magnetic field, some of these vortices will still be found in vortex cores.
%Only the stable $q=0$ vortices will be found in cores, because
%the unstable vortices can break up (with the fragments
%possibly remaining confined to the cores) in a way that lowers 
%the net kinetic energy.

Absolute stability of a vortex with charge $\Gamma$ implies that if
$\Gamma$ is forced to break up into the fragments $\Gamma_i$, then
the energy of the
fragments grows as they separate from one another.
%out whether the energy of any set of vortices it can break up into is lower.
% it can break up into (remembering that
%the analogue of charge conservation for nonabelian vortices is that
%the group element corresponding to a vortex is the product of the group elements of the vortices it breaks up into), and seeing whether any of
%these have a lower total energy.  
The energy of the fragments can be found by applying Eq. (\ref{eq:nerffusion}).
Note that $\prod_i\Gamma_i=\Gamma$ by conservation of charge.
If $\Gamma$ has phase and rotation
angles $\theta$ and $\alpha$
then
\begin{equation}
E_{fragments}\approx\frac{n_0\hbar^2}{4\pi m}
[\sum_i(\theta_i^2+2\alpha_i^2)-(\theta^2+2\alpha^2)]\ln\frac{L}{a}+cnst.
\label{eq:nerffusionexpanded}
\end{equation} 
%where we have assumed that $L'$ is much less than the distance between
%cluster being studied and any vortex, and where
%$E_{collapsed}$ is the energy of the configuration after the vortices 
%$(x_i,\alpha_i)$ have collapsed into a single vortex $(x,\alpha)$.
Thus, if $\sum_i(\theta_i^2+2\alpha_i^2)<(\theta^2+2\alpha^2)$, the
energy decreases as the vortices move apart, so the fragments
will move apart by themselves the rest of the way. 
On the other hand, the vortex is absolutely stable if
\begin{alignat}{1}
&\sum_i (\frac{\theta_i}{2\pi})^2+2(\frac{\alpha_i}{2\pi})^2>(\frac{\theta}{2\pi})^2+2
\frac{\alpha}{2\pi}^2\nonumber\\
& \mathrm{for\ }every\mathrm{\ set\ of\ }
\Gamma_i=
(e^{-i\frac{\alpha_i\mathbf{\sigma}\cdot\mathbf{\hat{n}}_i}{2}},e^{i\theta_i})
\mathrm{\ such\ that\ }\prod_i\Gamma_i=\Gamma 
\label{eq:absolutestability}
\end{alignat}
Any tetrahedral vortex not satisfying this ``absolute stability criterion"
can break up into a lower-energy state, and we will assume that this break-up
happens spontaneously for these \emph{component} vortices.
%  (We will show in Section \ref{sec:metastable} that 
%vortex \emph{molecules} can be long-lived even if there is a lower-energy
%configuration of vortices with the same \emph{net} charge.)
For example, the vortex $(id,4\pi)$ should break up into two $(id,2\pi)$'s, 
halving the energy.
(In fact, these two singly quantized vortex can break up even further.)
Generalizing this example,
any vortex with charge $(g,\theta)$ whose circulation $\theta$ is
bigger than $2\pi$ can
break up into
$(g,\theta-2\pi)$ and $(0,2\pi)$ since $\theta^2>(2\pi)^2+(\theta-2\pi)^2$.
This leaves only finitely many vortices that have the possibility
of being stable: all the ones with phase winding numbers
not more than
$2\pi$.  Some of
the vortices with $|\theta|\leq 2\pi$ are also unstable; trial and error
finds decay processes such as:
%\footnote{In some of these
%examples, there are other decompositions of the same vortex
%which involve rotations about other axes (of the same order);e.g., $C$ can be
%replaced by $A$ in the first example.}. 
%The $SU_2$ winding ``number" of each vortex
%is indicated using notation based on the illustration of the tetrahedron and its symmetry
%axes, Fig. \ref{fig:bottler}. $S,P,Q,R$ refer to $120^{\circ}$ counterclockwise rotations,
%as viewed facing towards the tetrahedron from the tip of the appropriate arrow.  $A,B,C$ refer
%to counterclockwise $180^{\circ}$ rotations.

\begin{enumerate}
\item$(-id,0)\rightarrow(C,0)*(C,0)$ or $(R,2\pi/3)*(R,2\pi/3)*(R,-4\pi/3)$. 
%($-id$, an element
%of $SU_2$, is the $2\pi$ rotation about an arbitrary axis.)
\item$(-id,2\pi)\rightarrow 
(R,\frac{2\pi}{3})*(R,\frac{2\pi}{3})*(R,\frac{2\pi}{3})$
\item$(R^2,4\pi/3)$ can break up into $(R,2\pi/3)*(R,2\pi/3)$ and also 
$(S^{-1},4\pi/3)*(C,0)$
\item$(R^2,-2\pi/3)$ might break up into  
 $(S^{-1},-\frac{2\pi}{3})*(C,0)$.
\item$(C,2\pi)$ could break up into 
$(R,\frac{2\pi}{3})*(S^{-1},\frac{4\pi}{3})$.
%\item$(0,4\pi)$ breaks up into $(R,2\pi/3)^6$,  $(0,2\pi)^2$ and (the nonabelian
%specialty) $(P^{-1},4\pi/3)(Q^{-1},4\pi/3)(R^{-1},4\pi/3)$.
\end{enumerate}
All of these decays lower the energy index. There are also two vortices
which can break up without the energy indices changing:
\begin{enumerate}
\item $(0,2\pi)$ has the same charge as
 $(R,\frac{2\pi}{3})*(R^{-1},\frac{4\pi}{3})$.
\item $(R,-\frac{4\pi}{3})$ has the same charge as
$(P^{-1},-\frac{2\pi}{3})*(Q^{-1},-\frac{2\pi}{3})$.
\end{enumerate}
The energy may either increase or decrease after one of these processes,
the logarithmic term which dominates the energy (see Eq. (\ref{eq:nerffusion}))
does not change, so the remainder term needs to be calculated
to determine whether these break-ups raise or lower the 
energy.  A point vortex whose
charge is $(0,2\pi)$ is probably unstable if $\alpha\gg\beta$
because a pure phase vortex
has to have an empty core (with energy density of
order $\alpha n_0^2$), while the two fragments it breaks up into
just have non-cyclic cores (energy of order $\beta n_0^2$).  
\footnote{ 
The break-up only lowers the net energy 
by a \emph{finite} amount,
about $\pi\frac{n_0\hbar^2}{m}\ln\sqrt{\frac{\alpha}{\beta}}$ and
the fragments interact with a short-range repulsion, $\propto \ln r/r^3$
perhaps; when the energy index of the fragments actually
decreases the force is
$\propto \frac{1}{r}$.  Whether $(R,-\frac{4\pi}{3})$ is stable or not we do
not know, and the answer may
depend on $c$.}

We can now enumerate the vortices which are stable, noting
that some vortices (such as $(P^2,\frac{4\pi}{3})$) can
break up similarly to the ones just listed on account of symmetry.
The only vortex types that have not been eliminated are the vortex with
one-third circulation, $(R,2\pi/3)$ (energy index $\frac{1}{3}$)
and the currentless
vortex $(C,0)$ (energy index $\frac{1}{2}$)
as well as possibly $(R,-\frac{4\pi}{3})$ 
(energy index $\frac{2}{3}$), and also the inverses and conjugates of these.
%This last vortex can break up into 
%$(R,2\pi/3)(R^{-1},4\pi/3)$, but this does not lower
%the energy index.  
%The latter pair of vortices has an energy that is lower by
%a \emph{finite} amount.
% The repulsive force
%between the break-up products is short-ranged (seemingly $\propto \ln r/r^3$)
%\footnote{
%Whether the break-ups
%$(R^2,-\frac{2\pi}{3}\rightarrow (R,-\frac{4\pi}{3})(R,\frac{2\pi}{3})$ 
%and $(C,2\pi)\rightarrow (C,0)(id,2\pi)$ lower the energy also
%cannot be determined by the energy-index
%stability criterion. However, each of the initial
%vortices can break up into alternative sets of vortices which lower the
%energy index more.}.

\subsection{\label{sec:molecules}Vortex Molecules at $q\neq0$ and their
Spin Textures}
In this section we will describe a qualitative wave function
for the example in Section \ref{sec:doublestar} to illustrate
how the field of a vortex molecule deforms in response to the anisotropy
energy as one leaves the region
containing the two vortices.  We will then give the 
binding criteria
which describe how to use group theory to check which sets of the 
stable
tetrahedral vortices (from the previous section)
can combine together to form a vortex molecule
at a nonzero magnetic field.  

A vortex with a charge that is not compatible with the magnetic field has 
an energy that grows proportionally to the area of the condensate.
Eq. (\ref{eq:compatibility}) shows that the vortices in a cluster
can avoid this energy cost if they have 
a net charge corresponding to a rotation around the
$z$ axis.  More specifically,
a set of tetrahedral vortices, with topologies $\Gamma_i$, has to have
a net topology of the form
\begin{equation}
\prod_i\Gamma_i=(R^m,\frac{2\pi m}{3}+2\pi n).
\label{eq:screened}
\end{equation}
The ``molecule" can then have the rotational and phase
windings
$(\alpha,\theta)_3=(\frac{2\pi m}{3}+4\pi j,2\pi(\frac{m}{3}+n))_3$ at infinity,
for any $j$. 
%is compatible with the quadratic Zeeman field but its component vortices
%are not, these vortices will form a vortex molecule (provided no subset of
%these vortices can lower their energy by coallescing).
One example of a molecule for the $\mathcal{M}_{q3}$ phase
is the  composite vortex discussed at the beginning, $(A,0)(A,0)$, 
which partner up
to make a compound vortex with $Q=(2\pi,0)_3$ in the field-aligned condensate.

\begin{figure*}
\includegraphics[width=\textwidth]{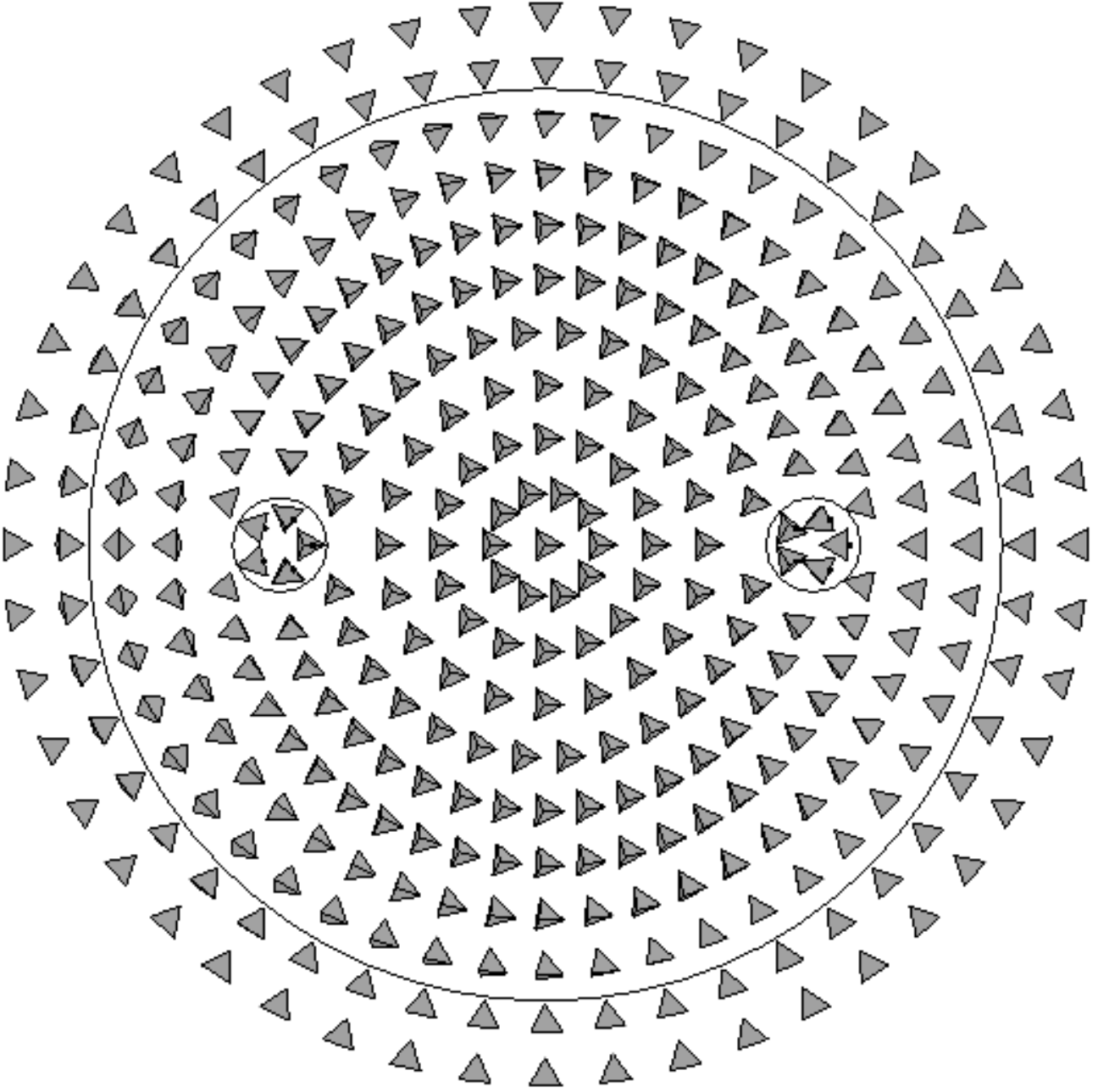}
\caption{\label{fig:suntiles} An illustration of the composite vortex 
$(A,0)*(A,0)$.  %The texture illustrated shows
%how a pair of vortices based on order 2 axes not compatible with
%the magnetic field can ``screen" one another in such a way
%that the tetrahedra have the proper orientation at infinity.  
The magnetic field is perpendicular to the figure, so the tetrahedra
prefer to be oriented with a face or a vertex facing up (since $c>4$).
The two small circles enclose the $180^{\circ}$ vortices, with symmetry
axes indicated by black dots.  The large circle indicates the transition
region where the axis of the $360^{\circ}$ rotation changes relative
to the tetrahedra, from the $A$ to the $R$ axis.  The
figure uses a functional form with a rapid
jump for $\mu(r)$
(unlike in Eq. (\ref{eq:random})) for simplicity, so that
the transition region does not overlap
the cores.  For the probably more
realistic form given by Eq. (\ref{eq:random}), the tetrahedra are
already tipped at the centers of the vortices so that the local rotation
axes $\bm{\hat{n}_i'}$ do not have the standard orientations illustrated in
Fig. \ref{fig:bottler}a.}
\end{figure*}

Writing a wave function that describes this example even qualitatively
is a little more complicated
than it sounded in Sec. \ref{sec:doublestar}.  A possible
wave function is illustrated in Fig. \ref{fig:suntiles}.  We can build
this wave function up in stages.
 The most obvious attempt at writing a wave function fails to eliminate the 
quadratic energy cost:
\begin{equation}
\psi_{misalign}=
e^{-\frac{i}{2}(\phi_1+\phi_2)
(\sqrt{\frac{2}{3}}F_x+\frac{1}{\sqrt{3}}F_z)}\sqrt{n_0}\chi_3,
\label{eq:bigsubscript}
\end{equation}
where $\phi_i=\mathrm{arctan}\frac{y-y_i}{x-x_i}$ is the polar angle
measured with respect to the location $(x_i,y_i)=\pm(\frac{L}{2},0)$ 
of the $i^{\mathrm{th}}$ vortex.  
The two $180^{\circ}$ rotations about $A$
combine into a $2\pi$ rotation on a circle surrounding both vortices,
but not about the magnetic field axis. \emph{This} wave function takes the form
$\psi\approx e^{-i\phi\frac{1}{\sqrt{3}}(\sqrt{2}F_x+F_z)}\sqrt{n_0}\chi_3$ 
at infinity
where $\phi_1$ and $\phi_2$ approach $\phi$. The
tetrahedra rotate around a tilted axis, so they do not stay aligned
appropriately with the magnetic field except on the $x$-axis and 
the $y$-axis (where the tetrahedra are reversed,
but this is still
a ground-state). The Zeeman
energy of $\psi_{misalign}$ is still infinite. 
This wave function agrees pretty
well with the one illustrated in the figure \emph{within} the region surrounded
by the large circle.  The vortex cores are cordoned off by the small
circles, where the $180^{\circ}$ symmetry axes are marked by dots.  Note
that the tetrahedra three rows inside the large circle behave like
the tetrahedra at infinity described by Eq. (\ref{eq:bigsubscript}); they
rotate through
$360^{\circ}$ about the order-two axis bisecting their right edge, 
producing all sorts
of arbitrary orientations. 

Another attempt, which uses the $R$ axis instead of the $A$ axis
to eliminate the tilting of the tetrahedra,
\begin{equation}
\psi_{discontinuous}=
e^{-\frac{i}{2}(\phi_1+\phi_2)
F_z}\sqrt{n_0}\chi_3,
\label{eq:biggersubscript}
\end{equation}
is a complete fiasco, since this function is not continuous along
the line connecting the two cores. (The $R$ axis has the wrong symmetry,
so as $(x,y)$ circles around $(x_2,y_2)$, $\psi_{discontinuous}$  
changes from $\chi_3$ to $e^{-i\pi F_z}\chi_3\neq\chi_3$.)

Luckily, since a $2\pi$ rotation around one axis can be deformed
to any other ($2\pi$ rotations all correspond to $-id\in SU_2$)
one can produce a spin texture which has a finite Zeeman
energy and is continuous.
A hybrid of Eqs. (\ref{eq:bigsubscript}) and (\ref{eq:biggersubscript})
which achieves this, illustrated in Fig. \ref{fig:suntiles}, is:
 \begin{alignat}{1}
\psi=&e^{-i\phi(F_x\sin \mu(r)+F_z\cos\mu(r))}
e^{i\phi\frac{\sqrt{2}F_x+F_z}{\sqrt{3}}}\nonumber\\
&\times\ \ \ \ \ \ e^{-\frac{i}{2}(\phi_1+\phi_2)
(\sqrt{\frac{2}{3}}F_x+\frac{1}{\sqrt{3}}F_z)}\sqrt{n_0}\chi_3
\label{eq:tailoring}
\end{alignat}
where  $\mu(r)$ must satisfy
\begin{eqnarray}
&&\mu(0)=\arccos\sqrt{\frac{1}{3}}\label{eq:1/3}\\
&&\mu(\infty)=0\label{eq:0}
\end{eqnarray}
For example, we could define
\begin{equation}
\cos\mu(r)=\sqrt{\frac{r^2+D^2}{r^2+3D^2}}
\label{eq:random}
\end{equation}
where $D$ is a variational parameter. (Probably $D\sim L$ is optimal.)

%Without the first two rotational factors, this wave function would 
%be the simple juxtaposition of two $(A,0)$ vortices:

%composite vortex which uses the $R$ axis instead of the $A$ axis is not
%continuous along the line connecting the two cores:

%The value of $\phi_2$ jumps by $2\pi$, but $e^{-i\pi F_z}\psi_3\neq\psi_3$.
%probably cannot be solved by using an axis different from the $A$ axis,
%because the axis needs to be an order $\pi$ symmetry of the tetrahedron
%so it has to be an order 2 axis of the tetrahedron if the wave function is 
%continuous around each of the two components.

%Although the first two factors seem to describe
%a field with a discontinuity near the origin, as a function
%of $\phi$ like in a vortex,
%the discontinuous parts of the two factors cancel
%on account of Eq. (\ref{eq:1/3}).  This continuity is apparent
%in Fig. \ref{fig:suntiles}, where the tetrahedra near the origin all have
%roughly the same orientation. 
This field has two vortices at $(\pm\frac{D}{2},0)$.
The texture varies continuously,
except at the at the very centers of the vortices.
(In the figure, the cores are inside the small circles.  The tetrahedra's
orientations change rapidly around the circles' centers, but each
tetrahedron inside has a similar orientation to
the tetrahedron next to it outside the circle.)
Around the cores,
the tetrahedra rotate around a tilted $\bm{\hat{n}'}$ axis which
is aligned with the symmetry axis of the tilted tetrahedra around
the vortex; the
tilting is carried out by the rotations described by the first two factors in
Eq. (\ref{eq:tailoring}), just as discussed in Section \ref{sec:doublestar}.
%the tetrahedra are tilted so
%that this axis coincides with an order three symmetry axis.  
Near vortex 1, the wave function is approximately
\begin{equation}
\psi\approx e^{-\frac{i\phi_1}{2}(\sqrt{\frac{2}{3}}F_x+\sqrt{\frac{1}{3}}F_z)}
\sqrt{n_0}\chi_3\label{eq:leaningtower}
\end{equation}
 and near vortex 2, 
\begin{multline}
\psi\approx e^{-i\pi(F_x\sin\mu(\frac{L}{2})+F_z\cos\mu(\frac{L}{2}))}
\\e^{-\frac{i\phi_2+i\pi}{2}(\sqrt{\frac{2}{3}}F_x+\sqrt{\frac{1}{3}}F_z)}\sqrt{n_0}\chi_3\label{eq:ofpisa}.
\end{multline}
(The local symmetry axis $\bm{\hat{n}'}$ is the rotation of the standard
axis
$\sqrt{\frac{2}{3}}\bm{\hat{x}}+\sqrt{\frac{1}{3}}\bm{\hat{y}}$, through
a half-turn about $\bm{\hat{x}}\sin\mu(\frac{L}{2})+
\bm{\hat{y}}\cos\mu(\frac{L}{2})$, according to Eq. (\ref{eq:lathes}).)

Though the two tetrahedra are rotated into misaligned positions inside
the large circle, 
the 
excursions from $\mathcal{M}_{q3}$ to $\mathcal{M}$ are brief 
and therefore cost only a finite amount of
energy. Indeed, the first two factors
fix the field up at infinity by applying a continuously varying
rotation to the overall texture.  These factors change
the axis from $A$ to $R$ at large $r$
as one can see by replacing $\phi_1\approx\phi_2$ by
$\phi$ and using Eq. (\ref{eq:0}):
\begin{equation}
\psi\rightarrow e^{-i\phi F_z}\sqrt{n_0}\chi_3.
\label{eq:2piatinfinity}
\end{equation}
The $360^{\circ}$ axis
changes from the $A$ axis to the $R$ axis as one crosses through
the transition region indicated by the large circle
in Fig. \ref{fig:suntiles}.  As you follow a radius outward past the circle,
the tetrahedra are tipped by different amounts. (The tetrahedra on
the negative $x$-axis have to be 
tipped the most, though their original orientation
is compatible with the magnetic field!  The face which is on top
is changed.  This reorientation is required to make the amount
of tipping continuous: the tipping angle increases more and more
as $\phi$ goes from $0$ to $\pi$.)

The first two factors manage to fix the orientation at infinity
without introducing discontinuities like in Eq. (\ref{eq:biggersubscript}).
Although they seem to have
a vortex-like discontinuity at the origin, 
their discontinuous parts cancel
on account of Eq. (\ref{eq:1/3}).  (Hence the tetrahedra
near the origin in Fig. \ref{fig:suntiles} all have roughly the
same orientation.) In short, though $(A,0)^2=(-id,0)$, 
the cancellation of the Zeeman energy at infinity is not automatic!  
The topological classification just implies that the field in Eq. 
(\ref{eq:bigsubscript})
\emph{can} be deformed as $r\rightarrow\infty$ so that the Zeeman
energy is small.

%corresponds to $-1$, independent
%of the axis.  Thus, it is possible to deform the axis of rotation
%so that it becomes the magnetic field axis instead of
%$\frac{1}{\sqrt{3}}(-\sqrt{2},0,1)$: 
%If $L>>d$, then $\psi$ even agrees with $\psi_{misalign}$ out to the
%location of the two vortices, but since rotations around different axes
%are coupled by the kinetic energy term, it is likely that $L$ is of order
%$d$. This makes it harder to picture the composite vortex.
%The only
%defects are the $(A,0)$ vortices of the third factor which are located at $\pm(\frac{d}{2},0)$.

The kinetic energy of this composite vortex, according to Eq. 
(\ref{eq:nerffusion}),
is
\begin{alignat}{1}
E_K\approx \frac{\hbar^2\pi n_0}{m}&[2I_E((A,0))-I_E((-id,0))]\ln\frac{L}{a_c}\nonumber\\
&+
\frac{\pi\hbar^2n_0}{m}I_E((-id,0))\ln\frac{R}{a_c}.
\label{eq:keamolecule}
\end{alignat}
Since $I_E((A,0))=\frac{1}{2}$ and $I_E(-id,0)=2$, the coefficient of the
first term is negative, so the vortices
are driven apart in order to decrease the kinetic energy.  This
effect competes with the anisotropy energy, which tries to bring
the vortices together.
The net energy is therefore
\begin{equation}
E\approx f(D/L)\frac{q^2D^2}{\beta}+\pi \frac{n_0\hbar^2}{m}(2\ln\frac{R}{L}+\ln\frac{L}{a}).
\label{eq:barrier2}
\end{equation}
This is a more complete version of Eq. (\ref{eq:barrier}).  $f$  summarizes
 the dependence of the
anisotropy energy Eq. (\ref{eq:effective}) on $D$ and $L$,
but is too complicated for us to figure out! 

The competition between the attractive Zeeman-term
and the repulsive kinetic-energy term determines the size
of the vortex.  Assuming $D=L$ and minimizing over $L$ 
gives 
\begin{equation}
L\sim L_q=\frac{\sqrt{\frac{n_0\hbar^2\beta}{m}}}{q}.
\label{eq:collie}
\end{equation}
This composite core-size is also the scale for the decay of tipping of
tetrahedra due to the competition of anisotropy and kinetic energy
(see Sec. \ref{sec:cubeshape}), just as the core size of a spin vortex
in a spin 1 condensate is equal to the magnetic healing length.
Vortex molecules will have a size of the same
order of magnitude (even if there are more
than $2$ subvortices) except when the component vortices
have a short-range repulsion, like the two examples at the end of Section
\ref{sec:list}.  (Such components form smaller molecules;  vortices never
form larger molecules.  If a pair of tetrahedral vortices is
stretched beyond $L_q$, a ``cord" of tipped-tetrahedra forms between them,
a simple version of the string imagined to hold the quarks together
in a rapidly-rotating baryon.) 

To find the total energy of the molecule,
substitute $L$ into Eq. (\ref{eq:barrier2})
and simplify
using $a_c=\hbar\sqrt{\frac{4\pi}{n_0\beta m}}$:
\begin{equation}
E_{vortex}\sim 2\pi \frac{n_0\hbar^2}{m}\ln R \frac{\sqrt{mq}}{\hbar}
\label{eq:energy2pies}
\end{equation}
We have dropped the contribution from the quadratic Zeeman term since
it adds something that is independent of $q$.

This formula can be compared to the energy one expects for
a \emph{simple} vortex with the same charge $(2\pi,0)_3$:
\begin{equation}
E_{vortex}=I_E((2\pi,0)_3)\pi \frac{n_0\hbar^2}{m}\ln\frac{R}{a_c'}+\epsilon_c
\label{eq:defcoreenergy}
\end{equation}
which includes a core energy $\epsilon_c$ and allows for a physical
definition of the core size.  If we take $a_c'$ to be the size of the molecule,
$L_q=\frac{1}{q}\sqrt{\frac{n_0\hbar^2\beta}{m}}$,
then the core energy has to be
\begin{equation}
\epsilon_c\approx\pi \frac{n_0\hbar^2}{m}\ln\frac{n_0\beta}{q}.
\end{equation}
The core energy becomes large as $q\rightarrow 0$ because the
kinetic energy of the component vortices diverges as the logarithm of the core
size $L_q$.

%All these examples have composite core sizes as given by Eq. (\ref{eq:compositecoresize}).
It makes sense to regard this composite vortex as a molecule because
Coulomb
repulsion keeps the vortices in the core separate.  The previous
discussion can be generalized by listing a set of 
binding criteria; these ensure that a set of tetrahedral vortices will
form a stable or metastable
composite vortex:
\begin{enumerate}
\item Each component vortex is one of the stable $q=0$ vortices
from Section \ref{sec:list}.
\item The kinetic energy is not decreased
when any subset of the component vortices coalesces
into a single vortex.
\item There is no way for the component vortices to form submolecules
that can break apart.
This would occur if the components could be rearranged and
then partitioned into $r$ sets $\{
\Gamma_1,\Gamma_2,\dots,\Gamma_{j_1}\},
\{\Gamma_{j_1+1},\Gamma_{j_1+2},\dots,\Gamma{j_1+j_2}\},\dots,\\
\{\Gamma_{j_1+j_2+\dots+j_{r-1}+1},\Gamma_{j_1+j_2+\dots+j_{r-1}+2},\dots,
\Gamma_{j_1+j_2+\dots+j_{r}}\}$ such that each
subset forms a molecule \emph{that is compatible with the magnetic field} (i.e.,
$\prod_{i=j_1+\dots+j_{k-1}+1}^{j_1+\dots+j_k}\Gamma_i=
(R^m_k,2\pi(\frac{m_k}{3}+n_k))$)
and such that \emph{the sum of the energy indices of these submolecules is
less than the energy index of the original molecule}.
\end{enumerate}
The vortex molecule $(A,0)*(A,0)$ clearly satisfies all these conditions:
Condition 1 is satisfied because $(A,0)$ is one of the stable vortices
found in Section \ref{sec:list}.  Condition 2 is satisfied because the
vortices repel each other. 
Condition 3 is easy to check for a diatomic molecule
like this one, since it
 can only break up into individual ``atoms"; neither
of the fragments $(A,0)$
is  compatible with the magnetic field. 
 
\subsection{\label{sec:metastable}Metastable Vortices and How They Decay}

%When $q\neq 0$, we can use Eq. (\label{eq:absolute}),
%now applied only to the group elements appropriate
%for $q\neq 0$ (see Eq. (\ref{eq:order3vortices})), to 
Not all of the vortex molecules satisfying the three conditions
above are \emph{absolutely} stable.
The analogue of the absolute
stability condition, Eq. (\ref{eq:absolutestability}), 
also selects a finite set of \emph{aligned}
vortex types when $q\neq0$,
 this time from among the group
elements listed in Eq. (\ref{eq:type3}).  
The absolutely stable charges
are\begin{equation}
\pm(\frac{2\pi}{3},\frac{2\pi}{3})_3,\  
\pm(\frac{2\pi}{3},-\frac{4\pi}{3})_3.
\label{eq:stable3}
\end{equation}
For any \emph{other}
vortex topology $Q=(\alpha,\theta)_3$, 
one can find vortex topologies $Q_i$ such
that $\prod_i Q_i=Q$ and
\begin{equation}
I_E(Q)>\sum_i I_E(Q_i).
\label{eq:degradation}
\end{equation}
\emph{Point} vortices with such a topology $Q$ 
would likely break apart spontaneously.

There is actually another pair of charges that could be absolutely
stable, but the energy index
estimate is not accurate enough to decide the issue:
%\footnote{\label{fn:fn}One
%cannot determine whether these vortices 
%are stable without considering
%their core structure because they are topologically equivalent to sets
%of vortices with the same total energy index (namely 
%$(\pm\frac{2\pi}{3},\pm\frac{2\pi}{3})_3*
%(\mp\frac{2\pi}{3},\pm\frac{4\pi}{3})_3$
%and $(\pm\frac{2\pi}{3},\pm\frac{2\pi}{3})_3
%*(\pm\frac{2\pi}{3},\mp\frac{4\pi}{3})$ respectively).}:
\begin{equation}
(0,\pm2\pi)_3,(\pm\frac{4\pi}{3},\mp\frac{2\pi}{3})_3
\label{eq:dicey}
\end{equation}
These vortices can break up into pairs without changing the net energy indices,
reprising the ambiguous behavior of the two vortices at the end
of Sec. \ref{sec:list}. (See Section \ref{subsec:bound} for an answer.)

\begin{figure}
\psfrag{A}{a)}
\psfrag{B}{b)}
\psfrag{pi}{$Q$}
\psfrag{pi1}{$Q_1$}
\psfrag{pi2}{$Q_2$}
\psfrag{pi3}{$Q_3$}
%\psfrag{D0}{$\pi \frac{n_0\hbar^2}{m}I_E(Q)$}
%\psfrag{D1}{$\pi \frac{n_0\hbar^2}{m}I_E(Q_1)$}
%\psfrag{D2}{$\pi \frac{n_0\hbar^2}{m}I_E(Q_2)$}
%\psfrag{D3}{$\pi \frac{n_0\hbar^2}{m}I_E(Q_3)$}
\psfrag{D0}{}
\psfrag{D1}{}
\psfrag{D2}{}
\psfrag{D3}{}
\includegraphics[width=.45\textwidth]{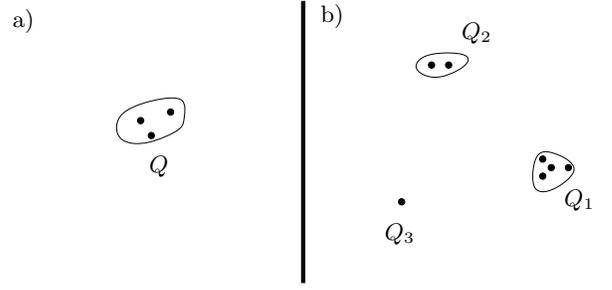}
\caption{\label{fig:tc}Illustration of the absolute stability criterion.
 A composite vortex (a) and a possible set of composite vortices (b) it can
break up into.  Even if $Q=Q_1Q_2Q_3$ and
$I_E(Q)>I_E(Q_1)+I_E(Q_2)+I_E(Q_3)$ so that the energy would decrease, the
break-up might not occur spontaneously.
If the
component vortices in (b) are different from the component vortices in (a),
then the vortices making up $Q_1$, $Q_2$, and $Q_3$ would have to be produced
in a ``chemical" reaction from the components of $Q$.
}
\end{figure}

There are some \emph{composite}
vortices of other charges besides the four listed in Eq. (\ref{eq:stable3})
which are long-lived.
The absolute stability criterion misses this possibility
because it ignores the details of the vortex
cores, drawing all its conclusions from the topology of the vortices
far away:
Suppose the
initial vortex $Q$ is a cluster of vortices with
topologies $\Gamma_1,\Gamma_2,\dots,\Gamma_r$ (see Fig. \ref{fig:tc}). 
The decay products discussed in the previous paragraph, $Q_i$,
may be the combined topology of other
vortex clusters. 
%a molecule and its possible break-up products are indicated in Fig. 
%\ref{fig:tc}.  
The energy of the vortices after the reaction (Fig. \ref{fig:tc}b)
is smaller if
\begin{alignat}{1}
E_{init}&=I_E(Q)\pi \frac{n_0\hbar^2}{m}\ln\frac{R}{L}+k\ln\frac{L}{a_c}\nonumber\\
&>\sum_i [I_E(Q_i)\pi \frac{n_0\hbar^2}{m}\ln\frac{R}{L}+k_i\ln\frac{L}{a_c}]
=E_{fin};
\label{eq:degrad2}
\end{alignat}
The terms proportional to $\ln\frac{R}{L}$
stand for the kinetic energies outside the composite cores,
and the terms proportional to $\ln\frac{L}{a_c}$ stand
for the energies within the composite cores; the latter contributions
do not matter once the composite vortices are far apart ($R\gg L$). 
Therefore, the energy-index relation, Eq. (\ref{eq:degradation}),
implies that the energy
decreases when 
$Q\rightarrow Q_1*Q_2*\dots Q_r$.

However, at zero temperature, a vortex molecule satisfying
the binding criteria \emph{cannot} break up, even
if the total energy would end up smaller. 
Although the metamorphosis of
Fig. \ref{fig:tc}a into Fig. \ref{fig:tc}b lowers the energy, the process
will not occur spontaneously.  According to 
condition 3, the vortices in Fig. \ref{fig:tc}b
are different than those in Fig. \ref{fig:tc}a. And according
to conditions 1 and 2, there are no spontaneous chemical reactions
that can occur to make the components in Fig. \ref{fig:tc}b out  
of those in Fig. \ref{fig:tc}a.
Thus, at zero temperature, composite vortices besides 
the ones with charges listed in Eq. (\ref{eq:stable3}),
whose energy indices seem to be too big, can still be stable.  
%(
%Section \ref{subsec:double} shows that
%a doubly-quantized phase vortex can be stable if its core
%is made out of the right components.)
%the vortex cluster, without changing, vortex \emph{molecules}, because of their
%extended composite cores, may be stable at zero temperature.  
At \emph{nonzero} temperature, such a molecule will 
only be metastable
because it can decay after reactions inside
its core produce the vortex types that appear in Fig. \ref{fig:tc}b.
These reactions are prevented by energy barriers, so the
decay will occur only after a long time.
%with a lower energy, without
%changing the topology at infinity, but there will be an energy barrier
%discouraging such a break-up.  
%can be broken up into a set of vortices with a lower energy,
%in a way that is consistent with topology.
%Some of the other vortices still have metastable forms.

The vortex molecule made up of $(A,0)*(A,0)$
is an example of a metastable vortex; its charge $(2\pi,0)_3$ did not
appear in Eq. (\ref{eq:stable3}) because it is the same as
the net charge of the three \emph{field-aligned}
 point vortices $(\frac{2\pi}{3},\frac{2\pi}{3})_3*
(\frac{2\pi}{3},-\frac{4\pi}{3})_3*(\frac{2\pi}{3},\frac{2\pi}{3})_3$. The
energy index of the molecule is $2(1^2)+0^2$ (see Eq. (\ref{eq:index}))
while the
energy index of the three point vortices is smaller, 
$\frac{1}{3}+\frac{2}{3}+\frac{1}{3}$.  Nevertheless, 
since these three vortices are not present in the core of the original vortex 
the decay cannot occur spontaneously.  
%\section{Second Example: Rioting Vortices and Energy Barriers}

A vortex with a composite core can break up by a combination of vortex-fusion
and vortex fragmentation.  Some
subsets of the original component vortices fuse
and the fused vortices each break up into some other vortices. 
These regroup into
clusters each of which has a charge compatible with the magnetic field. Then
each cluster goes its own way. (The fusion and fragmentation
 steps might sometimes happen more
than once.)  Conditions 1 and 2 ensure that at least one of these steps
will be opposed by the Coulomb potential, 
but the \emph{total} energy will decrease if the
energy index decreases.  The vortex molecule will be long-lived
because its components do not know that the hard effort of fusing will
allow them to change into vortices which can separate.
%If the melding of the subsets is not energetically
%favored, then there is an energy barrier to this process, and
%if also the final set of vortices has a lower energy than the initial
%vortex, then the composite vortex will be metastable, but not absolutely
%stable. 

%some of which are very close, can be estimated by nesting;

%The three nearby vortices can be thought
%of as the core of a vortex whose winding numbers are the resultant winding
%numbers of the three vortices as illustrated in the second image.
%Away from the three clustered vortices, the field is determined by
%the combined winding number, so the energy of the white region in the
%second figure is determined by the composite winding numbers.  Within 
%the "core" of the imaginary vortex, the three vortices can be resolved.

Trial and error yields a couple of ways in which
the $(A,0)*(A,0)$ bound state can break up.
One possibility begins with the two component vortices coalescing,
\begin{equation}
(A,0)*(A,0)\rightarrow (-id,0) \rightarrow (R,\frac{2\pi}{3})*(R,\frac{2\pi}{3})
*(R,-\frac{4\pi}{3})
\label{eq:fusion}
\end{equation}
The other begins when one of the components breaks up,
\begin{alignat}{1}
(A,0)*(A,0)&\rightarrow [(R^{-1},-\frac{2\pi}{3})*(Q,\frac{2\pi}{3})]*(A,0)\nonumber\\
&\rightarrow (R^{-1},-\frac{2\pi}{3})*[(Q,\frac{2\pi}{3})*(A,0)].
\label{eq:fission}
\end{alignat}
In the first process, the two vortices
come together, increasing the kinetic energy in accordance with
Condition 2 (as shown by calculating
the $I_E$'s and substituting into Eq. (\ref{eq:nerffusion})). The resulting vortex breaks up into three vortices
which can separate from each other because they are compatible
with the magnetic field.
 The increase in energy during the first stage is given by 
$E_{s1}-E_{init}=\pi\frac{n_0\hbar^2}{m}\ln\frac{L_q}{a_c}$ where
$E_{s1}=\pi\frac{n_0\hbar^2}{m}I_E(-id,0)\ln\frac{R}{a_c}$ is the energy of
the intermediate vortex.  Thermal
fluctuations have a chance of 
driving the $(A,0)$'s together, in spite of the energy increase
$E_{s1}-E_{init}$.

In the second process, Eq. (\ref{eq:fission}), 
the $(A,0)$ vortex first splits up into two vortices.
The first of
these, $(R^{-1},-\frac{2\pi}{3})$ is compatible with the Zeeman field
and can leave. The remaining two vortices form a new molecule which
cannot break up because $Q$ is a rotation around the wrong order $3$
axis.
For \emph{this} process Condition 1 requires
 that the energy increases during the initial fragmentation.
To check this, note that the energy of 
the intermediate state is
\begin{alignat}{1}
E_{s2}=\pi\frac{n_0\hbar^2}{m}& \{[I_E(R^{-1},-\frac{2\pi}{3})
+I_E(Q,\frac{2\pi}{3})
+I_E(A,0)]\ln\frac{L}{a_c}\nonumber\\
&+I_E(-id,0)\ln\frac{R}{L}\}
\end{alignat}
and the energy barrier is
\begin{alignat}{1}
E_{s2}&-E_{init}\nonumber\\&
=\pi\frac{n_0\hbar^2}{m}[(I_E(R^{-1},-\frac{2\pi}{3})+
I_E(Q,\frac{2\pi}{3})-I_E(A,0)]\ln\frac{L}{a}\nonumber\\
&=\frac{\pi n_0\hbar^2 m}{6}\ln\frac{L}{a}.
\label{eq:fissionbar}
\end{alignat}
This energy barrier is lower than $E_{s1}-E_{init}$, so Eq. (\ref{eq:fission})
is a more common break-up route.  (In a finite condensate,
thermally excited break-ups can be observed only if a vortex molecule
is somehow prevented from wandering to the boundary of the
condensate and annihilating before it can decay.)

These two examples illustrate the meaning of the stability conditions.
Conditions 1 and 2 ensure that fragmentation and fusion processes 
cannot happen spontaneously.
The third condition simply points out that 
vortex clusters like $(A,0)^2*(B,0)^2$ will not be stable because
the components can sort themselves into field-aligned
groups and break up without any thermal assistance.
The second condition can be difficult to check for a 
composite vortex with three or more
sub-vortices.
One must consider subsets of every size and check that they
cannot lower their energy by collapsing all at once into one vortex. 
Just knowing that
any two vortices of the subset \emph{repel} each other does not guarantee that
the set of vortices do not \emph{collectively attract} each other! 
An example is
the set of three vortices $(A,0)^3$. 
Any two of these vortices
would combine to form a vortex with a $360^{\circ}$ rotation, $(-id,0)$,
whose index $2$ is higher than the sum, $\frac{1}{2}+\frac{1}{2}$, of the
indices of the collapsing pair.  On the other hand
all three vortices 
could form a vortex $(A^{-1},0)$
with energy index 
$\frac{1}{2}<3I_E(A,0)$, so the three vortices 
can collapse simultaneously to lower the energy.   
An even more counterintuitive complication is that, 
because of the noncommutative
behavior of the combination rules, more complicated
fusion processes can occur. A vortex can change its type by
circling around one vortex so that it can fuse with another vortex.  (See
Appendix \ref{app:catstring}.)  For a bound state of many
vortices there will
be many possibilities for 
how the vortices meander around each other before some
of them
fuse. To test Condition 3, one also has to enumerate all possible wanderings.

A mathematical statement of our results
is that there will be (at least) one solution to
the time-independent Gross-Pitaevskii equation for any set of vortex
topologies that satisfy the three conditions.
If $q$ is small, the energy at the top of the barrier is greater, by
a logarithmically large amount, than the energy of an initial
variational state like the approximate wave function
Eq. (\ref{eq:tailoring}).
A solution to the Gross-Pitaevskii equation
should result if one
starts from this
qualitative texture
and lets it relax to a local minimum of the energy.  There
is not enough energy for the 
wave function to get over the energy barrier, 
so the wave function
should get stuck in 
a local minimum. (The energy of the intermediate state is not known precisely
because of the rough estimates we have made of the Zeeman energy and
the kinetic energy, but these errors are small compared to the height of the
barrier.)

%and then checked to hold even without any minus sign with these expressions:
%\begin{eqnarray}
%A&=&e^{-i\pi(\sqrt{2}{3}\frac{\sigma_x}{2}+\sqrt{\frac{1}{3}}\frac{\sigma_z}{2}}
%\\
%R&=&e^\frac{\pi i\sigma_z}{3}\\
%Q&=&e^{-\frac{2\pi i}{3}(\sqrt{\frac{2}{3}}\frac{\sigma_x}{2}\\
%\end{eqnarray}
%-\sqrt{\frac{2}{3}}\frac{\sigma_y}{2}+\frac{1}{3}\frac{\sigma_z}{2})}
\section{\label{sec:sampler}Additional Examples}
\begin{table*}
\begin{tabular}{c|c|c|c|c}
&Components&Net Charge&c&Stable?\\\hline 1&
$(A,0)*(A,0)$&$(2\pi,0)_3$&$c>4$&Metastable\\\hline 2&
$(P^{-1},\frac{4\pi}{3})*(Q^{-1},\frac{4\pi}{3})*(R^{-1},\frac{4\pi}{3})$&
$(0,4\pi)_2$&$c<4$&Metastable\\\hline 3&
$(Q,\frac{2\pi}{3})*(A,0)$&$(-\frac{4\pi}{3},\frac{2\pi}{3})_3$&$c>4$&Stable\\
\hline 4&
None&$(4\pi,0)_{2,3}$&Any value&Metastable\\\hline
\end{tabular}
\caption{\label{table:ledger}Examples of vortex molecules.  The tetrahedral
charges of the components
of the molecules and the net aligned charge are given. The
condition on $c$ determines how the tetrahedra are oriented
far from the vortex, due to the magnetic field. 
The final column indicates whether the vortex molecule is expected
to have the absolute minimum energy of all vortices with a given net aligned
charge.  The second molecule
might actually not be bound--see the text.}
\end{table*}
Now we can construct some other, more interesting, examples.
We will use 
the algebra of the group
of vortex charges to find molecules whose net charge is interesting
in different ways, and we will use the energy index to test whether 
they are stable.

The parameter $c$ will
be less than 4 for some of these examples. If $c<4$ the
ground state orientation of the tetrahedron will be as in
Fig. \ref{fig:bottler}b so the $z$-axis is an order 2 axis.  The aligned
topologies have the forms $(\pi n,2\pi m)_2$, as described in 
Section \ref{sec:namingofcats}.  Of these, the only
absolutely stable topologies are $\pm(\pi,0)_2$, $\pm(0,2\pi)_2$ 
(and possibly
 $(\pm\pi,\pm 2\pi)_2$).
\subsection{\label{subsec:double}A Doubly Quantized Pure Phase Vortex}
First let us find a vortex molecule whose phase
winds by $4\pi$.  In single component condensates, such vortices
are usually unstable; one has been observed to break up, maybe into an entwined
pair of $2\pi$ vortices\cite{shinvortexsplit}. If phase and spin
textures were completely independent of one another, doubly quantized
vortices would not be any more stable in the cyclic condensates; but 
fractional circulations are ``bound" to certain spin 
textures (see Sec. \ref{sec:namingofcats}).
\emph{If} we assume the vortex $(R,-\frac{4\pi}{3})$ is stable 
(at the end of Sec. \ref{sec:list} we could not decide), then
a doubly-quantized vortex
can occur in a cyclic condensate when
$c>4$. 
 It consists of the three parts
\begin{equation}
(P^{-1},\frac{4\pi}{3})(Q^{-1},\frac{4\pi}{3})(R^{-1},\frac{4\pi}{3}).
\end{equation}
The phase changes by $4\pi$ while the orientation of the tetrahedron
does not change at infinity
as we can check 
using the coordinate system from
Fig. \ref{fig:bottler}b.   
The three group elements are
\begin{eqnarray*}
P^{-1}=\frac{1}{2}(1+i(\sigma_x-\sigma_y-\sigma_z))\\
Q^{-1}=\frac{1}{2}(1+i(\sigma_x+\sigma_y+\sigma_z))\\
R^{-1}=\frac{1}{2}(1+i(\sigma_z-\sigma_x-\sigma_y))
\end{eqnarray*}
and their product is the identity. 

Let us discuss the conditions for binding.  We have not
checked Condition 1; it is not easy to check
because $(P^{-1},\frac{4\pi}{3})$
has the same charge and energy index as 
$(R,\frac{2\pi}{3})*(Q,\frac{2\pi}{3})$;  
an accurate solution
for the spin texture around this pair of
vortices is needed.  Besides, $(P^{-1},\frac{4\pi }{3})$ might be stable
for some ranges of $c$ values, but not others.  Let us therefore hope
that condition 1 is satisfied.
Condition 3 is clear.
To check condition 2,
let us first consider whether one of the \emph{pairs} of vortices in the
trio can coalesce.
Using conservation of topological charge helps to
avoid enumerating all the ways
the vortices can braid around each other.
If the first two vortices
have coalesced into a vortex
$(X,\frac{8\pi}{3})$ (after some permutation)
and the third vortex, by winding around the
other two vortices as they collapsed, has changed to 
$(Y^{-1},\frac{4\pi}{3})$, then 
\begin{equation}
(XY^{-1},4\pi)=(id,4\pi)
\end{equation}
by conservation of charge.
Hence $X=Y$.  Also, braiding one vortex between other vortices can only
\emph{conjugate}
 its group element. Therefore, $Y$, like $R$, is a counterclockwise
rotation through
$120^{\circ}$.
 Since $X=Y$, the rotation part of the
coalesced vortex $(X,\frac{8\pi}{3})$ also is a
$120^{\circ}$ turn and thus
the energy index of this coalesced vortex is
$2\times(1/3)^2+(4/3)^2=2$, which is \emph{greater} than the sum of the energy
indices of
the two vortices which formed it. Therefore the two
vortices cannot coalesce spontaneously. (This argument can be generalized
to any trio of
vortices $\Gamma_1,\Gamma_2,\Gamma_3$ each of which commutes with the
net charge $\Gamma$. Fusing two of the vortices gives the same result
(up to conjugacy) no matter how the vortices are mixed around first; so
braiding cannot make a repulsive interaction between two vortices
into an attractive one.)  
Finally, the three vortices cannot
coalesce simultaneously because $I_E(0,4\pi)>I_E(P^{-1},\frac{4\pi}{3})+
I_E(Q^{-1},\frac{4\pi}{3})+I_E(R^{-1},\frac{4\pi}{3})$.

\subsection{\label{subsec:bound}A Vortex Molecule which \emph{is} Stable}
Returning to the original assumption,
$c>4$, where the ground
state orientation is illustrated by 
Fig. \ref{fig:bottler}a, we can show that the second charge in Eq. 
(\ref{eq:dicey}) \emph{does} correspond to a completely stable vortex molecule.
In fact, consider
\begin{equation}
(Q,\frac{2\pi}{3})(A,0).
\end{equation}
This molecule, one of the decay products in Eq. (\ref{eq:fission}),
has the topology $(R^{-2},\frac{2\pi}{3})$ or (using the notation
appropriate for the field-aligned tetrahedra outside
the composite core), 
$(-\frac{4\pi}{3},\frac{2\pi}{3})_3$. The energy of
this molecule is approximately
\begin{alignat}{1}
\frac{\pi n_0\hbar^2}{m}[I_E(R^{-2},\frac{2\pi}{3})&\ln\frac{R}{L_q}+I_E(Q,\frac{2\pi}{3})\ln\frac{L_q}{a_c}+I_E(A,0)\ln\frac{L_q}{a_c}]\nonumber\\
&=\frac{n_0\pi\hbar^2}{m}( \ln\frac{R}{a_c}-\frac{1}{6}\ln\frac{L_q}{a_c}).
\end{alignat}
where $L_q$ is the size of the composite core, given
by Eq. (\ref{eq:collie}).

This molecule answers a question from Section \ref{sec:metastable}.
Are there stable vortices with charge $(-\frac{4\pi}{3},\frac{2\pi}{3})_3$?
The two vortices 
$(-\frac{2\pi}{3},-\frac{2\pi}{3})_3*(-\frac{2\pi}{3},\frac{4\pi}{3})_3$,
have the same net topology as a vortex of charge 
$(-\frac{4\pi}{3},\frac{2\pi}{3})_3$, and they
have the \emph{same} net
energy index.  Now we can
check that the composite vortex $(Q,\frac{2\pi}{3})(A,0)$ is a stable
realization for the charge $(-\frac{4\pi}{3},\frac{2\pi}{3})_3$. 
Its energy is lower \emph{by a finite amount}
than the energy 
$\frac{\pi n_0\pi\hbar^2}{m}(\frac{1}{3}+\frac{2}{3})\ln\frac{R}{a_c}$
of the pair of vortices.
This finite binding energy is
$\frac{\pi n_0\hbar^2}{6m}\ln\frac{L_q}{a}$. 

To take another point of view,
the minimum-energy spin texture with
the topology $(-\frac{4\pi}{3},\frac{2\pi}{3})_3$ imposed far away
has an asymmetric structure:  it
has two ``singularities" with topologies
$(Q,\frac{2\pi}{3})$ and $(A,0)$ at a distance of order $L_q$.  By contrast,
when the topology imposed at a boundary corresponds to unstable vortices,
the ground state has singularities whose spacing is on the order of the
size of the system $R$.  E.g., in a scalar
one might try to impose
$\psi(R,\phi)=\sqrt{n_0}e^{2i\phi}$.  The spacing
of the vortices in the
energy minimizing wave function grows with $R$,
reflecting the fact that these vortices
would repel each other to infinity in an infinite condensate.
%The topology
%$(-\frac{4\pi}{3},\frac{2\pi}{3})_3$
%classification (see footnote \ref{fn:fn})

%When $c<4$,
%energy indices did not give any answer about
%whether any vortex with charge $(-\frac{4\pi}{3},\frac{2\pi}{3})_3$  would
%be stable.
%It has the same charge as the vortices $(-\frac{2\pi}{3},-\frac{2\pi}{3})_3$
%and $(-\frac{2\pi}{3},\frac{4\pi}{3})$ whose energy indices have a total
%value equal to the energy index of the original vortex. The molecule
%fragment that is produced from the decay Eq. (\ref{eq:fission}) gives the
%answer.  Its topology at infinity is $(-\frac{4\pi}{3},\frac{2\pi}{3})_3$
%and its energy
%is $\approx\frac{n_0\pi\hbar^2}{m}( \ln\frac{R}{a}-\frac{1}{6}\ln\frac{L}{a})$ 
%where $L$ is 
%given by Eq. (\ref{eq:collie}); the subtracted term lowers its energy
%relative to its potential fragments, $(-\frac{2\pi}{3},-\frac{2\pi}{3})_3$
%and $(-\frac{2\pi}{3},\frac{4\pi}{3})$. So the list of absolutely
%stable vortices when $c<4$ now includes
%$\pm(-\frac{4\pi}{3},\frac{2\pi}{3})_3$ as well.  The situation with 
%$(0,2\pi)_3$ is still uncertain, but it is probably unstable.

%\section{Third Example: A vortex without vortices}
\subsection{\label{subsec:coreless}A ``Bound State" of No Vortices}
The final example shows that point vortices are not necessary to hold
a core together--there is a ``composite" vortices without any components!
In other words we can construct a vortex for which the order parameter
stays
in $\mathcal{M}$. There is still a ``composite core" where
the tetrahedra leave $\mathcal{M}_q$ and
are no longer aligned with the field axis.
The trick is that the amount of rotation in a texture around
a vortex is defined only modulo $4\pi$, in
the absence of a magnetic field (because of the $SU_2$ charge
classification).  When $B$ is turned on, the spin
part of the order parameter space $\mathcal{M}_q$
has the same topology as a circle, so each additional winding by $2\pi$
changes the topological charge.  (A texture which rotates by $4\pi$ about
the field axis
can relax only by using 
axes perpendicular to the magnetic field.)  Thus a $4\pi$-rotation-vortex is 
stable
in a magnetic field, but since it has zero \emph{tetrahedral} charge, it 
does not have to have point
vortices inside of it. (Another way to say this: Eq. 
(\ref{eq:screened}) does not uniquely determine the aligned topology,
because $R^6=id$.  Hence a
$(4\pi,0)_3$-vortex can be made from
$6$ $R$ vortices (i.e., some $(R,\frac{2\pi}{3})$'s and $(R,-\frac{4\pi}{3})$'s)
or out of no vortices at all!)

Such vortices occur for both
the $c>4$ and $c<4$ cases. A variational wave function
can be constructed using the formula that shows how
a $4\pi$ rotation-vortex can relax in the absence of a magnetic
field:
\begin{equation}
\psi(\phi;t)
=e^{-i\phi(F_x\sin\pi t+F_z\cos\pi t)}e^{-iF_z\phi}\sqrt{n_0}\chi_0,
\label{eq:rubiksolution}
\end{equation}
where $\chi_0$ is an arbitrary cyclic spinor.
At each moment of time $t$, the expression describes an $r$-independent
 texture as a 
function of $\phi$.  When $t=0$, there is a vortex which is a full rotation
through $720^\circ$. By the time $t=1$, this vortex has completely
dissipated.  When $q\neq 0$ a $4\pi$ vortex cannot relax in
this way because
the tetrahedra rotate
away from the orientation preferred by the magnetic field before
returning to the preferred orientation at the end. 
But Eq. (\ref{eq:rubiksolution}) has a reincarnation as the description of
a $(4\pi,0)_{2,3}$ vortex. 
We replace the time coordinate by a function of
the radius to give a spin texture
that winds through $4\pi$ at infinity but does not have any singularities
at $0$:
\begin{equation}
\psi=\psi(\phi;\frac{1}{1+(\frac{r}{L})^2}).
\end{equation}
If $\chi_0=\chi_2$ or $\chi_3$, then 
this wave function, at large $r$'s, has the winding number $(4\pi,0)_{2,3}$. 
At small $r$'s, the
wave function is $\phi$-independent, giving a continuous and ``coreless" 
wave-function.  (The exact solution
not only has a more complicated $r$-dependence, but
also a less-symmetrical $\phi$-dependence.)
 The region $r\lesssim L$ is the composite core of
this vortex in the sense that $\psi\in\mathcal{M}$ rather than $\mathcal{M}_q$.
The optimal size $L$ of this region is again $L_q$, as balancing
the kinetic and Zeeman energies shows.

This vortex cannot disappear because the classification of
vortices at \emph{nonzero} $q$ implies that $\alpha=4\pi$ is conserved.
Furthermore, though it does not satisfy the \emph{absolute} stability 
criterion,
since two $(2\pi,0)_{2,3}$'s
have a smaller energy, it is obviously metastable--there are no vortices
in the core to break apart!
The vortex can only break up if 
thermal energy causes a vortex-antivortex
pair to nucleate in the core.  Suppose a
pair involving rotations through $2\pi$ in \emph{opposite} directions
appears.  These vortices initially
attract each other but if the thermal fluctuations pull them
to opposite sides of the core the 
nonlinear coupling  with the background field switches this force from
attractive to repulsive, and the vortices can separate the rest of
the way by themselves.

%Thinking of halted vortex-breakups as metastable vortices with asymmetrical
%cores: LIST A COUPLE OF VORTICES AND HOW THEY SHOULD BREAK UP . refer
%to previous example discuss why it is metastable and estimate energy barrier
%give footnote on 6 vortex example maybe (giving variational form). GIVE
%variational form for another noncommutative example like the three
%order threes. GIVE comment on why it is hard to figure out stability.

%Interestingly, even vortices which seem like they should be able to break up
%in the presence of the magnetic field sometimes do not (Can I argue that this
%REQUIRES nonabelian symmetry group). This is because an energy barrier to the break-up is produced.  Because the symmetry group is nonabelian, there are two ways for the vortex to break up--if it breaks up one way, the magnetic field
%prevents the break up from completing.  However, the different subvortices have to be brought back together again before the break-up reaction can proceed differently. (show this by trying to break each sub-vortex up, and pull one
%of its pieces over toward the other) As example for this, use
%(-1/3A,2/3)(-1/3B,2/3)(-1/3C,2/3), maybe.  Calculate the height of the
%barrier for some of the examples.

\section{\label{sec:fieldwork}Creating and Observing Vortex Molecules}

Let us discuss the conditions under which the Zeeman-effect bound states
might be observed and the methods one can use for observing them.  
First of all, we must assume that $q\ll\beta n_0$ 
in order to justify neglecting $q$ near the tetrahedral vortices
and to 
justify the perturbation theory of Sec. \ref{sec:cubeshape}.  This
is not just a technical assumption: above a certain magnetic field
the component vortices probably merge. 
%The magnetic field must be weak enough for us
%to make the approximation $q=0$ inside the vortex cores.  In
%we need to ensure (as in Eq. (\ref{eq:smallfield})
%that the energy gain $qn_0$ from appropriately
%orienting 
%the cyclic-state tetrahedra is much less than the energy gain from choosing
%a cyclic state $\beta n_0^2$, so that the hierarchical spin textures
%of Eqs. (\ref{eq:Manifold}) and (\ref{eq:alphax}) exist.
%Because the tetrahedral phase may be difficult to study, it
%is useful to work out what this criterion means
%while keeping other spins (such as spin 3) and phases 
%in mind.
To estimate the maximum magnetic field
note that $q$ is related to the
hyperfine splitting $A_{HF}$ via
\begin{equation}
|q|=\frac{\mu_B^2B^2}{8A_{HF}},
\label{eq:qzhyperfine}
\end{equation}
for rubidium and sodium atoms\cite{pethickbook}, and similar relations
hold for other atoms.  Also note that the spin independent interaction is
\begin{equation}
\alpha=\frac{4\pi\hbar^2}{ma}
\label{eq:alphaa}
\end{equation}
where $a\sim 50$ \AA\  is an \emph{average}
 of the scattering lengths corresponding to
 different net spins
and that the spin dependent interaction is
\begin{equation}
\beta=\frac{4\pi\hbar^2\Delta a}{m}
\label{eq:betada}
\end{equation}
where $\Delta a\sim 1$ \AA\cite{chang1and2,schmaljohann} 
depends on the \emph{differences} between the 
scattering lengths\cite{ciobanu}.
%$\alpha=\frac{4\pi\hbar^2 a}{m}$ (where $a$ is some average of
%the scattering lengths, see \cite{ciobanu} for the spin 2 case).  
%and spin-dependent interactions can be parameterized by
%which represents the cost of changing the ratios of the spinor components
The condition for our analysis to be applicable, $q<<n_0\beta$,
therefore implies
\begin{equation}
B<<B_{Max}\sim \frac{1}{\mu_B}\sqrt{A_{HF}\frac{\hbar^2 n_0\Delta a}{m}},
\label{eq:minimumfield}
\end{equation}
about $.1$ G for a condensate of rubidium atoms
with density $n_0=5\times10^{14}/$cc.

In order to observe vortex bound states, one might start with a condensate
prepared with a spin order other than the ground state
and then watch it evolve as in Ref. \cite{tiedye}. Thermal (and less importantly
quantum) noise will produce perturbations that grow exponentially, producing
complicated patterns.
%(Such random textures have been discussed in theoretical
%articles, including \cite{quenchedvortices,quantumquenches} 
%(on the statistics of the spin
%fluctuations and vortices
%produced from this random evolution), 
%\cite{cherng1} on instabilities and \cite{ferrodynamics,lewentropy}
%on the dynamics of spinor condensates.  The experiments
%described in \cite{vengdipole} show that the patterns that evolve
%in Rubidium condensates are probably affected by dipole-dipole interactions,
%though we are not considering these.)
If the magnetic
field is small enough, vortex bound states might be found after some time.
To test whether these vortex bound states behave in the way we have
been describing, one would have
to identify the topological charges of the vortices.
One could then check that vortex sets whose net charge is compatible
with the magnetic field have a size on the order of the theoretical
value, $L_q$.  One may have to use statistical correlations if
too many vortices stay around.
(One could also take a more deliberate approach, choosing vortex types
and imprinting them as in
\cite{phaseimprint} or \cite{twistvortex}.  One can then observe the
subsequent evolution of the vortices to see whether they bind together.)

In fact, identifying the vortices that appear in a spinor condensate is
difficult;
%To identify bound states, one
%has to identify
%the component vortices and demonstrate
% that they are actually bound to one another.
vortex cores in a spinor condensate are not empty like the vortices
in an ordinary condensate; they have nearly the same density as the
rest of the condensate\cite{Isohmi}.  One thus
has to measure something about the spins
to observe
the vortices.  Two possibilities have already been invented.
One can either measure the  magnetization field
as in \cite{tiedye} or 
use Stern-Gerlach separation to measure the density of the different
spin species.

Measuring the magnetization as a function of position
 is less informative for a \emph{nonmagnetic}
phase like the cyclic phase than for the ferromagnetic phase
studied in Ref. \cite{tiedye}.  The magnetization outside the
core of a vortex, where the spinor state is approximately a rotation of the
unmagnetized cyclic state
 will be close to zero (see Eq. (\ref{eq:sphere?1})), but
inside the core, where the order parameter leaves the ground-state
space $\mathcal{M}$, the magnetization can be nonzero.  Measuring
the magnetic moment in the core of a vortex helps to determine the
topological charge of the vortex.
(The magnetization
will not provide any \emph{direct} evidence 
of the rotating orientation of the tetrahedral order parameter, though.)
Any vortex one might have to identify involves a rotation about
an arbitrary axis $\bm{\hat{n}'}$ as in Eq. (\ref{eq:lathes}), or  Eq. 
(\ref{eq:lathes2}) which is more convenient for understanding
what a vortex will look like.   The latter description
starts with a vortex whose rotation axis is special--say it 
is parallel to
$\bm{\hat{z}}$, and applies some overall rotation to it.

For example a vortex
of type $(R,\frac{2\pi}{3})$ can be obtained from a vortex whose
axis is
$\bm{\hat{n}}=-\hat{z}$.
Eq. (\ref{eq:lathes2}) implies that the vortex is described by
\begin{eqnarray}
\psi(r,\phi)&=&D(R)e^{\frac{i}{3}(1+F_z)\phi}
\sqrt{n_0}\bigg(f(r)\sqrt{\frac{1}{3}},0,0,g(r)\sqrt{\frac{2}{3}},0\bigg)^T\nonumber\\
&=&D(R)\sqrt{n_0}\bigg(f(r)e^{i\phi}\sqrt{\frac{1}{3}},0,0,g(r)\sqrt{\frac{2}{3}},0
\bigg)^T,
\label{eq:bump}
\end{eqnarray}
where $f(r)$ and $g(r)$ are appropriate functions approaching $1$ at
infinity and $R$ is a rotation that moves $\bm{\hat{n}}$ to
$\bm{\hat{n'}}$. (The phase, $\xi$, does not matter.)
If $\bm{\hat{n'}}=\bm{\hat{n}}=-\bm{\hat{z}}$, then $R$ is the identity,
so $m_x(r)=m_y(r)=0$ and
\begin{eqnarray}
m_z(r,\phi)&=&\psi(r,\phi)^{\dagger}
F_z\psi(r,\phi)\nonumber\\
&=&\frac{2}{3}[f(r)^2-g(r)^2]n_0.
\label{eq:magnetization}
\end{eqnarray}
The magnetization is parallel to the symmetry axis and is given
by 
$\bm{m}=\frac{2}{3}n_0[g(r)^2-f(r)^2]\bm{\hat{n}}$.
Applying an arbitrary reorientation $R$ changes the magnetization
axis and the symmetry axis in the same way, so the general
result is
\begin{equation}
\bm{m}=\frac{2}{3}n_0[g(r)^2-f(r)^2]\bm{\hat{n}'}.\label{eq:coremag}
\end{equation}
%The magnetization for an arbitrary $R$ 
%can be found using Eq. (\ref{eq:transfnrule}); 
%one finds that $\bm{m}(r)=R(0,0,\frac{2}{3}[f(r)^2-g(r)^2])^T$,
%or\footnote{Interestingly, the direction
%of the magnetization, $\bm{\hat{n}}$ gives
%information about the orientations of the tetrahedra
%in the vicinity of the core, since $\bm{\hat{n}}$ points toward
%the vertex that is fixed as the tetrahedra are being rotated.}
%(using Eq. (\ref{eq:anotherpictureonanotherwall}))
%\begin{equation}
%\bm{m}(r)=\frac{2}{3}[g(r)^2-f(r)^2]\hat{n}.
%\label{eq:coremag}
%\end{equation}
Far from the core, the magnetization vanishes.  Inside the core,
the magnetization can be found by noting that
the top component of the vortex Eq. (\ref{eq:bump}) must vanish at $r=0$
in order 
to be continuous:
\begin{equation}
 f(0)=0.
\label{eq:dimple}
\end{equation}
Since $\alpha$ is much larger than $\beta$ and $\gamma$, the
density of atoms will be almost uniform across the whole vortex and
hence $\frac{1}{3}f(r)^2+\frac{2}{3}g(r)^2\approx 1$.  
Eq. (\ref{eq:dimple}) therefore implies
\begin{equation}
g(0)\approx\sqrt{\frac{3}{2}}.
\label{eq:gbump}
\end{equation} 
Hence the magnetization, Eq. (\ref{eq:coremag}), is approximately
$n_0\bm{\hat{n}'}$ in the core; the atoms have a single
unit of hypefine spin in the direction of the vector from the center
to the \emph{fixed} vertex of the rotating tetrahedra near vortex
the core\footnote{Hence, the direction
of the magnetization, $\bm{\hat{n}}$, helps determine the orientation
of the
tetrahedra near each vortex core, in spite of Eq. 
(\ref{eq:sphere?}).  If there are several vortices
nearby one can try to guess how the tetrahedron fields around them
fit together.}.  The inverse vortex, $(R^{-1},-\frac{2\pi}{3})$,
has the same magnetization (it does not change sign).  On
the other hand, similar arguments show that $(R,-\frac{4\pi}{3})$
will have a magnetization approximately equal to $-2n_0\bm{\hat{n}}$
at the core center because it is the $m=-1$ component of the spinor in
the analogue of Eq. (\ref{eq:bump}) that has
the phase winding for this case.  The third stable vortex type, $(A,0)$,
will not have any magnetization in its center and would be hard to
see using this method.  Measuring
the magnetization reveals vortices of order three
but does not distinguish between vortices
and antivortices
and does not even indicate the presence of an order two vortex. (One
\emph{can} observe the composite
vortex described in Section \ref{subsec:double}.)  Another deficiency is that
the cores are only about $1\ \mu$m across, so the vortices might be hard to
observe directly by this method.  However, one could first
allow the condensate
to expand in the transverse direction so that the atomic interactions
decrease.  The vortex cores would expand;
as in experimental observations of vortices
in single-component condensates, the depleted region in $f$ or $g$ 
(whichever corresponds to the component of 
the transformed spinor with the phase winding)
would fly apart and the magnetized core would become much larger. A magnetized
ring would form at the edge of the core where the atoms of one magnetization
accumulate more than the atoms of the other.

\begin{figure}
\psfrag{25}{.25}
\psfrag{5}{.5}
\psfrag{zero}{0}
\psfrag{2pi}{$2\pi$}
\psfrag{1}{$n_1$}
\psfrag{-2}{$n_{-2}$}
\psfrag{0}{$n_0$}
\psfrag{2}{$n_2$}
\psfrag{-1}{$n_{-1}$}
\psfrag{phi}{$\phi$}
\includegraphics[width=.45\textwidth]{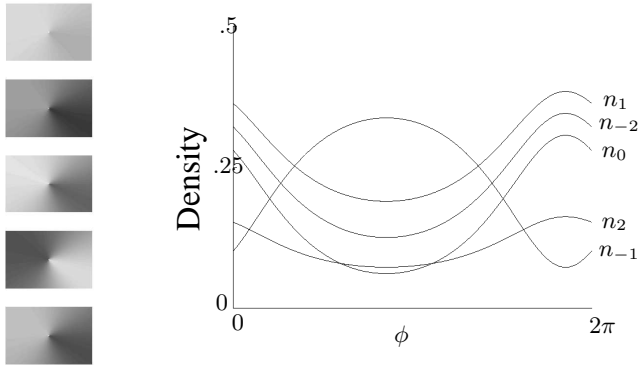}
\caption{
\label{fig:shed}Densities in the five spinor components around
an order $3$ vortex, with a randomly oriented local axis $\bm{\hat{n}'}$.  
On the left is the pattern one might
observe experimentally; the lighter regions correspond
to regions with fewer atoms.  On the right are plotted the percentage
of atoms for each value of $F_z$ at some fixed distance from the
vortex core.  The phases and amplitudes of these oscillations
should help to determine the direction of the local axis.}
\end{figure}

The Stern-Gerlach method gives more information about the
vortices.  Though the density is depleted at the center,
the field \emph{around}
a vortex in a single-component condensate
is not observable, unless one reconstructs the
phase variation of the condensate, perhaps using the
technique described in Ref. \cite{meiserreconstruction}. 
But in a spinor condensate, the spin vortices produce observable
patterns in the condensates' Stern-Gerlach images.  These images
capture separately the density
of atoms in each of the five components of the spinor as functions of position.
In these density profiles 
each vortex (aside from pure phase vortices with $g= id$)
will be ornamented by radiating 
density ripples as illustrated in
Fig. \ref{fig:shed}.
For example, according to Eq. (\ref{eq:bump}), the density
of atoms with $F_z=m$ is given by
\begin{eqnarray}
\frac{n(r,\phi,m)}{n_0}&=&|D_{m2}(R)\sqrt{\frac{1}{3}}f(r)e^{i\phi}+
D_{m,-1}(R)\sqrt{\frac{2}{3}}g(r)|^2\nonumber\\
&=&a_m+b_m\cos(\phi-\phi_m).
\label{eq:joseph}
\end{eqnarray}
where $a_m,b_m$ are constants outside the vortex cores, since $f(r)$
and $g(r)$ approach 1.
While a vortex in a condensate of a single type of atom
does not show any density modulation (unless the
condensate interferes with a second condensate, see e.g. Ref.
\cite{vortexfringe}), angular density ripples 
do result for a spinor vortex as a result of the interference
between the $f$ and $g$ components of the spinor produced by of the
unitary transformation changing the quantization axis
from the 
vortex's rotation axis
$\bm{\hat{n}'}$ to the magnetic field direction.
If
the $\bm{\hat{n}'}$ axis happens to line up
exactly with the axis of the Stern-Gerlach field, then
there are no radial ``interference fringes," but only the axes of
\emph{point}
vortices with aligned charges will tend to line up 
with $\mathbf{B}$.  This is illustrated by 
the qualitative wave function in Section \ref{sec:lavoisier}. 
(See Eqs. (\ref{eq:leaningtower}), (\ref{eq:ofpisa}).)

Both the order three and order two vortices will be visible based
on the images of the five spin components.
One can determine the types of the vortices and their
axes $\bm{\hat{n}'}$ (which are encoded in $D(R)$) 
from the average magnitude of the densities
$a_m$ together with
the amplitudes $b_m$ and offsets $\phi_m$ of the density modulations.
(An order
$2$ vortex will have $\cos2\phi$ and $\sin2\phi$ Fourier modes in addition
to the terms given in Eq. (\ref{eq:joseph}).)  A possible difficulty
with this method arises because,
once the five spin components are separated in space, 
the density oscillations
in each of them are no longer stable.  The ensuing dynamics
in the clouds could mix the atoms up.

Distinguishing among vortices with the same
rotation but different phase winding numbers
$\theta$ is not possible with this method without
resolving the cores. For example, the vortices $(R,\frac{2\pi}{3})$
and $(R,-\frac{4\pi}{3})$ have the same density patterns,
since they differ only by an overall phase $e^{i\phi}$.
%(With enough resolution (submicron)
%the functions $f(r)$ and $g(r)$ can maybe distinguish between the two
%types since $f(0)=0$ for $(R,\frac{2\pi}{3})$ and $g(0)=0$ for 
%$(R,-\frac{4\pi}{3})$.)
%Since the Stern-Gerlach method
%is destructive anyway, one can imagine manipulating the condensate
%by interfering it with other condensates or applying magnetic fields 
%or $RF$ fields to make more details of the vortices show up. 

One would also hope to check some predictions about the size
and charges of the bound states.
One can select
 clusters of vortices in an image of the condensate
(if there are not too many vortices) and use the methods just
discussed to identify the vortex charges and check that each
cluster satisfies Eq. (\ref{eq:screened}).
Additionally,
a sign that the vortex clusters are actually
 \emph{bound} states is that
the bound state size depends in the right way on the magnetic field.
Now atoms whose ground state is cyclic may be difficult to find ($^{87}$Rb 
is likely to be polar\cite{chang1and2}, 
though it may be possible to adjust the interaction
parameters by applying light fields.).  The general
considerations of this article also apply to spin 3 condensates 
(see Ref. \cite{griesmaier05}), as
well as to spin 1 condensates and pseudospin $\frac{1}{2}$ condensates
as already studied by 
\cite{isoshima,kasamatsu}.  So we will
estimate the bound state sizes
for the more generic phases as well
as for the cyclic phase.
The cyclic phase is unique because 
the misalignment energy is a second order effect (see Sec. \ref{sec:cubeshape}).
For other phases where the misalignment energy is a first order effect of
the Zeeman energy, the effective potential would be on the order of $qn_0$.
Hence 
\begin{equation}
L_q\sim\frac{\hbar}{\sqrt{mq}}\mathrm{\ for\ phases\ with\ }V_{eff}\propto q.
\label{eq:stbernard}
\end{equation}
In contrast, for the cyclic phase, Eq. (\ref{eq:collie})
 can be expressed in terms of the scattering
lengths as
\begin{equation}
L_q\sim\sqrt{n_0\Delta a}\frac{\hbar^2}{mq}\mathrm{\ for\ cyclic\ phase}
\label{eq:babycollie}
\end{equation}
Since $q=\frac{\mu_B^2B^2}{8A_{HF}}$, the size of the 
molecules in the cyclic phase is proportional to $\frac{1}{B^2}$ and
the size of molecules in phases with $V_{eff}\propto q$
is proportional to $\frac{1}{B}$.

The size of the condensate must be large enough hold an entire vortex molecule.
Substituting $B_{Max}$
from Eq. (\ref{eq:minimumfield}) into Eqs. (\ref{eq:stbernard})
and (\ref{eq:babycollie})
one finds that $L_q(B_{Max})\sim\frac{1}{\sqrt{\Delta a n_0}}\sim a_c$ 
(for either
phase).  This size is the magnetic healing length of the
condensates (and the size of a vortex core)
and is on the order of $1\ \mu$m. The condensate
should be narrow in one direction (so that the
behavior of the order parameter is two-dimensional) but
at least several times wider than the magnetic healing length in
the other two directions;  in order to measure the field dependence of
the molecule sizes,
one should be able to decrease the magnetic field by some
factor below $B_{Max}$ without the molecules leaving the
condensate.
%In order to observe composite vortices, the magnetic field should be
%strong enough to make the composite vortex size $L$ small compared to the width
%of the condensate.  At a magnetic field of $.1\ $G, the size of a molecule
%is several microns.  

%COMMENT THAT CONDENSATE HAS TO BE SOMEWHAT WIDER THAN MAGNETIC HEALING
%LENGTH SO THAT THERE IS A RANGE OF BOUND STATE SIZES

\begin{figure}
\includegraphics[width=.45\textwidth]{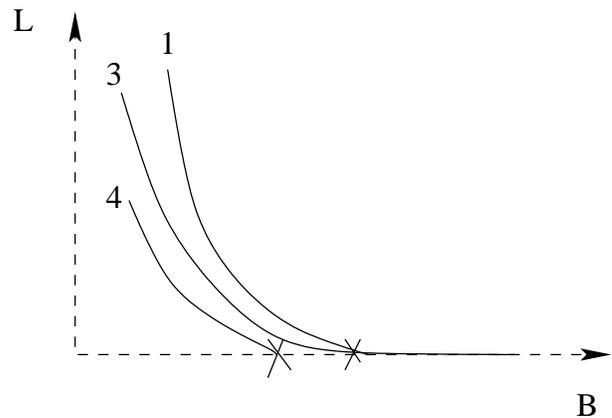}
\caption{\label{fig:whiskers} The evolution of vortex molecules as the magnetic
field is increased.  The three curves illustrate how the sizes of the molecules
from Table \ref{table:ledger}, for $c>4$, might change as the strength of the
magnetic field is increased. The sizes decrease as $\frac{1}{B^2}$. At
a certain magnetic field, an absolutely stable vortex will become
rotationally symmetric and the component cores will coincide (molecule 3).
Metastable vortices will become unstable when a certain magnetic field
is reached, indicated by the x's terminating the curves corresponding
to molecules $1$ and $4$.}
\end{figure}

As a side-comment, vortex molecules probably undergo
transitions at fields close to $B_{Max}$
(see Fig. \ref{fig:whiskers}) since the component vortices overlap
at larger fields.
Absolutely stable vortex molecules, like example 3
in Table \ref{table:ledger}, will be compressed so that the
cores coincide and the vortex becomes
rotationally symmetric at a finite field. 
Once the components' cores overlap a little bit, 
being slightly offset might not lead to significant savings in
kinetic energy. On the other hand, when the vortices in
a \emph{metastable} molecule
are squeezed together, they form an unstable tetrahedral vortex.
Metastability occurs only when the ``Coulomb" force keeps the component
vortices apart.

\section{Conclusion}
We have shown that vortex molecules can be understood reasonably
well based only on simple group theory and rough energy estimates.
Some of these vortex molecules are actually metastable, and
we can study their possible break-up ``channels," reminiscent
of some of the decay processes in nuclear physics.
(In practice, the molecules will probably escape through 
the surface of the condensate before any kind
of ``ultracold fusion" can happen!)

More accurate calculations of the vortex fields and energies could
address other interesting questions.  As is well-known, unlike
the Coulomb interaction between vortices in an ordinary scalar
superfluid, the interaction energy cannot be written
as a sum of two-vortex interaction terms,
as indicated by our estimate in terms of the energy index.  
A more accurate understanding of the kinetic energy landscape might show
that the vortices in a molecule can have several 
spatial arrangements in the core.
Another problem that requires more detailed calculations of vortex
fields is determining whether a vortex with charge
$(R,-\frac{4\pi}{3})$ is stable: the energy index
estimate shows that it can break up into vortices 
with only a finite change in energy, but whether the
energy \emph{increases} or \emph{decreases} 
is not clear yet.

\emph{Acknowledgments} We wish to thank R. Barnett, M. Greiner,
H. Y. Kee, J. Moore, K. Sengstock, 
D. Stamper-Kurn,  and W. P. Wong for useful conversations.
We also acknowledge financial support from CUA, DARPA, MURI, AFOSR 
and NSF grant 0705472.
\appendix
\section{\label{sec:feelingofdealing}Appendix:  Finding the Effective Action}
Eq. (\ref{eq:rotatedfield}) is not as difficult to minimize as it appears,
because of the special symmetry of $\chi_2$. We must substitute
$\tilde{\psi}=\sqrt{n_0}\chi_2+\delta\tilde{\psi}$, where
\begin{equation}
\delta\tilde{\psi}=d\chi_2+aF_x\chi_2+bF_y\chi_2+cF_z\chi_2+(e+if)\chi_{2t}
\label{eq:tilde}
\end{equation}
into the energy, Eq. (\ref{eq:rotatedfield}).  The perturbation 
$\delta\tilde{\psi}$ is the deformation of the tetrahedron, measured
relative to its body axes.
Let us figure out how many powers of the coefficients $a,b,\dots$ to
keep at each stage of calculating $V$.  It helps to
complete the square in Eq. (\ref{eq:rotatedfield})  to get
\begin{alignat}{1}
V_{tot}(\psi)=\frac{1}{2}\alpha&(\tilde{\psi}^{\dagger}\tilde{\psi}-n_0)^2+
\frac{1}{2}\beta(\tilde{\psi}^{\dagger}\mathbf{F}\tilde{\psi})^2
+\frac{1}{2}\gamma|\tilde{\psi}_t^{\dagger}\tilde{\psi}|^2\nonumber\\
&-\frac{1}{2}\alpha n_0^2
-q\sum_{i,j=1}^3\cos\alpha_i\cos\alpha_j\tilde{\psi}^{\dagger}F_iF_j\tilde{\psi}
,\label{eq:rotatedfield2}
\end{alignat}
where we note that the chemical potential for the cyclic
state is $\mu=\alpha n_0$ and define $\gamma=c\beta$.  We also use
$F_{1,2,3}$ to stand for $F_{x,y,z}$.  We need to find the minimum
of this energy only to quadratic order in $q$. At the end
we will find that $a,b,c,\dots$ are each \emph{linear} in $q$. 
Since each
of the squared quantities in $V$ vanishes when $a,b,c,\dots=0$,
just the linear contributions
from $a,b,c,\dots$ give the potential to
quadratic order in $q$.
The quadratic Zeeman term, since it is multiplied by $q$, 
also is not needed 
beyond linear order in $a,b,c,d,e,f$.

Next find which matrix elements of $\chi_2$
need to be calculated to evaluate
all these contributions to the energy.
Only the cross terms between the unperturbed part 
$\sqrt{n_0}\chi_2$ and the perturbation give linear functions of $a,b,c,d,e,f$.  For example, one cross-term contained in the quadratic
Zeeman contribution is
\begin{alignat}{1}
\tilde{\psi}^{\dagger}&F_xF_y\tilde{\psi}\approx n_0\chi_2^{\dagger} F_xF_y\chi_2
\nonumber\\
&+2\sqrt{n_0}\Re \chi_2^{\dagger}F_xF_y[(aF_x+bF_y+cF_z+d)\chi_2\nonumber\\
&\ \ \ \ \ \ \ \ \ \ \ \ \ \ \ \ \ \ \ \ \ \ \ \ +(e+if)\chi_{2t}]
.\end{alignat}
Expanding this gives a sum of matrix elements such as 
$\chi_2^{\dagger}F_xF_y\chi_2$ and $\chi_{2t}^{\dagger}F_xF_yF_z\chi_2$. 
We need only the matrix elements of products of at most three $F$'s.
Many of these  (e.g., $\chi_{2}^{\dagger}F_iF_jF_k\chi_2$
when $i,j,k$ are not all different, and $\chi_{2t}^{\dagger}F_iF_j\chi_2$
when $i$ and $j$ are different)
are equal to zero
because of the $180^{\circ}$ symmetries of $\chi_2$ around the coordinate
axes. The numerical values of the few remaining ones can be worked out quickly.
Using these matrix elements to
calculate all the terms in Eq. (\ref{eq:rotatedfield2}) produces 
an expression $V_{tot}(a,b,c,d,e,f,\cos\alpha_1,\cos\alpha_2,\cos\alpha_3)$.
Along the way, one notices that each of the variables
$a,b,c,\dots$ contributes to only one term in the $q=0$ potential (the
first line of Eq. (\ref{eq:rotatedfield2})).
The variables $a,b,c$ determine the magnetization, $d$ determines the density
perturbation and $e$ and $f$ determine the singlet-amplitude $\theta$. E.g.,
\begin{eqnarray}
&&n=n_0+2\sqrt{n_0}d\nonumber\\
&&M_x=4\sqrt{n_0}a\label{eq:nah}\\
&&\Re [\theta]=2\sqrt{n_0}e\nonumber.
\end{eqnarray}

Finally, minimize the potential.
It can be written as a sum of independent
quadratic functions of $a,b,c,d,e,$ and $f$:
\begin{alignat}{1}
V&=(-\frac{1}{2}\alpha n_0^2+2qn_0)+2\alpha n_0 d^2
-4\sqrt{n_0}qd\nonumber\\
&\ \ \ \ +8\beta n_0(a^2+b^2+c^2)+4\sqrt{3n_0}q(a\cos\alpha_2\cos\alpha_3\nonumber\\
&\ \ \ \ \ \ \ \ \ \ +b\cos\alpha_1\cos\alpha_3+c\cos\alpha_1\cos\alpha_2)\nonumber\\
&\ \ \ \ +2\gamma n_0(e^2+f^2)\nonumber\\
&\ \ \ \ \ \ \ \ \ \ +2q\sqrt{n_0}[e(\cos^2\alpha_1+\cos^2\alpha_2-2\cos^2\alpha_3)\nonumber\\
&\ \ \ \ \ \ \ \ \ \ +\sqrt{3}f(\cos^2\alpha_1-\cos^2\alpha_2)].
\label{eq:boardmembers}
\end{alignat}
(Note that the second term, $2qn_0$, is the first-order contribution of
the Zeeman energy, 
which is independent
of orientation.)

Minimizing each quadratic (which gives 
$a=-\frac{\sqrt{3}q}{4\beta\sqrt{n_0}}\cos\alpha_2\cos\alpha_3,\dots$)
and combining the results together with the 
help of Eq. (\ref{eq:pythagoras})
%of the minima are $a=-\frac{q}{2\beta}\sqrt{\frac{3}{2n_0}},
%d=\frac{q}{\alpha\sqrt{n_0}},
%f=-\frac{q}{2\gamma\sqrt{n_0}}(\cos^2\alpha_1-\cos^2\alpha_2),\dots$),
%and combining the results
%together with the help of
%$\cos^2\alpha_1+\cos^2\alpha_2+\cos^\alpha_3=1$ gives:
gives
\begin{alignat}{1}
V_{eff}=-\frac{1}{2}&\alpha n_0^2+2qn_0-2\frac{q^2}{\alpha}+\frac{q^2}{\gamma}
-\frac{3q^2}{4\beta}\nonumber\\
&+(\frac{3q^2}{4\beta}-\frac{3q^2}{\gamma})
(\cos\alpha_1^4+\cos\alpha_2^4+\cos\alpha_3^4);
\end{alignat}
all the constant terms can be dropped to give Eq. (\ref{eq:effective}).
Note that the magnetization varies with the orientation of the tetrahedron
(as can be checked by substituting the optimal
values for $a,b,c$ into the magnetization, Eq. (\ref{eq:nah})).  In particular,
the $c>4$ ground state with 
$\cos\alpha_1,\cos\alpha_2,\cos\alpha_3=\pm\frac{1}{\sqrt{3}}$ has a 
small magnetization, $\bm{m}=\mp\frac{q}{\beta\sqrt{3}}(1,1,1)$; since
this has been calculated from
$\tilde{\psi}$, it is the magnetization
\emph{relative to
the body axes} tetrahedron.  Comparing this to the magnetic field relative
to the body axes,
$B(\cos\alpha_1,\cos\alpha_2,\cos\alpha_3)=\frac{B}{\sqrt{3}}(1,1,1)$, shows
that
the state will become magnetized either parallel
or antiparallel to the magnetic field.

The point of the effective potential
is that it allows us to eliminate the $6$ most rigid
degrees of freedom corresponding
to $a,b,c,d,e,$ and $f$; then it is easier to understand vortex textures
by concentrating on the rephasing and rotation angles
as a function of position. The wavefunction in Eq. (\ref{eq:Manifold})
can be parameterized in terms of Euler angles for the rotation, e.g.,
$\psi=\sqrt{n_0}e^{i\sigma F_z}e^{i\tau F_x}e^{i\rho F_z}e^{i\theta}\chi_2$.
The first angle, $\sigma$, does not come into the angles $\alpha_i$ that
describe the orientation of the magnetic field ($B\bm{\hat{z}}$)
relative to the
tetrahedron.  (The tetrahedron can be rotated around the $z$-axis
without changing these angles.)  One can check that $\cos\alpha_1=\sin\tau\sin\rho,\cos\alpha_2=-\sin\tau\cos\rho,$ and $\cos\alpha_3=\cos\tau$.
Now, working out the kinetic energy and combining it with the effective
potential gives the ``phase-and-rotation-only" energy functional
\begin{alignat}{1}
\mathcal{E}_{eff}&=\iint d^2\mathbf{r} \frac{n_0\hbar^2}{2m}[2(\nabla\rho)^2+2
(\nabla\tau)^2+2(\nabla\sigma)^2+\nonumber\\
&\ \ \ \ \ \ (\nabla\theta)^2+
4\cos\tau\nabla\sigma\cdot\nabla\rho]\nonumber\\
&+(c-4)\frac{3q^2}{4c\beta}[\cos^4\tau+\sin^4\tau(\sin^4\rho+\cos^4\rho)].
\label{eq:anotherlandslide}
\end{alignat}
This can be solved (in principle) to give the textures
around sets of vortices and the relative positions of the
vortices in equilibrium. Each vortex type implies a certain type
of discontinuity in the four angles as the core is encircled.  
This expression does not
seem too easy to use, but at least
it shows just the two effects we have been balancing
against one another (kinetic and anisotropy energy).  
The size of a vortex molecule can be estimated by assuming that
the two terms are comparable, 
$\frac{n_0\hbar^2}{mL_q^2}\sim|c-4|\frac{q^2}{c\beta}$.  If distances
are rescaled by $L_q$, we then find that the energy function has a form
that depends only on the \emph{sign} of $c-4$.  Therefore,
in a molecule with three vortices, the angles of the
triangle they form will be independent of all
the parameters, including $c$, even though it is dimensionless.

Eq. (\ref{eq:anotherlandslide}) is derived from Eq. 
(\ref{eq:howtousebatteries}) by determining how the tetrahedra are
distorted by the quadratic Zeeman effect. 
But the kinetic
energy also causes distortions of the wave function from the perfect tetrahedral
forms, and it seems possible that these distortions could lead to
kinetic effects in the anisotropy term and anisotropy effects in the kinetic
energy term.  However, at the lowest order, treating the two terms independently
seems correct.
A simple argument for this (neglecting the kinetic energy when finding the
anisotropy potential) is that the
distortion due to the Zeeman term is linear in $q$ while
the distortion due to the kinetic energy is quadratic in $q$.
To see this, think of an ordinary
scalar vortex, where the density varies as $n_0(1-\frac{a_c^2}{r^2})$ far
from the core\cite{pethickbook}. The amount of ``distortion" is 
$-\frac{a_c^2}{r^2}n_0$.  In the cyclic state, distortion (i.e.,
perturbations to the spinor components
that take it out of $\mathcal{M}$) implies
changes in the magnetization as well as the density.  But we may
assume that these distortions are still of 
order $\frac{a_c^2}{r^2}n_0$.
The majority of the ``pulp" in
a molecule's core consists of points whose distance
is of order $L_q$ from the actual vortex cores (the ``seeds"), 
so the amount
of distortion can be found by substituting $r=L_q$ from Eq. 
(\ref{eq:betterlate}).
Using the relation between $a_c$ and $\beta$ shows that the
fractional distortion in the ``pulp" regions is of order 
$\frac{a_c^2}{r^2}n_0\sim(\frac{q}{n_0\beta})^2$, 
to be compared with the deformations
of order $\frac{q}{\beta n_0}$ that result from minimizing Eq. 
(\ref{eq:boardmembers}).  
%A more rigorous approach is to express $\psi(x,y)$
%in terms of $a(x,y),b(x,y),c(x,y),d(x,y),e(x,y),f(x,y)$ 
%and an arbitrary position-varying rotation and rephasing
%in the Gross-Pitaevskii equations of a spin 2 condensate. 
%If $a,b,c,\dots$ can be determined
%locally in terms of the rotation and rephasing, by neglecting the
%kinetic energy, then these variables may be eliminated, giving the
%simplified model described by Eq. (\ref{eq:}).
\section{\label{app:catstring}Appendix: Noncommutativity of vortex charges}

%G2's charge doesn't change, because nothing passed under it and got
%tangled in its tether!

%Using relation between body-axis n and regular nG2's charge doesn't change, because nothing passed under it and got
%tangled in its tether!

To give a complete description of charge conservation when the charges
are described by the noncommuting rotations of a tetrahedron, one
needs to give a rule for how to multiply the charges of a set of vortices
together to get the net charge.  %(In order to be consistent
%we will have to make some other conventions as well.)
A convention we used is to multiply the 
topological charges together in order of increasing x-coordinates.  

\begin{figure*}
\psfrag{O}{$O$}
\psfrag{G1}{$\Gamma_1$}
\psfrag{1}{1}
\psfrag{2}{2}
\psfrag{3}{3}
\psfrag{G2}{$\Gamma_2$}
\psfrag{G3}{$\Gamma_3$}
\psfrag{G2'}{$\Gamma_1^{-1}\Gamma_2\Gamma_1$}
\includegraphics[width=\textwidth]{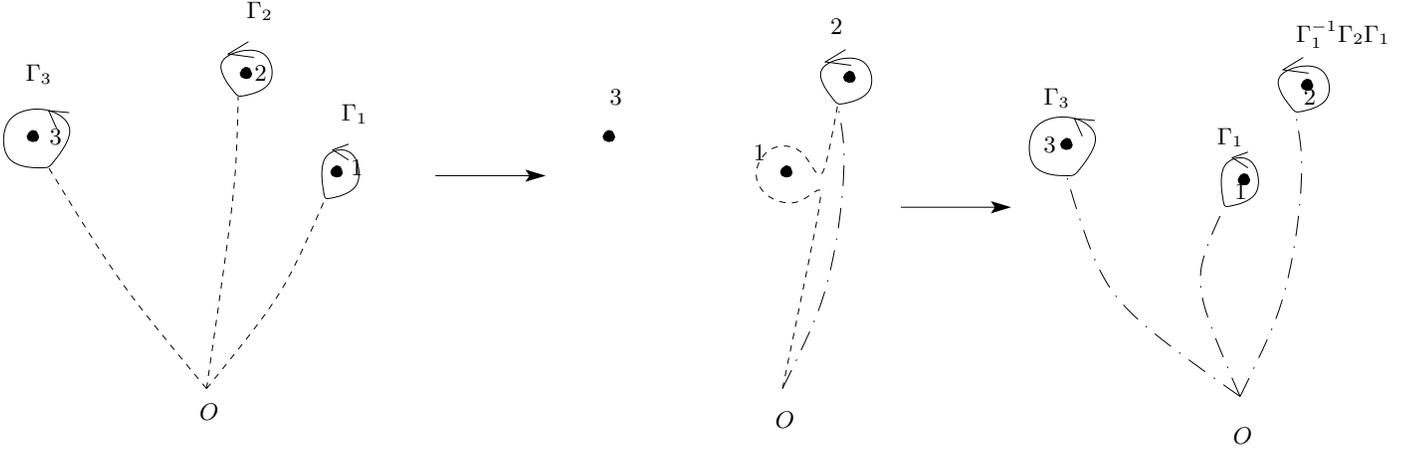}
\caption{\label{fig:marigolds}%Vortex charges change but their net
%charge does not when the vortices are rearranged.  This can be
a) The convention for assigning vortex charges.
For a set of vortices, tethers are drawn
directly from the origin to an anchor just below each vortex.
As long as the tethers are moved continuously, the correspondence
between the tetrahedral state at the anchor and the standard
tetrahedral state does not change.
b,c) These show what happens when vortex 1 is moved below
vortex 2.
The dashed lines in a)
and the dash-dot-dash lines in c)
are the tethers before and after vortex 1 moves.  Part b) focuses
on the tether of vortex 2 , showing how the original tether gets pushed to
the side by vortex 1 and is replaced by a new tether.  Continuing
the labelling of the tetrahedron vertices around the original
path changes the labelling of the tetrahedron just below
vortex $2$.  Hence
the charge of vortex 2 is identified differently in c).
}
\end{figure*}

It seems that this definition has an awkward consequence:
does the net
vortex charge jump suddenly when two of the vortices
are reordered, because of the noncommutativity
of the group of charges?  In
Fig. \ref{fig:marigolds} vortices $1$ and $2$ are interchanged between
frames a) and c), which suggests that the net charge
changes from $\Gamma_3\Gamma_2\Gamma_1$ to $\Gamma_3\Gamma_1\Gamma_2$.
But this deduction is incorrect, and
the net vortex charge \emph{is} actually conserved.

The resolution of the paradox has to do with the fact that
the charge of a vortex can
only be determined up to conjugacy, unless one introduces a systematic 
convention.
For example,  the charge
of a $120^{\circ}$-rotation vortex is ambiguous--the rotation
could be
either $P$, $Q$, $R$,
or $S$, and there is no way to distinguish between these because the
four vertices of the tetrahedra are indistinguishable.  (Abstractly speaking,
the four rotations are conjugate elements of the group.) In order to identify
the charge of each vortex, we have to choose a routine for labelling the
vertices of the tetrahedra nearby.  Here is a convention that is consistent
with the rule for ordering the vortex charges.
Take a point $O$ far below all the vortices in the system and
connect it with lines to points just below the vortices (see Fig.
\ref{fig:marigolds}a).
Now identify the \emph{base} tetrahedron at $O$ with
the standard tetrahedron in
Fig. \ref{fig:bottler}a, making a choice from
among the twelve possible ways. The labelling at $O$ can be communicated
to the tetrahedron at the end-point of each line, by copying the labelling
from $O$ to a nearby tetrahedron on the line, and then continuing to copy
the labelling until the end of the line is reached.  Now the charges
of the vortices can be identified by using the labelling of the nearby
tetrahedron to assign a letter to
the rotation axis.  

Now that we have a consistent convention
for assigning vortex charges, we can show that the net charge
of a set of vortices does not change when two of them are interchanged.
The trick is that the charges of the \emph{individual}
 vortices do change in
such a way that the product charge does not change!  Between
Fig. \ref{fig:marigolds}a and Fig. \ref{fig:marigolds}c, vortex
$2$ is moved over vortex $1$.  Because vortex $2$'s tether gets tangled up
with vortex $1$ when vortex $1$ passes below it, 
its charge gets redefined, as $\Gamma_1^{-1}\Gamma_2\Gamma_1$.  The other two
vortices' charges do not change.
The net charge, obtained by multiplying the vortex charges from left
to right, is
\begin{equation}
\Gamma_3\Gamma_1(\Gamma_1^{-1}\Gamma_2\Gamma_1)=\Gamma_3\Gamma_2\Gamma_1.
\nonumber
\end{equation}
Thus the net charge does not change.  On the other hand,
there is a sudden jump in the charge
of vortex $2$, but this does not mean that the fields of tetrahedra are 
changing suddenly; the vortex has just been reclassified, with a vortex
charge that is conjugate to the original charge.

\begin{figure}
\psfrag{A}{$B$}
\psfrag{B}{$B$}
\psfrag{P}{$P$}
\psfrag{P'}{$BPB^{-1}$}
\psfrag{B'}{$B'=P^{-1}BP$}
\includegraphics[width=.45\textwidth]{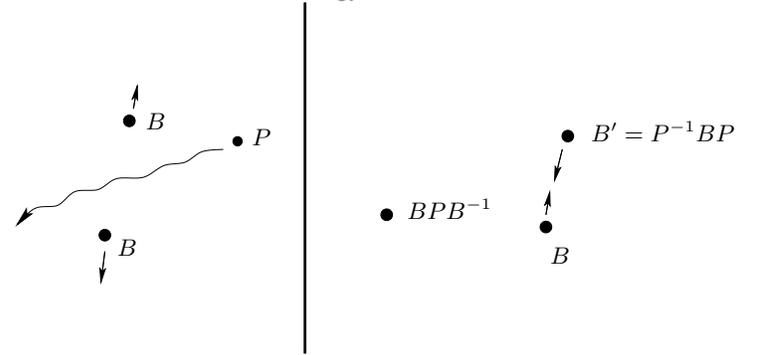}
\caption{\label{fig:halfmagic}
Catalysis by conjugation; the vortex on the right moves between
the other two vortices changing their repulsion to attraction.}
\end{figure}

Here is an interesting consequence of the noncommuting charges: the force between a pair
of vortices
changes from repulsive to attractive if a third vortex 
wanders between them.  In Fig. \ref{fig:halfmagic} one
vortex (the $P$ one)
catalyzes a reaction without touching the other two vortices involved.
(Assume the phases are
$0$ for the two $B$ vortices and $\frac{2\pi}{3}$ for the $P$ vortex.)
To figure out what happens, keep track of when a vortex's connection
to the reference point (below the figure) is interrupted by another
vortex. The charge of the vortex passing underneath is not changed,
and the charge of the vortex on top changes to keep the total charge
the same.  This information is sufficient for working out all the
charges:
When $P$ passes below the $B$ on the right, 
the latter vortex changes to a $B'=P^{-1}BP$, so
that the net charge is still the same, even though $P$ has moved.
(This can be used to work out the charges: the net charge of
the two vortices which have switched has to be the same, so
$PB'=BP$ so $B'=P^{-1}BP$.)
Next, when $P$ passes \emph{above} the $B$ on the left, 
the \emph{former} vortex changes
to a $BPB^{-1}$ vortex.  Now the $SU_2$ matrices for $B$ and $P$
are $-i\sigma_y$ and $\frac{1-i\sigma_x+i\sigma_y+i\sigma_z}{2}$
(using the axes associated with $\chi_2$).
Multiplying out the charges shows that $B'=A^{-1}$.  
The force between the original
pair of vortices (say the $P$ is
far away at the beginning and end) 
is $\frac{n_0\hbar^2 \pi}{m L}(I_E(B^2,0)-I_E(B,0)-I_E(B,0))$ and the force
between the vortices they turn into is 
$\frac{n_0\hbar^2\pi}{mL}(I_E(BA^{-1}),0)-I_E(A^{-1},0)-I_E(B,0))$.
($L$ is the distance between the vortices.) The vortices repel
each other at first, but after $P$'s intervention, they attract each other,
as one sees by checking that $BA^{-1}=C$,
and that $B^2$ is a $2\pi$ rotation with energy index $2$ while the other
reacting charges are all $\pi$ rotations, with index $\frac{1}{2}$.

We have been assuming that $q=0$, but a similar reaction could also
happen in the core of a composite
vortex when $q\neq 0$; that is why one has to check all the possible
ways for vortices to wander between one another before coalescing
or before dividing into groups.
%This is illustrated in Fig. \ref{fig:halfmagic}; the $P$
%vortex (short for $(P,\frac{2\pi}{3})$) starts far on the right,
%passes between the other two vortices, and then goes off to the left.ABP and APB have different energies, and anyway, that's not what happens
%when P moves under B.  The vortices change to A P (P^-1BP).
%Now move P over A.  The vortices change to (APA^{-1}) A (P^{-1}BP)
%(we know this because vortex A doesn't change since nothing goes under
%it, and the total charge has to stay the same.)  The neat thing is that,
%while $A$ and $B$ repelled each other, $A$ and $P^{-1}BP$ attract
%one another,as you can check using energy indices.  So P catalyzes a chemical
%reaction--that's the trick mentioned in Chapter \ref{ch:}.

%\bibliography{bibli}

\end{document}